\documentclass{jfm}
\usepackage{graphicx}
\usepackage{epstopdf,epsfig}
\usepackage{newtxtext}
\usepackage{newtxmath}
\usepackage{natbib}
\usepackage{hyperref}
\hypersetup{
    colorlinks = true,
    urlcolor   = blue,
    citecolor  = black,
}

\newcommand{\RomanNumeralCaps}[1]
\linenumbers
\usepackage{color}
\newcommand{\firstrev}[1]{{\textcolor{black}{#1}}}
\newcommand{\secondrev}[1]{{\textcolor{black}{#1}}}
\newcommand{\thirdrev}[1]{{\textcolor{black}{#1}}}
\newcommand{\fourthrev}[1]{{\textcolor{black}{#1}}}
\usepackage{subcaption}
\usepackage{tikz}

\title{Numerical Simulation of an Idealised Richtmyer-Meshkov Instability Shock Tube Experiment}

\author{Michael Groom\aff{1}
  \corresp{\email{michael.groom@sydney.edu.au}},
 \and Ben Thornber\aff{1}}

\affiliation{\aff{1}School of Aerospace, Mechanical and Mechatronic Engineering, University of Sydney, Sydney, NSW 2006, Australia}

\begin{document}
\maketitle

\begin{abstract} 
The effects of initial conditions on the evolution of the Richtmyer--Meshkov instability (RMI) at early to intermediate time are analysed, using numerical simulations of an idealised version of recent shock tube experiments performed at the University of Arizona (Sewell et al., \emph{J. Fluid Mech.} (2021), \textbf{917}, A41). The experimental results are bracketed by performing both implicit large-eddy simulations (ILES) of the high-Reynolds number limit as well as direct numerical simulations (DNS) at Reynolds numbers lower than those observed in the experiments, both using the \texttt{Flamenco} finite-volume code. Various measures of the mixing layer width $h$, known to scale as $\sim t^\theta$ at late time, based on both the plane-averaged turbulent kinetic energy (TKE) and volume fraction (VF) profiles are used to explore the effects of initial conditions on $\theta$ and are compared with the experimental results. The decay rate $n$ of the total fluctuating kinetic energy is also used to estimate $\theta$ based on a relationship that assumes self-similar growth of the mixing layer. The estimates for $\theta$ range between 0.44 and 0.52 for each of the broadband perturbations considered and are in good agreement with the experimental results. \thirdrev{Decomposing the mixing layer width into separate bubble and spike heights $h_b$ and $h_s$ shows that, while the bubbles and spikes initially grow at different rates, their growth rates $\theta_b$ and $\theta_s$ have equalised by the end of the simulations indicating that the mixing layer is approaching self-similarity.} Anisotropy of the Reynolds stresses is also analysed and is shown to persist throughout each of the simulations. Outer-scale Reynolds numbers and various key length scales are calculated for the DNS cases, showing that fully developed turbulence is not obtained due to the challenges associated with performing DNS for broadband initial conditions. Overall the results demonstrate important differences between broadband and narrowband surface perturbations, as well as persistent effects of finite bandwidth on the growth rate of mixing layers evolving from broadband perturbations. \thirdrev{Good agreement is obtained with the experiments for the different quantities considered, however the results also show that care must be taken when using measurements based on the velocity field to infer properties of the concentration field, as well as when it is appropriate to assume the mixing layer is growing self-similarly with a single growth rate $\theta$.}
\end{abstract}


\section{Introduction}
\label{sec:intro}
This paper analyses the effects of initial conditions on the evolution of the Richtmyer--Meshkov instability (RMI), which occurs when an interface separating two materials of differing densities is accelerated impulsively, typically by an incident shock wave \citep{Richtmyer1960,Meshkov1969}. The instability evolves due to the deposition of baroclinic vorticity at the interface, caused by a misalignment of density and pressure gradients during the shock--interface interaction. This occurs either from surface perturbations on the interface, or when the shock wave is non-uniform or inclined relative to the interface. The baroclinic vorticity that is deposited on the interface leads to the growth of surface perturbations and the development of secondary shear layer instabilities, which drive the transition to a turbulent mixing layer. Unlike the closely related Rayleigh--Taylor instability (RTI), the RMI is induced for both light to heavy and heavy to light configurations. In both cases the initial growth of the interface is linear in time and can be described by analytical expressions \citep{Richtmyer1960,Meyer1972,VDM}. However, as the amplitudes of modes in the perturbation become large with respect to their wavelengths the growth becomes nonlinear, whereby numerical simulation is required to calculate the subsequent evolution of the mixing layer. Another key difference between RTI and RMI is that, for the RMI, baroclinic vorticity is only deposited initially and not continuously generated, compared to the \thirdrev{(classical)} RTI where the interface is continuously accelerated. For a comprehensive and up-to-date review of the literature on both RTI, RMI and the Kelvin-Helmholtz instability (KHI), the reader is referred to \citet{Zhou2017a,Zhou2017b,Zhou2021}, \thirdrev{as well as \cite{livescu2020} for an excellent review on variable-density turbulence more generally.}

The understanding of mixing due to RMI is of great importance in areas such as inertial confinement fusion (ICF) \citep{Lindl2014}, where a spherical capsule containing thermonuclear fuel is imploded using powerful lasers with the aim of compressing the contents to sufficient pressures and temperatures so as to initiate nuclear fusion. The compression is performed using a series of strong shocks, which trigger hydrodynamic instabilities at the ablation front due to capsule defects and drive asymmetries \citep{Clark2016}. The subsequent mixing of ablator material and fuel that ensues can dilute and cool the hotspot, which reduces the overall efficiency of the implosion. As a contrast to ICF, in high-speed combustion such as in a scramjet or rotating detonation engine, RMI due to weak shocks improves the mixing of fuel and oxidiser leading to more efficient combustion \citep{Yang1993,Yang2014}. An understanding of mixing due to RMI is also important for many astrophysical phenomena such as supernovae and the dynamics of interstellar media \citep{Arnett2000}. \thirdrev{Note that in such applications RTI usually occurs alongside RMI and in general it is impossible to separate the effects of both instabilities. However, there is still great value in studying RMI independently, particularly when comparing with shock tube experiments that have been designed to isolate its effects using an RT-stable configuration.}

In the applications mentioned above, the most important statistical quantity one would like to know is typically the mixing layer width, denoted by $h$. At late time $h$ scales as $\sim t^2$ for RTI and $\sim t^\theta$ for RMI where the exponent $\theta\le 1$ has been shown to depend on initial conditions \citep{Youngs2004,Thornber2010}. Various approaches have been taken to define $h$, which fall into one of two categories. The first is to consider the distance between two cutoff locations based on a particular threshold of some spatially-averaged profile in the direction normal to the mixing layer (i.e. the direction of the \firstrev{shock-induced acceleration}). Examples include the visual width \citep{Cook2001} based on the 1\% and 99\% locations of the mean volume fraction profile (the choice of a 1\% threshold is somewhat arbitrary; see \citet{Zhou2019a} for a comparison of different thresholds in the context of RTI). Such measures have the advantage of being easily interpretable but can be sensitive to statistical fluctuations. The second approach is to define an integral measure by integrating a particular spatially-averaged profile in the normal direction, for example the integral width \citep{Andrews1990}. Integral measures are less susceptible to statistical fluctuations but are also less interpretable, as different profiles can give the same integrated value. The recently proposed mixed mass \citep{Zhou2016a} and integral bubble and spike heights \citep{Youngs2020b} are attempts to combine the best aspects of both approaches. 

Over the last few decades, both shock tube experiments and numerical simulations have been performed in order to better understand the fundamentals of RMI, such as the value of $\theta$ at late time.  Previous numerical studies have typically used large-eddy simulation (LES) or implicit LES (ILES) to predict mixing at late time in the high Reynolds number limit \citep{Youngs1994,Hill2006,Thornber2010,Lombardini2012,Tritschler2014,Thornber2017,Soulard2018}. Key findings include the dependence of $\theta$ on the type of surface perturbation used to initiate the instability \citep{Youngs2004,Thornber2010}. Narrowband perturbations, which include only a small, annular band of modes in wavenumber space, have been found to give values of $\theta$ at late-time between 0.25 \citep{Soulard2022} and 0.33 \citep{Youngs2020a} whereas perturbations including additional long wavelength modes, known as broadband perturbations, have been found to give values of $\theta$ as high as 0.75 \citep{Groom2020}. \thirdrev{Studies of the effects of initial conditions in RTI have found similar results for the growth rate $\alpha$ when additional long wavelength modes were included in the initial perturbation \citep{Ramaprabhu2005,Banerjee2009}. When only short wavelength perturbations are present the growth rate of RTI is limited by the nonlinear coupling of saturated short wavelength modes (bubble merger), while additional long wavelength perturbations cause the growth rate to become limited by the amplification and saturation of long wavelength modes (bubble competition). \fourthrev{Futhermore, \cite{Aslangil2020a} considered the case of RTI where the applied acceleration is completely withdrawn after initial development. The resulting mixing layer is closely related to an RMI-induced mixing layer, differing only by the mechanism of the initial acceleration, with the growth rate exponent for narrowband initial conditions shown to be within the bounds of 0.2 to 0.28 suggested by \citet{Weber2013}.}}

Early shock tube experiments made use of membranes to form the initial perturbation between the two gases \citep{Vetter1995}, however these tended to leave fragments that dampened the subsequent instability growth, inhibited mixing and interfered with diagnostics. In order to circumvent this, modern shock tube experiments use membraneless interfaces, for example by forming by a shear layer between counter-flowing gases \citep{Weber2012,Weber2014,Reese2018,Mohaghar2017,Mohaghar2019}, using a gas curtain \citep{Balakumar2008,Balasubramanian2012} or by using loudspeakers to generate Faraday waves at the interface \citep{Jacobs2013,Krivets2017,Sewell2021}. 

These methods of interface generation typically result in the formation of a broadband surface perturbation and as such these experiments have obtained values of $\theta$ that are higher than the 0.25--0.33 expected for narrowband initial conditions. For example \citet{Weber2012,Weber2014} measured $\theta$ in the range 0.43--0.58, while later experiments on the same facility by \citet{Reese2018} obtained $\theta=0.34\pm0.01$ once the concentration field was adjusted to remove larger-scale structures from the mixing layer prior to averaging in the spanwise direction. \citet{Jacobs2013} found that their measurements of mixing layer width prior to reshock could be partitioned into two groups with different power law exponents. The particular diagnostic used was the mixing layer half width, found by taking the distance between the 10\% and 90\% average concentration locations and halving this. Prior to reshock, both groups initially had growth rates close to 0.5 ($\theta=0.51$ and $\theta=0.54$), while at later times the growth rates were smaller but also more different ($\theta=0.38$ and $\theta=0.29$ respectively). \citet{Krivets2017} also found a wide range of $\theta$ for the integral width prior to reshock, ranging from $\theta=0.18$ to $\theta=0.57$, using a similar experimental setup. During these experiments the timing of the arrival of the shock wave relative to the phase of the forcing cycle was not controlled, which resulted in large variations in the initial amplitudes of the perturbation. More recent experiments by \citet{Sewell2021} took this into account and divided the results into a low-amplitude and high-amplitude group. Using a measure for the mixing layer width based on 5\% threshold locations of the turbulent kinetic energy profile, they found $\theta=0.45\pm0.08$ and $\theta=0.51\pm0.04$ for the low- and high-amplitude groups prior to reshock. 

In this paper, both ILES and direct numerical simulations (DNS) are performed of 3D RMI with narrowband and broadband perturbations, using a setup that represents an idealised version of the shock tube experiments performed at the University of Arizona \citep{Jacobs2013,Krivets2017,Sewell2021} to investigate the effects \firstrev{of} long wavelength modes in the initial perturbation. A similar study was performed in \citet{Groom2020} but the main aim in that paper was to approximate the regime where there are always longer and longer wavelength modes in the initial condition that are yet to saturate (referred to as the infinite bandwidth limit). Of primary interest here is to explore the impacts of finite bandwidth broadband perturbations on the mixing layer growth over the length and time scales of a typical shock tube experiment and compare the results with those of both narrowband perturbations and broadband perturbations in the infinite bandwidth limit. While the main aim is not to match the experiments as closely as possible, it is anticipated that the results generated in this study could in principle be verified experimentally. Direct comparisons are also still able to be made through appropriate non-dimensionalisations, which has previously been difficult to do when comparing results between simulations and experiments. An assessment will also be made as to the validity of using measurements based on the velocity field to draw conclusions about the concentration field (and vice versa). 

The paper is organised as follows. In \S\ref{sec:setup}, an overview of the governing equations and numerical methods employed to solve these equations is given, as well as a description of the computational setup and initial conditions. This section also gives a brief discussion on some of the challenges associated with performing DNS with broadband surface perturbations. \S\ref{sec:results} details an analysis of many of the same quantities presented in \citet{Sewell2021}, including turbulent kinetic energy profiles and spectra as well as various measures of the mixing layer width that are used to estimate the growth rate $\theta$. The evolution of key length scales and Reynolds numbers is also given for the DNS cases. Finally, \S\ref{sec:conclusions} gives a \firstrev{summary} of the main findings, as well as directions for future work on this problem.

\begin{figure}
	\centering
	\begin{tikzpicture}
		\node[anchor=south west,inner sep=0] at (0,0) {\includegraphics[width=\textwidth]{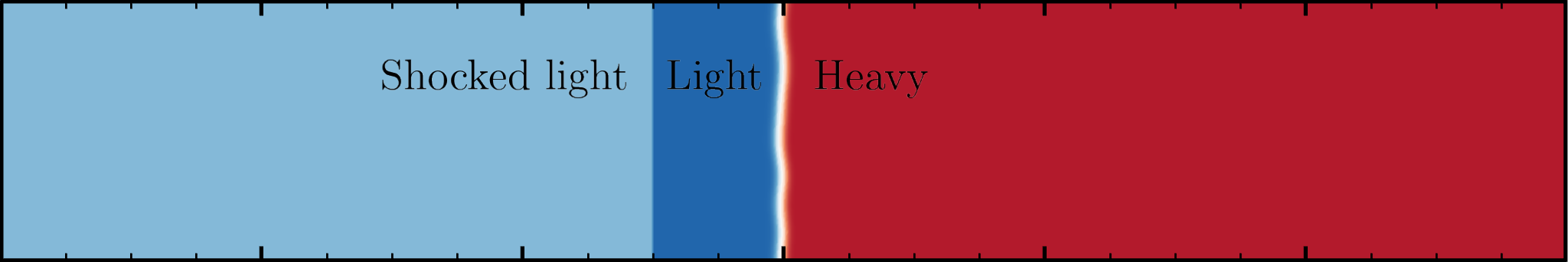}};
		\node[anchor=south west,inner sep=0] at (0.1,0.1) {\includegraphics[width=0.1\textwidth]{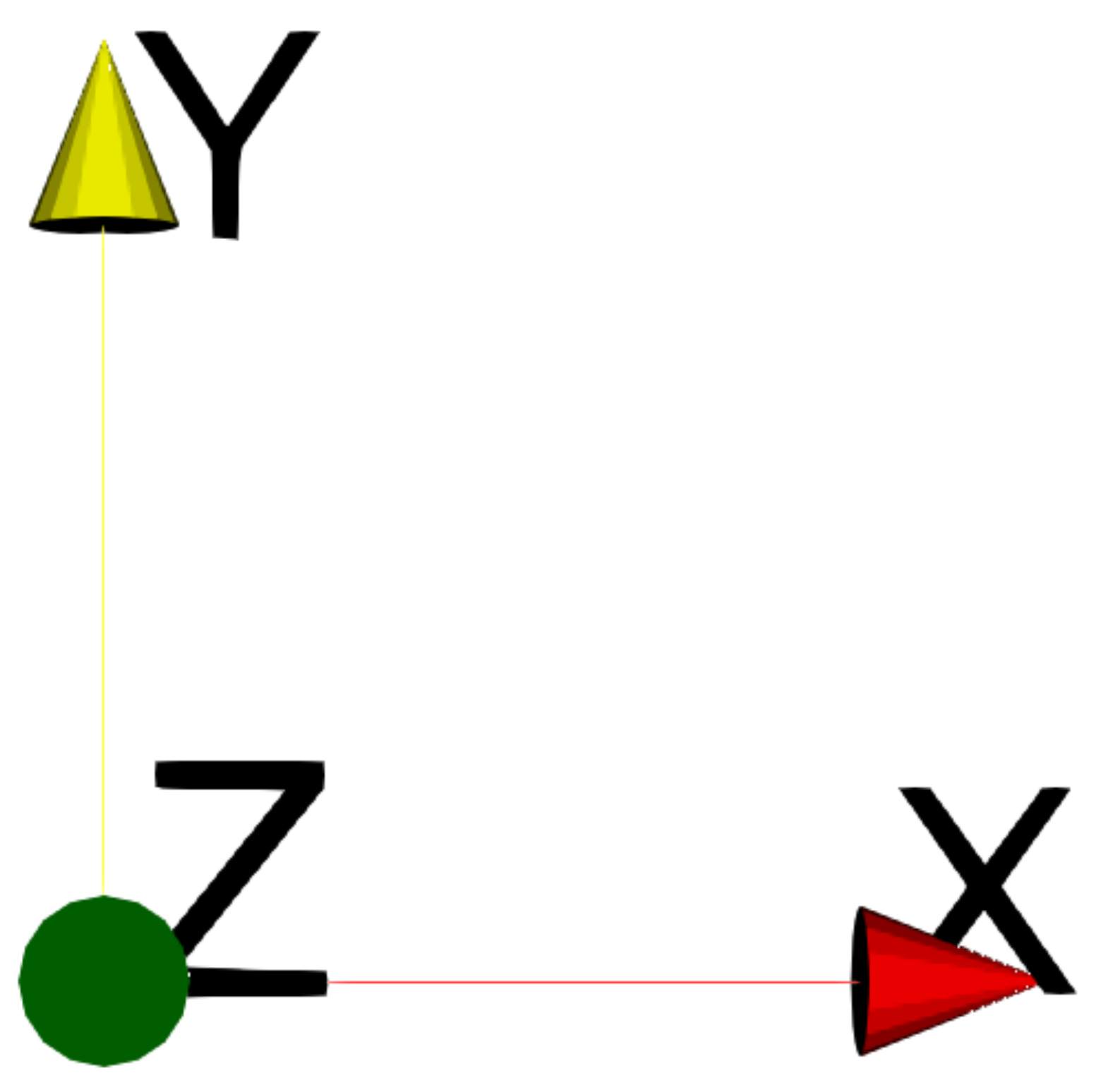}};
	\end{tikzpicture}
	\caption{A schematic of the problem setup. The major ticks correspond to a grid spacing of $\Delta x=1.0$ m. The interface is initially located at $x=3.0$ m and the shock is initially located at $x=2.5$ m in the light fluid and travels from light to heavy. \label{fig:schematic}}
\end{figure}

\section{Computational Setup}
\label{sec:setup}
\subsection{Governing Equations}
\label{subsec:equations}
The computations presented in this paper all solve the compressible Navier--Stokes equations extended to a five-equation, quasi-conservative system of equations based on volume fractions rather than the conventional four-equation, fully-conservative model based on mass fractions for multicomponent flows. This ensures that pressure and temperature equilibrium is maintained across material interfaces when upwind discretisations are used and the ratio of specific heats varies across the interface, as is the case for air and SF$_6$, which greatly improves the accuracy and efficiency of the computation \citep{Allaire2002,Massoni2002}. This is a well-established approach for inviscid computations and was recently extended to include the effects of species diffusion, viscosity and thermal conductivity by \citet{Thornber2018}, enabling accurate and efficient DNS to be performed for this class of problems. The full set of equations for binary mixtures is

\begin{subeqnarray}
	\centering
	\frac{\partial \rho}{\partial t}+\boldsymbol{\nabla}\bcdot(\rho\boldsymbol{u}) & = & 0 \label{subeqn:5eqn1} \\
	\frac{\partial \rho \boldsymbol{u}}{\partial t}+\boldsymbol{\nabla}\bcdot(\rho \boldsymbol{u}\boldsymbol{u}^t+p\boldsymbol{\delta}) & = & \boldsymbol{\nabla}\bcdot\boldsymbol{\sigma} \label{subeqn:5eqn2}\\
	\frac{\partial \rho e}{\partial t}+\boldsymbol{\nabla}\bcdot\left(\left[\rho e+p\right]\boldsymbol{u}\right) & = & \boldsymbol{\nabla}\bcdot\left(\boldsymbol{\sigma}\bcdot\boldsymbol{u}-\boldsymbol{q}\right) \label{subeqn:5eqn3}  \\
    \frac{\partial \rho_1 f_1}{\partial t}+\boldsymbol{\nabla}\bcdot(\rho_1 f_1\boldsymbol{u}) & = & \boldsymbol{\nabla}\bcdot\left(\rho D_{12}\bnabla \frac{W_1f_1}{W}\right) \label{subeqn:5eqn4} \\
    \frac{\partial f_1 }{\partial t}+\boldsymbol{u} \bcdot \bnabla f_1 & = & \bnabla \bcdot (D_{12} \bnabla{f_1})-\mathcal{M}D_{12} \bnabla f_1 \bcdot \bnabla f_1+D_{12} \bnabla{f_1}\bcdot\frac{\bnabla N}{N}. \label{subeqn:5eqn5}
    \label{eqn:5eqn}
\end{subeqnarray}

In (\ref{eqn:5eqn}), $\rho$ is the mass density, $\boldsymbol{u}=[u,v,w]^t$ is the mass-weighted velocity vector, $p$ is the pressure, $f_n$ is the volume fraction of species $n$ and $e=e_i+e_k$ is the total energy per unit mass, where $e_k=\frac{1}{2}\boldsymbol{u\cdot u}$ is the kinetic energy and the internal energy $e_i$ is given by the equation of state. Note that only (\ref{subeqn:5eqn4}$e$) is in non-conservative form, hence the term quasi-conservative as conservation errors are negligible (only species internal energies are not conserved). All computations are performed using the ideal gas equation of state
\begin{equation}
e_i = \frac{p}{\rho(\overline{\gamma}-1)}
\label{eqn:eos}
\end{equation}
where $\overline{\gamma}$ is the ratio of specific heats of the mixture. For the five-equation model this is given by
\begin{equation}
\frac{1}{\overline{\gamma}-1} = \sum_n \frac{f_n}{\gamma_n-1}
\label{eqn:gamma}
\end{equation}
which is an isobaric closure (individual species temperatures are retained in the mixture). The viscous stress tensor $\boldsymbol{\sigma}$ for a Newtonian fluid is 
\begin{equation}
\boldsymbol{\sigma} = -\overline{\mu}\big[\boldsymbol{\nabla u}+(\boldsymbol{\nabla u})^t\big]+\frac{2}{3}\overline{\mu}(\boldsymbol{\nabla\cdot u})\boldsymbol{\delta}
\label{eqn:sigma}
\end{equation}
where $\overline{\mu}$ is the dynamic viscosity of the mixture. Note that in (\ref{eqn:sigma}) the bulk viscosity is assumed to be zero according to Stokes' hypothesis. The heat flux $\boldsymbol{q}=\boldsymbol{q}_c+\boldsymbol{q}_d$, with the conductive heat flux $\boldsymbol{q_c}$ given by Fourier's law
\begin{equation}
\boldsymbol{q}_c = -\overline{\kappa}\boldsymbol{\nabla}T
\label{eqn:heat-flux}
\end{equation}
where $\overline{\kappa}$ is the thermal conductivity of the mixture, and $T$ is the temperature. The thermal conductivity of species $n$ is calculated using kinetic theory as $\kappa_n=\mu_n\left(\frac{5}{4}\frac{\mathcal{R}}{W_n}+c_{p,n}\right)$, while the thermal conductivity of the mixture (as well as the mixture viscosity) is calculated using Wilke's rule. The enthalpy flux $\boldsymbol{q}_d$, arising from changes in internal energy due to mass diffusion, is given by
\begin{equation}
\boldsymbol{q}_d = \sum_{n}h_n\boldsymbol{J}_n
\label{eqn:enthalpy-flux}
\end{equation}
where $h_n=c_{p,n}T$ is the enthalpy of species $n$ and $c_{p,n}$ the specific heat at constant pressure. The diffusion flux on the RHS of (\ref{subeqn:5eqn4}$d$) invokes Fick's law of binary diffusion, written in terms of volume fraction. $W_n$ is the molecular weight of species $n$, $W$ is the molecular weight of the mixture and the binary diffusion coefficient $D_{12}$ is calculated by assuming both species have the same Lewis number  ($\mbox{\textit{Le}}_1=\mbox{\textit{Le}}_2=\mbox{\textit{Le}}$), such that
\begin{equation}
D_{12}=\frac{\overline{\kappa}}{\mbox{\textit{Le}}\rho \bar{c}_{p}}
\label{eqn:diffusivity}
\end{equation}
with $\bar{c}_{p}$ the specific heat at constant pressure for the mixture. Finally in (\ref{subeqn:5eqn5}$e$), $\mathcal{M}=\frac{W_1-W_2}{W_1 f_1+W_2 f_2}$ and $N=p/k_bT$ is the number density.

\subsection{Numerical method}
\label{subsec:numerics}
The governing equations presented in \S\ref{subsec:equations} are solved using the University of Sydney code \texttt{Flamenco}, which employs a method of lines discretisation approach in a structured, multiblock framework. Spatial discretisation is performed using a Godunov-type finite-volume method, which is integrated in time via a second-order TVD Runge-Kutta method \citep{Spiteri2002}. The spatial reconstruction of the inviscid terms uses a fifth-order MUSCL scheme \citep{Kim2005}, which is augmented by a modification to the reconstruction procedure to ensure the correct scaling of pressure, density and velocity fluctuations in the low Mach number limit \citep{Thornber2008b}. The inviscid flux component is calculated using the HLLC Riemann solver \citep{Toro1994}, while the viscous and diffusive fluxes are calculated using second-order central differences. Following \citet{Abgrall1996}, the non-conservative volume fraction equation is written as a conservative equation minus a correction term
\begin{equation}
	\frac{\partial f_1 }{\partial t}+\boldsymbol{\nabla}\bcdot(\mathcal{U}f_1)-f_1(\boldsymbol{\nabla}\bcdot\mathcal{U})=\bnabla \bcdot (D_{12} \bnabla{f_1})
	\label{eqn:volume-fraction}
\end{equation}
with $\mathcal{U}=\boldsymbol{u}+\mathcal{M}D_{12} \bnabla f_1-D_{12}\frac{\bnabla N}{N}$. The additional terms in $\mathcal{U}$ that arise from species diffusion must be included in the calculation of the inviscid flux component, as even though they are viscous in nature they modify the upwind direction of the advection of volume fraction in the solution to the Riemann problem at each cell interface. In the HLLC Riemann solver used in \texttt{Flamenco} this is achieved by modifying the wave speeds to incorporate the additional diffusion velocity, see \citet{Thornber2018} for further details. In the absence of viscosity \firstrev{and} thermal conductivity the governing equations reduce to the inviscid five-equation model of \citet{Allaire2002}, which has been used in previous studies of RMI \citep{Thornber2016,Thornber2017}. The numerical algorithm described above has been extensively demonstrated to be an effective approach for both ILES and DNS of shock-induced turbulent mixing problems \citep[see][]{Thornber2010,Thornber2011a,Groom2019,Groom2021}.

\subsection{Problem Description and Initial Conditions}
\label{subsec:IC} 
The computational setup is similar to previous studies of narrowband and broadband RMI by \citet{Groom2019,Groom2020} but with a few key differences that will be described here. A Cartesian domain of dimensions $x\times y\times z=L_x\times L\times L$ where $L=2\upi$ m is used for all simulations. The extent of the domain in the $x$-direction \thirdrev{is either $L_x=1.5\upi$ for the ILES cases or $L_x=0.75\upi$ for the DNS cases}. Periodic boundary conditions are used in the $y$- and $z$-directions, while in the $x$-direction outflow boundary conditions are imposed very far away from the test section so as to minimise spurious reflections from outgoing waves impacting the flow field. The initial mean positions of the shock wave and the interface are $x_s=2.5$ m and $x_0=3.0$ m respectively and the initial pressure and temperature of both (unshocked) fluids is $p=0.915$ atm and $T=298$ K, equal to that in the experiments of \citet{Jacobs2013}. All computations employ the ideal gas equation of state with a fixed value of $\gamma$ for each species. A schematic of the initial condition is shown in Figure \ref{fig:schematic}.

The shock Mach number is $M=1.5$, which is higher than the $M=1.2$ shock used in \citet{Jacobs2013,Krivets2017} and the $M=1.17$ shock used in \citet{Sewell2021}. This is so that the initial \firstrev{velocity jump} is larger, which makes more efficient use of the explicit time stepping algorithm, but not so large that it introduces significant post-shock compressibilty effects. Therefore the post-shock evolution of the mixing layer is still approximately incompressible in both the present simulations and the experiments in \citep{Jacobs2013,Krivets2017,Sewell2021}. The initial densities of air and SF$_6$ are $\rho_1=1.083$ kg/m$^3$ and $\rho_2=5.465$ kg/m$^3$ and the post-shock densities are $\rho_1^+=2.469$ kg/m$^3$ and $\rho_2^+=15.66$ kg/m$^3$ respectively. This gives a post-shock Atwood number of $A^+=0.72$, which is essentially the same as the value of $0.71$ given in \citet{Jacobs2013}, indicating that the effects of compressibilty are minimal. The variation in $\rho$ and $f_1$ across the interface are computed based on the surface perturbation described in (\ref{eqn:volume-fraction}) below. The evolution of the interface is solved in the post-shock frame of reference by applying a shift of $\Delta u=-158.08$ m/s to the initial velocities of the shocked and unshocked fluids. The initial velocity field is also modified to include an initial diffusion velocity at the interface, which is calculated as in previous DNS studies of RMI \citep{Groom2019,Groom2021}. To improve the quality of the initial condition, three-point Gaussian quadrature is used in each direction to accurately compute the cell averages required by the finite-volume algorithm.

\begin{table}
	\begin{center}
		\def~{\hphantom{0}}
		\begin{tabular}{lcc}
			Property & Air   & SF$_6$ \\[3pt]
			$W_l$   & 28.964 & 146.057 \\
			$\gamma_l$   & 1.4 & 1.1 \\
			$\mu_l$   & 1.836 & 1.535 \\
			$\Pran_l$  & 0.71 & 0.90 \\
			$\mbox{\textit{Sc}}_l$   & 0.71 & 0.90 \\
		\end{tabular}
    	\caption{The molecular weight $W_l$ (g/mol), ratio of specific heats $\gamma$, dynamic viscosities ($\times10^{5}$ Pa-s) and Prandtl and Schmidt numbers of air and SF$_6$.}
		\label{tab:properties}
	\end{center}
\end{table}

Table \ref{tab:properties} gives the thermodynamic properties of each fluid. The dynamic viscosities of both fluids are calculated using the Chapman--Enskog viscosity model at a temperature of $T=298$ K, while the diffusivities are calculated under the assumption of Lewis number equal to unity (hence $\Pran_l=\mbox{\textit{Sc}}_l$). In the DNS calculations, the actual values of viscosity used are much higher, so as to give a Reynolds number that is able to be fully resolved, but are kept in the same proportion to each other. This is so that the same domain width $L$ can be used for each calculation.

Based on the interface characterisation of the low-amplitude set of experiments performed in \citet{Sewell2021}, four different initial surface perturbations of a planar interface are considered which follow an idealised power spectrum of the form
\begin{equation}
	P(k)=Ck^m.
	\label{eqn:P-k}
\end{equation}
Three broadband initial conditions are simulated, containing length scales in the range $\lambda_{max}=L/2$ to $\lambda_{min}=L/32$ and with a spectral exponent $m=-1$, $-2$ and $-3$ respectively. The choice of bandwidth $R=\lambda_{max}/\lambda_{min}=16$ is based on estimates of the minimum initial wavelength performed in \citet{Jacobs2013} of $\lambda_{min}=2.9$ to $3.2$ mm, relative to a test section width of $L=8.9\times10^{-2}$ m. When scaled to the dimensions of the experiment, the perturbations in this study all have a minimum wavelength of $\lambda_{min}=2.8$ mm. Note also that the diagnostic \firstrev{spatial resolution} of the PIV method used in \citet{Sewell2021} is $1.98$ mm, resulting in attenuation of the measured scales that are smaller than this. The constant $C$ dictates the overall standard deviation of the perturbations and is set such that all initial amplitudes are linear and each perturbation has the same amplitude in the band between $k_{max}/2$ and $k_{max}$, specifically $a_{k_{max}}k_{max}=1$. See \citet{Groom2020} for further details, noting that unlike the broadband perturbations analysed in that study the perturbations considered here have different total standard deviations for the same bandwidth. 

\begin{figure}
	\centering
	\includegraphics[width=0.65\textwidth]{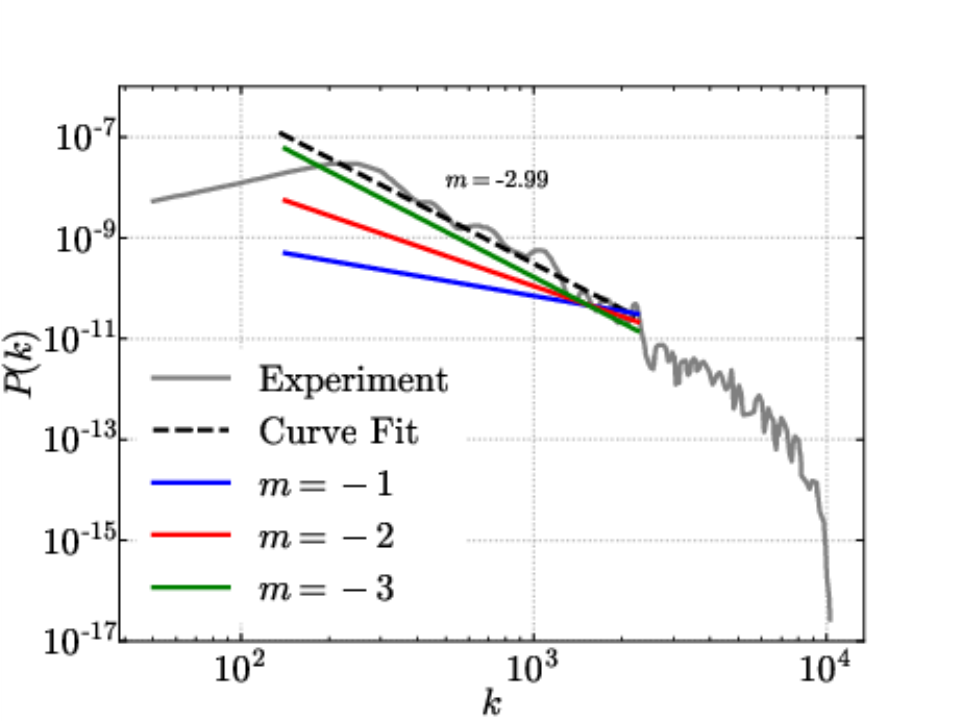}
	\caption{Power spectra of the broadband perturbations as well as the mean power spectrum of the low-amplitude experiments from \citet{Sewell2021}. Note that the spectra are scaled to match the dimensions of the experiment. \label{fig:P-k}}
\end{figure}

The power spectra for these three perturbations are shown in Figure \ref{fig:P-k}, along with the mean power spectrum of the low-amplitude experiments from \citet{Sewell2021}. 
In Figure \ref{fig:P-k} it can be seen that the $m=-3$ initial condition is the closest match to the experiments (with an estimated slope of $m=-2.99$ over the same range of modes), with the other perturbations included to study the effects of varying $m$. A fourth perturbation (not shown) is also considered; a narrowband perturbation with a constant power spectrum (i.e. $m=0$) and length scales in the range $\lambda_{min}=L/16$ to $\lambda_{max}=L/32$. This is used to study the effects of additional long wavelength modes in the initial condition and is essentially the same perturbation as the quarter-scale scale case in \citet{Thornber2017}, however the initial amplitudes are larger and are defined such that $a_{k_{max}}k_{max}=1$, which is at the limit of the linear regime. Note that in the experiments of \citet{Jacobs2013}, $a_{k_{max}}k_{max}$ ranged between 2.82 and 3.14, which is much more nonlinear. The choice of restricting the mode amplitudes such that all modes are initially linear is made so that the results may be easily scaled by the initial growth rate and compared with the results of the previous studies. 

\begin{figure}
	\centering
	\begin{subfigure}{0.45\textwidth}
		\begin{tikzpicture}
			\node[anchor=south west,inner sep=0] at (0,0) {\includegraphics[width=\textwidth]{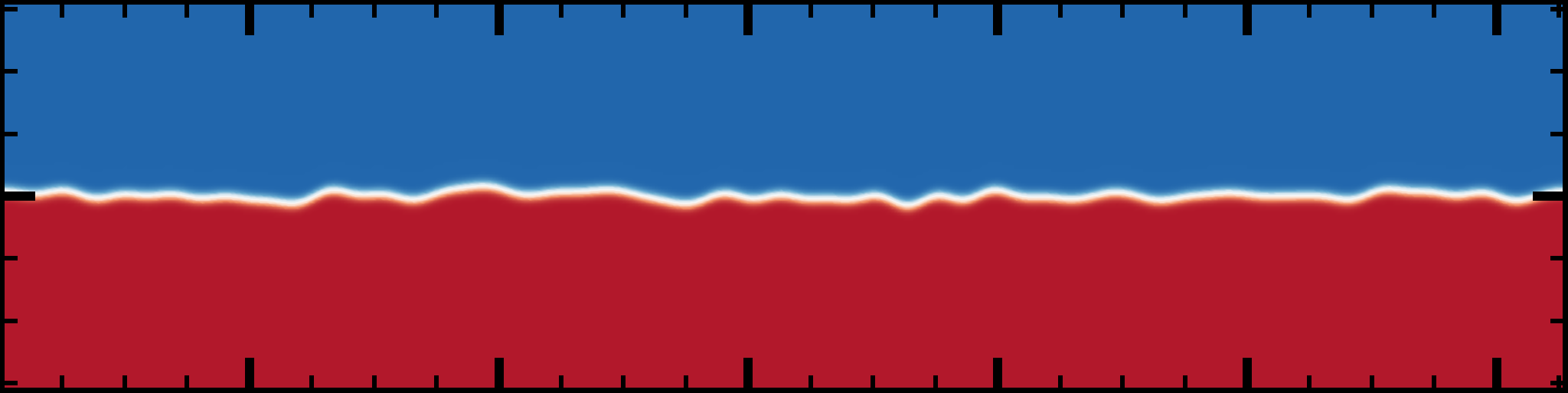}};
			\node[anchor=south west,inner sep=0] at (0.15,0.85) {\includegraphics[width=0.08\textwidth]{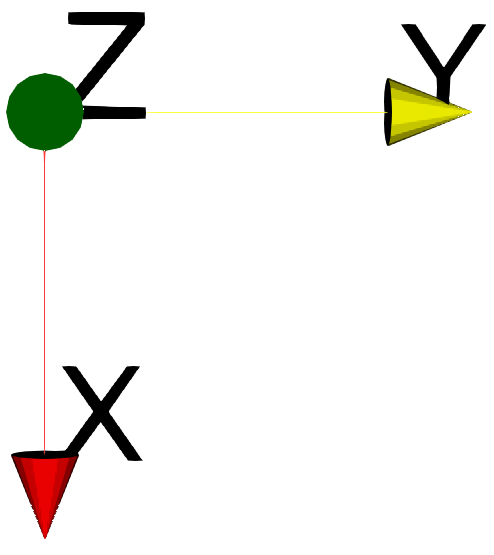}};
		\end{tikzpicture}
		\caption{$m=-1$.}
	\end{subfigure}
	\begin{subfigure}{0.45\textwidth}
		\begin{tikzpicture}
			\node[anchor=south west,inner sep=0] at (0,0) {\includegraphics[width=\textwidth]{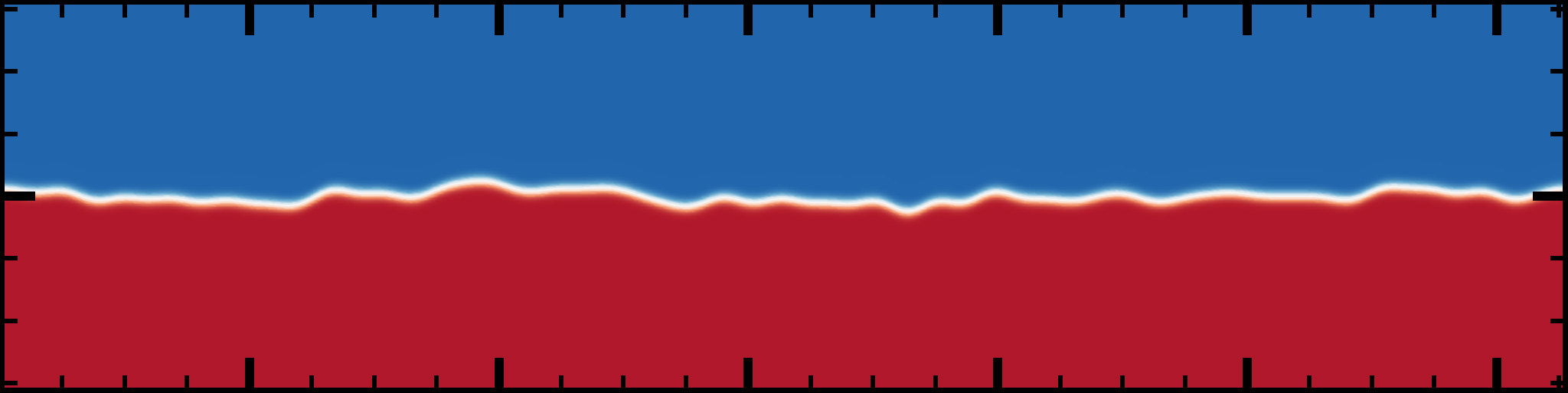}};
			\node[anchor=south west,inner sep=0] at (0.15,0.85) {\includegraphics[width=0.08\textwidth]{Figures/fig3e.png}};
		\end{tikzpicture}
		\caption{$m=-2$.}
	\end{subfigure}
	\begin{subfigure}{0.45\textwidth}
		\begin{tikzpicture}
			\node[anchor=south west,inner sep=0] at (0,0) {\includegraphics[width=\textwidth]{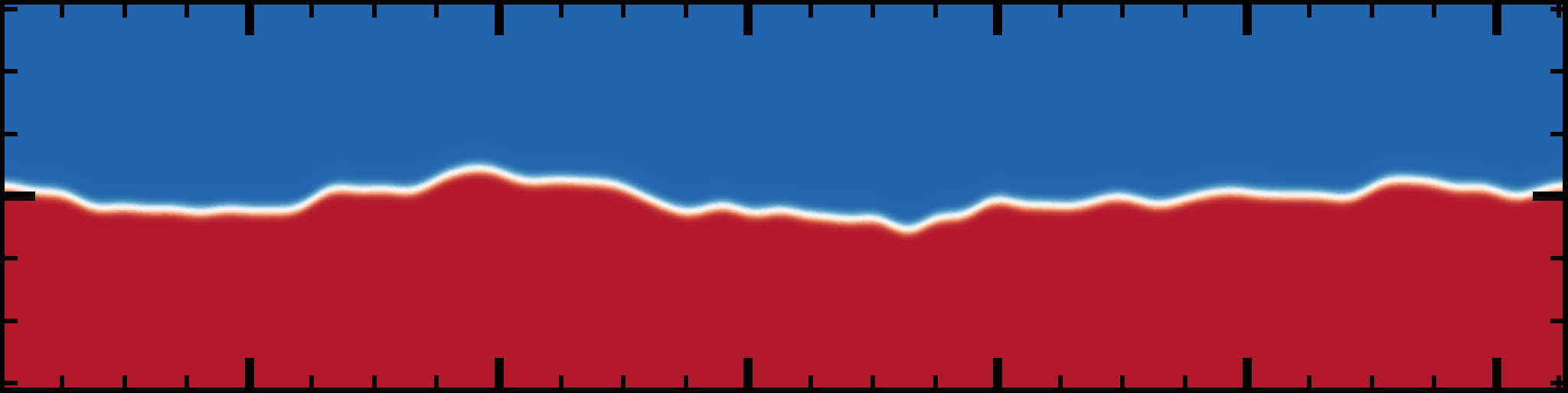}};
			\node[anchor=south west,inner sep=0] at (0.15,0.85) {\includegraphics[width=0.08\textwidth]{Figures/fig3e.png}};
		\end{tikzpicture}
		\caption{$m=-3$.}
	\end{subfigure}
	\begin{subfigure}{0.45\textwidth}
		\begin{tikzpicture}
			\node[anchor=south west,inner sep=0] at (0,0) {\includegraphics[width=\textwidth]{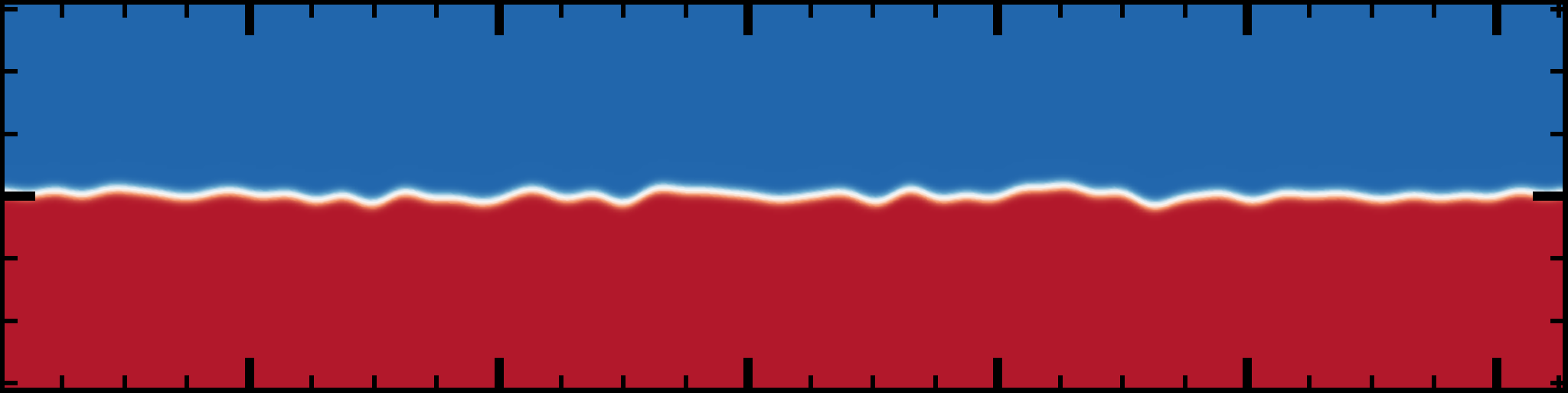}};
			\node[anchor=south west,inner sep=0] at (0.15,0.85) {\includegraphics[width=0.08\textwidth]{Figures/fig3e.png}};
		\end{tikzpicture}
		\caption{$m=0$ ($R=2$).}
	\end{subfigure}
	\caption{Contours of volume fraction $f_1$ for the ILES cases at $t=0$ \firstrev{and $z=0$}. The major ticks on both axes correspond to a grid spacing of $\Delta x=\firstrev{\Delta y=} 1$ m. \label{fig:IC}}
\end{figure}

The amplitudes and phases of each mode are defined using a set of random numbers that are constant across all grid resolutions and cases, thus allowing for a grid convergence study to be performed for each case. The interface is also initially diffuse for this same reason, with the profile given by an error function with characteristic initial thickness $\updelta=\lambda_{min}/4$. The volume fractions $f_1$ and $f_2=1-f_1$ are computed as
\begin{equation}
f_1(x,y,z)=\frac{1}{2}\textrm{erfc}\left\{\frac{\sqrt{\upi}\left[x-S(y,z)\right]}{\updelta}\right\}
\label{eqn:f1}
\end{equation}
where $S(y,z)=x_0+A(y,z)$, with $A(y,z)$ being the amplitude perturbation satisfying the specified power spectrum and $x_0$ the mean position of the interface. The amplitude perturbation $A(y,z)$ is given by
\begin{eqnarray}
A(y,z) = \sum_{m,n=0}^{\firstrev{N_{max}}}  \big[ & a_{mn}&\cos(mk_0y)\cos(nk_0z)+b_{mn}\cos(mk_0y)\sin(nk_0z) \nonumber\\
+ & c_{mn}&\sin(mk_0y)\cos(nk_0z) + d_{mn}\sin(mk_0y)\sin(nk_0z) \big]
\label{eqn:amplitude}
\end{eqnarray}
where $\firstrev{N_{max}}=k_{max}L/(2\upi)$, $k_0=2\upi/L$ and $a_{mn}\ldots d_{mn}$ are selected from a Gaussian distribution. Crucially, the Mersenne Twister pseudorandom number generator is employed which allows for the same random numbers to be used across all perturbations. This facilitates grid convergence studies for DNS and ensures that the phases of each mode are identical when comparing across perturbations with different values of $m$; only the amplitudes are varied. For full details on the derivation of the surface perturbation see \citet{Thornber2010,Thornber2017} and \citet{Groom2020}. A visualisation of each initial perturbation is shown in figure \ref{fig:IC}. Whilst there is a noticeable difference between the narrowband and broadband surface perturbations, the differences between the $m=-1$ and $m=-2$ perturbations in particular are quite subtle. Nevertheless these subtle differences in the amplitudes of the additional, longer wavelengths are responsible for quite noticeable differences in the subsequent evolution of the mixing layer, as will be shown in the following sections. This highlights the importance of understanding the sensitivity to initial conditions in RMI-induced flows.

For each perturbation, the weighted-average wavelength can be defined as $\bar{\lambda}=2\upi/\bar{k}$, where
\begin{equation}
	\bar{k} = \frac{\sqrt{\displaystyle\int_{k_{min}}^{k_{max}}k^2P(k)\:\mathrm{d} k}}{\sqrt{\displaystyle\int_{k_{min}}^{k_{max}}P(k)\:\mathrm{d} k}}.
	\label{eqn:kbar}
\end{equation}
Similarly, the initial growth rate of the perturbation variance is given by
\begin{equation}
\dot{\sigma_0}=\sigma_0^+A^+\Delta u\bar{k}/\psi
\end{equation}
where $\sigma_0^+=C_V(1-\Delta u/U_s)\sigma_0$ is the post-shock standard deviation, $\sigma_0$ is the initial standard deviation and $\psi$ is a correction factor to account for the diffuse interface \citep{Duff1962,Youngs2020a}. Here $C_V=(A^-+C_RA^+)/(2C_RA^+)$ is an additional correction factor that is applied to the Richtmyer compression factor $C_R=(1-\Delta u/U_s)$ to give the impulsive model of \citet{VDM}. For the present gas combination and configuration, $C_V=1.16$ and is used to account for deficiencies in the original impulsive model of \citet{Richtmyer1960} for certain cases. \citet{Thornber2017} showed that for a Gaussian height distribution, the integral width $W=\int\langle f_1\rangle\langle f_2\rangle\:\mathrm{d} x$ is equal to $0.564\sigma$ and therefore $\dot{W_0}=0.564\dot{\sigma_0}$. For the DNS cases, the initial Reynolds number is calculated in line with previous studies as
\begin{equation}
\Rey_0=\frac{\bar{\lambda}\dot{W}_0\overline{\rho^+}}{\overline{\mu}}
\end{equation}
$\overline{\rho^+}=9.065$ kg/m$^3$ is the mean post-shock density. Table \ref{tab:ICs} gives the initial growth rate and weighted-average wavelength for each perturbation.

\begin{table}
	\begin{center}
		\def~{\hphantom{0}}
		\begin{tabular}{ccccc}
			Quantity & $m=0$   & $m=-1$ & $m=-2$ & $m=-3$  \\[3pt]
			$R$   & 2 & 16 & 16 & 16 \\
			$\bar{\lambda}$  & 0.278 & 0.463 & 0.785 & 1.33 \\
			$\dot{W_0}$  & 16.74 & 20.03 & 23.84 & 34.32 \\
		\end{tabular}
    \caption{The bandwidth, weighted-average wavelength (m) and initial growth rate of integral width (m/s) for each of the four perturbations.}
		\label{tab:ICs}
	\end{center}
\end{table}

\subsection{Direct Numerical Simulations}
\label{subsec:DNS}
Prior to presenting results for each perturbation, it is important to discuss some of the challenges present when performing DNS of RMI with broadband perturbations. Previous DNS studies of 3D multi-mode RMI have focussed exclusively on narrowband perturbations \citep{Olson2014,Groom2019,Wong2019,Groom2021} or perturbations with a dominant single mode \citep{Tritschler2014pre}. The present set of broadband DNS use a perturbation with $8\times$ the bandwidth of initial modes compared to the narrowband perturbation analysed in \citet{Groom2019,Groom2021}, but still require the same number of cells per initial minimum wavelength for a given Reynolds number in order to fully resolve the calculation. \thirdrev{To be considered fully resolved and thus qualify as "strict" DNS, grid convergence must be demonstrated for statistics that depend on the smallest scales in the flow, such as enstrophy and scalar dissipation rate. Of the previously cited studies, only \cite{Groom2019,Groom2021} fully resolve these gradient-dependent quantities and none of the studies mentioned (as well as the present study) resolve the internal structure of the shock wave.  Demonstration of grid convergence for enstrophy and scalar dissipation rate in the present set of DNS cases is given in Appendix \ref{app:convergence}, however this comes at the cost of limiting the Reynolds number that can be achieved, as discussed below.}

Regarding the Reynolds number, using the standard width-based definition $\Rey_h=\dot{h}h/\nu$ where the width $h\propto t^\theta$ then the Reynolds number, and hence the grid resolution requirements, can either increase or decrease in time depending on the value of $\theta$ since
\begin{equation}
	\Rey_h\propto\frac{\theta t^{\theta-1}t^\theta}{\nu}\propto t^{2\theta-1}.
\end{equation}
Therefore for $\theta<1/2$ the Reynolds number is decreasing and vice versa for $\theta>1/2$. \citet{Youngs2004,Thornber2010} showed that the value of $\theta$ depends on both the bandwidth and spectral slope $m$ of the initial condition, which was recently demonstrated in \citet{Groom2020} using ILES for perturbations of the form given by (\ref{eqn:P-k}) with $m=-1$, $-2$ and $-3$. For the largest bandwidths simulated, these perturbations gave values of $\theta=0.5$, $0.63$ and $0.75$ respectively, which for the $m=-1$ and $-2$ cases are quite close to the theoretical values of $\theta=1/2$ and $\theta=2/3$. What these results imply is that the Reynolds number of a broadband perturbation with $m\le -1$ will either be constant or increase with time as the layer develops, which make performing fully grid-resolved DNS more challenging than for a narrowband layer where $\theta\le 1/3$ \citep{Elbaz2018,Soulard2018}. 

For DNS of narrowband RMI the number of cells per $\lambda_{min}$ can be maximised, which sets the smallest scale that can be grid resolved and therefore the maximum Reynolds number that can be obtained on a given grid. For fully developed isotropic turbulence, it is well known that grid resolution requirements scale as $\Rey^{9/4}$ and the total number of floating point operations required to perform a simulation to a given time scales as $Re^3$ \citep{Pope2000}. For transitional RMI, empirically the scaling appears to be less severe (closer to $\Rey^2$), but available computing power still quickly limits the maximum Reynolds number that can be obtained. The simulations presented in \citet{Groom2021} represent the current state of the art in terms of maximum Reynolds number that can be achieved using the \texttt{Flamenco} algorithm. Even then, the highest Reynolds number simulation in that study was still short of meeting the mixing transition requirement for fully developed turbulence in unsteady flows \citep{Zhou2003b}. 

For DNS of broadband RMI, assuming the same grid resolution is used, the larger bandwidth necessitates a smaller Reynolds number since the number of cells per $\lambda_{min}$ required to resolve the shock-interface interaction and subsequent evolution is the same. This is before any considerations about whether additional grid resolution is required at later time due to increasing Reynolds number. The requirement that all initial amplitudes be linear also limits the initial \firstrev{velocity jump} (and hence the Reynolds number) that can be obtained, and the diffuse profile across the interface that is required to properly resolve the shock-interface interaction in DNS also dampens the initial \firstrev{velocity jump} (relative to if a sharp interface was used). All of this results in the fact that for the current maximum grid sizes simulated in this and previous studies (e.g. $2048^2$ cross-sectional resolution), DNS can be performed at either a moderate Reynolds number but small bandwidth (i.e. too narrow to be indicative of real surface perturbations) as in \citet{Groom2021} or a moderate bandwidth but low Reynolds number (i.e. too diffuse to be indicative of fully-developed turbulence) as in the present study. These observations are not exclusive to DNS of RMI but also apply to RTI, Kelvin-Helmholtz instability and other flows where the effects of initial conditions are important and realistic initial perturbations need to be considered.

In spite of all this, DNS is still a useful tool in the context of this study as it provides results that may be considered a plausible lower bound to the experimental results in a similar manner to which ILES results may be considered a plausible upper bound. It is also necessary for computing statistical quantities that depend on the smallest scales of motion being sufficiently resolved, such as the turbulent length scales and Reynolds numbers presented in \S\ref{subsec:length-scales} as well as many other quantities that are important for informing modelling of these types of flows (see \citet{Groom2021,Wong2022} for some examples). Comments on how some of the limitations mentioned above might be resolved are given in \S\ref{sec:conclusions}.

\section{Results}
\label{sec:results}
Using the initial conditions and computational setup described in \S\ref{sec:setup}, \thirdrev{six} simulations are performed with \texttt{Flamenco}. These consist of four ILES corresponding to the four different initial conditions as well as \thirdrev{two} DNS; one for the $m=-1$ initial condition \thirdrev{and} one for the $m=-2$ initial condition. The \thirdrev{viscosity} used in these DNS is $\overline{\mu}=0.3228$ Pa$\cdot$s, which corresponds to initial Reynolds numbers of $\Rey_0=261$ \thirdrev{and $\Rey_0=526$ for the $m=-1$ and $m=-2$ cases respectively}. While \thirdrev{this viscosity is} much higher than would occur experimentally, \thirdrev{it is} equivalent to using \thirdrev{a} much smaller value of $\overline{\lambda}$ to obtain the same Reynolds number due to the various simplifications employed in the governing equations, such as no variation in viscosity with temperature.  For each simulation, grid convergence is assessed using the methodology outlined in \citet{Thornber2017} for ILES and \citet{Groom2019} for DNS. The simulations were run up to a physical time of $t=0.1$ s, at which point some of the spikes were observed to have reached the domain boundaries in the $m=-3$ ILES case. The complete set of simulations is summarised in table \ref{tab:simulations}. 

\begin{table}
	\begin{center}
		\def~{\hphantom{0}}
		\begin{tabular}{lccccc}
			Case & $m$ & $\Rey_0$ & Simulation time (s) & Domain size (m$^3$) & Grid resolution \\[3pt]
			1 & 0  & -    & 0.1 & $1.5\upi\times 2\upi\times 2\upi$ & $384\times512^2$ \\
			2 & -1 & -    & 0.1 & $1.5\upi\times 2\upi\times 2\upi$ & $384\times512^2$ \\
			3 & -2 & -    & 0.1 & $1.5\upi\times 2\upi\times 2\upi$ & $384\times512^2$ \\
			4 & -3 & -    & 0.1 & $1.5\upi\times 2\upi\times 2\upi$ & $384\times512^2$ \\
			5 & -1 & 261  & 0.1 & $0.75\upi\times 2\upi\times 2\upi$ & $384\times1024^2$ \\
			6 & -2 & 526  & 0.1 & $0.75\upi\times 2\upi\times 2\upi$ & $384\times1024^2$ \\
		\end{tabular}
		\caption{The initial power spectrum slope, initial Reynolds number (DNS only), total simulation time, domain size and maximum grid resolution employed for each case.}
		\label{tab:simulations}
	\end{center}
\end{table}

\begin{figure}
	\centering
	\begin{subfigure}{0.45\textwidth}
		\begin{tikzpicture}
			\node[anchor=south west,inner sep=0] at (0,0) {\includegraphics[width=\textwidth]{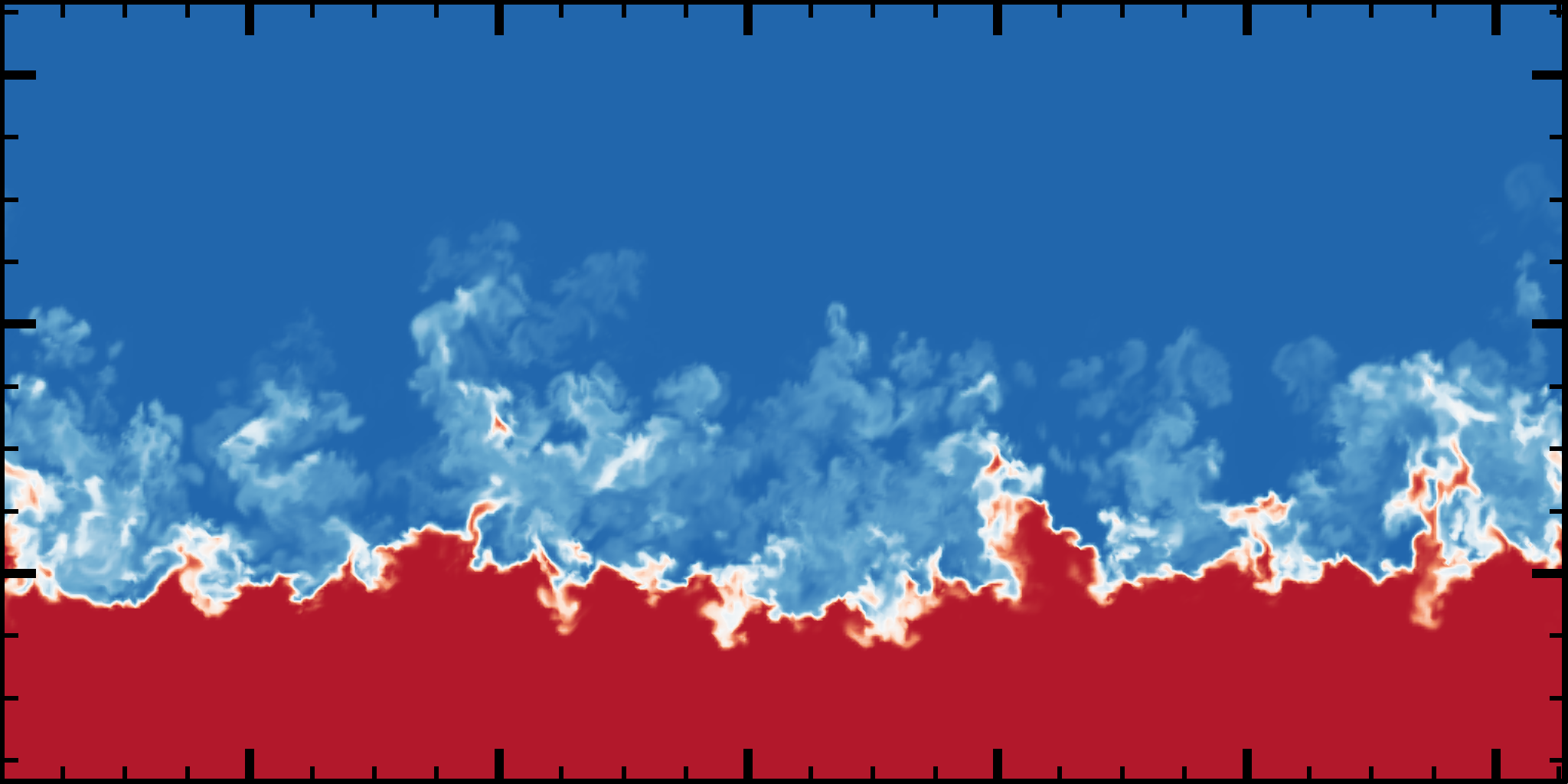}};
			\node[anchor=south west,inner sep=0] at (0.2,2.35) {\includegraphics[width=0.08\textwidth]{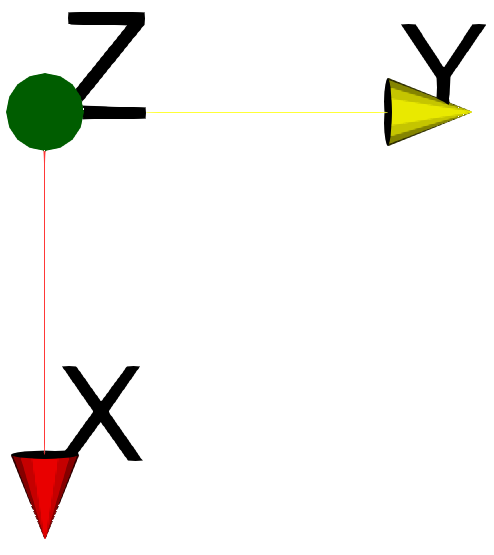}};
		\end{tikzpicture}
		\caption{$m=-1$.}
	\end{subfigure}
	\begin{subfigure}{0.45\textwidth}
		\begin{tikzpicture}
			\node[anchor=south west,inner sep=0] at (0,0) {\includegraphics[width=\textwidth]{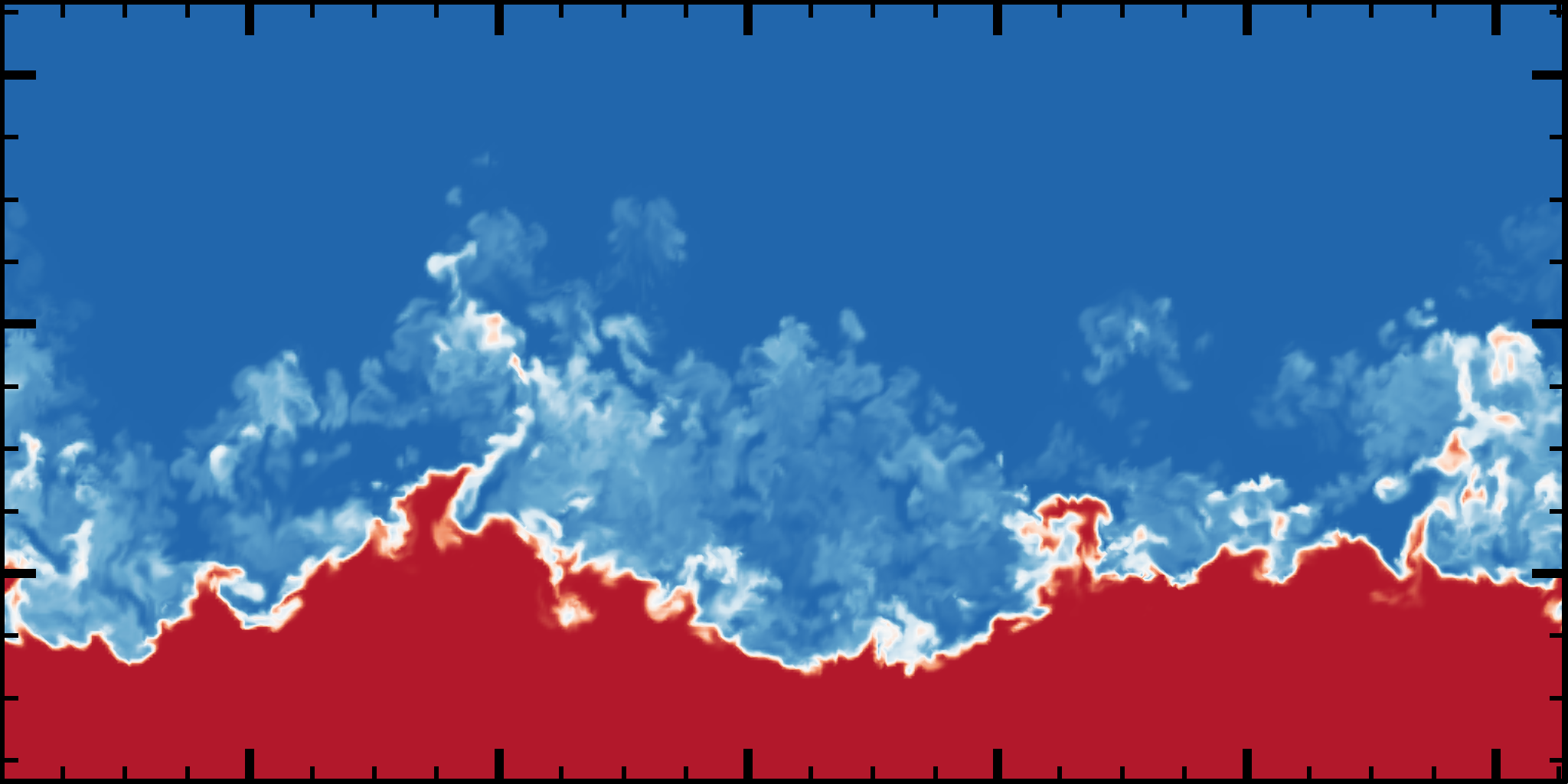}};
			\node[anchor=south west,inner sep=0] at (0.2,2.35) {\includegraphics[width=0.08\textwidth]{Figures/fig4e.png}};
		\end{tikzpicture}
		\caption{$m=-2$.}
	\end{subfigure}
	\begin{subfigure}{0.45\textwidth}
		\begin{tikzpicture}
			\node[anchor=south west,inner sep=0] at (0,0) {\includegraphics[width=\textwidth]{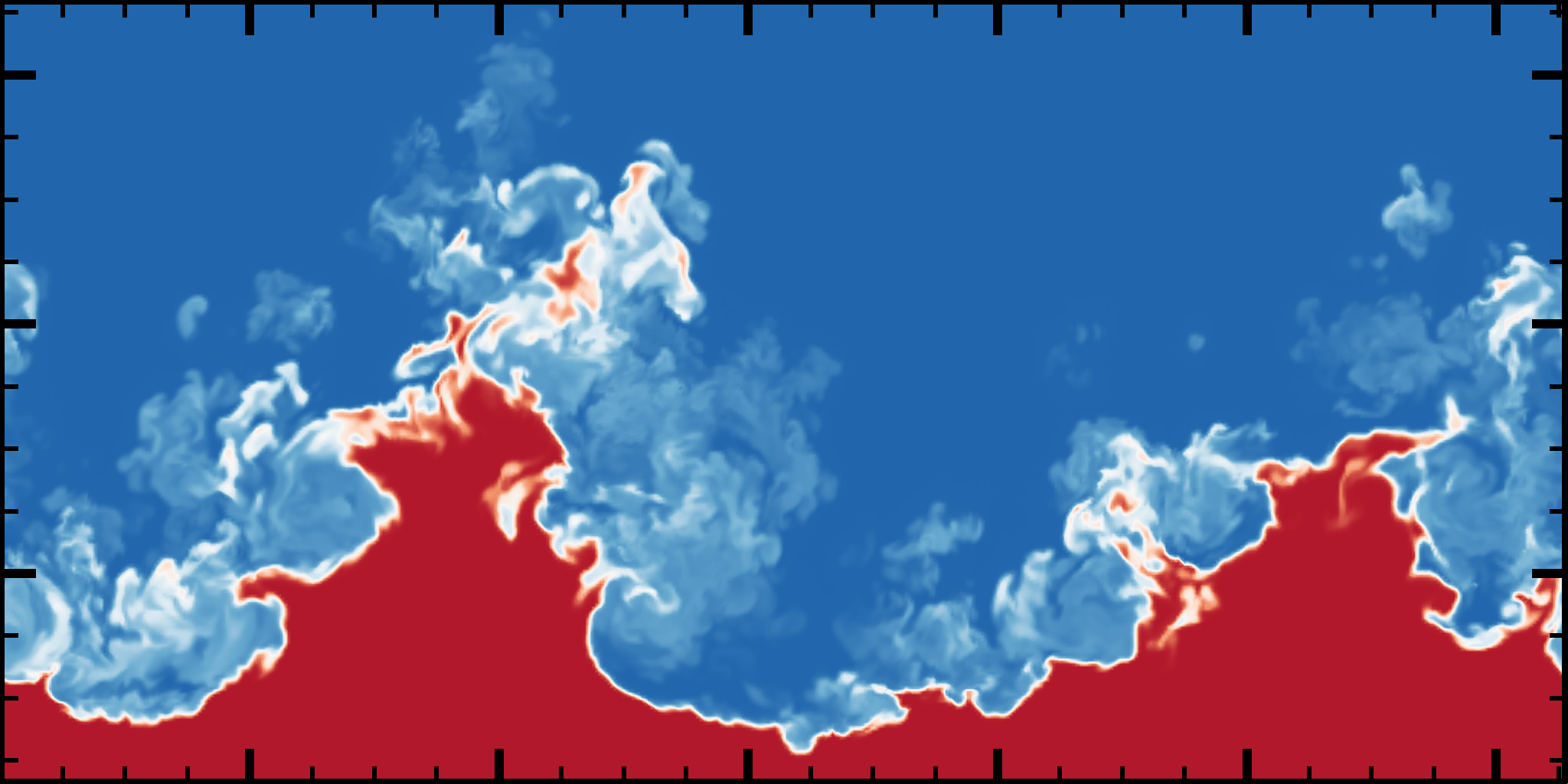}};
			\node[anchor=south west,inner sep=0] at (0.2,2.35) {\includegraphics[width=0.08\textwidth]{Figures/fig4e.png}};
		\end{tikzpicture}
		\caption{$m=-3$.}
	\end{subfigure}
	\begin{subfigure}{0.45\textwidth}
		\begin{tikzpicture}
			\node[anchor=south west,inner sep=0] at (0,0) {\includegraphics[width=\textwidth]{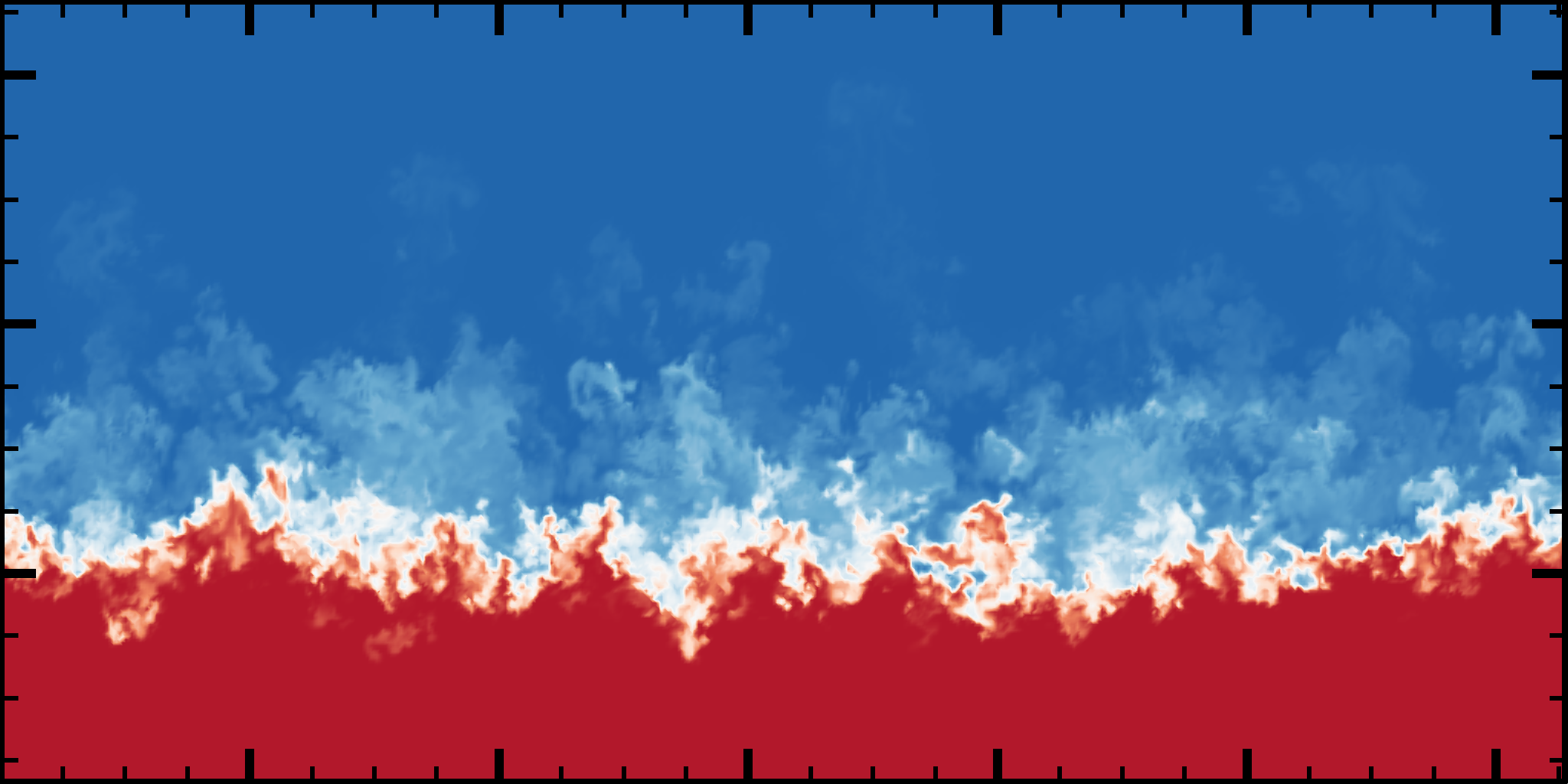}};
			\node[anchor=south west,inner sep=0] at (0.2,2.35) {\includegraphics[width=0.08\textwidth]{Figures/fig4e.png}};
		\end{tikzpicture}
		\caption{$m=0$ ($R=2$).}
	\end{subfigure}
	\caption{Contours of volume fraction $f_1$ for the ILES cases at $t=0.1$ \firstrev{s} \firstrev{and $z=0$}. The major ticks on both axes correspond to a grid spacing of $\Delta x=\firstrev{\Delta y=} 1$ m. \label{fig:f1-ILES}}
\end{figure}

Figure \ref{fig:f1-ILES} shows visualisations of the solution at the latest time of $t=0.1$ s for the four ILES cases. Bubbles of light fluid can be seen flowing into the heavy fluid on the lower side of the mixing layer, while heavy spikes are penetrating into the light fluid on the upper side. In the narrowband case the mixing layer has remained relatively uniform over the span of the domain, whereas in the broadband cases, particularly the $m=-2$ and $m=-3$ cases, large-scale entrainment is starting to occur at scales on the order of the domain width. Another noticeable phenomenon at this time is that in the narrowband case some spikes have \secondrev{penetrated much further away from the main mixing layer than in the broadband cases}. \secondrev{This is shown in greater detail in figure \ref{fig:f1-ILES-m0} where isosurfaces of volume fraction $f_1=0.001$ and $f_1=0.999$ are plotted for both the $m=0$ narrowband case and the $m=-2$ broadband case to highlight the differences in spike behaviour. Note that in the narrowband case there are taller structures on the spike side that in some instances have been ejected from the main layer. See also Figure 5 from \citet{Youngs2020b} for a similar visualisation at a lower Atwood number.} A plausible explanation for this is that the slower but more persistent growth of the low wavenumber modes in the broadband cases cause the main mixing layer to eventually disrupt the trajectory of any spikes that were initially ejected \secondrev{from high wavenumber modes}. \secondrev{Future work will study this comparison of spike behaviour between narrowband and broadband mixing perturbations at higher Atwood numbers that are more relevant to ICF.}

\begin{figure}
	\centering
	\begin{subfigure}{0.45\textwidth}
		\begin{tikzpicture}
			\node[anchor=south west,inner sep=0] at (0,0) {\includegraphics[width=\textwidth]{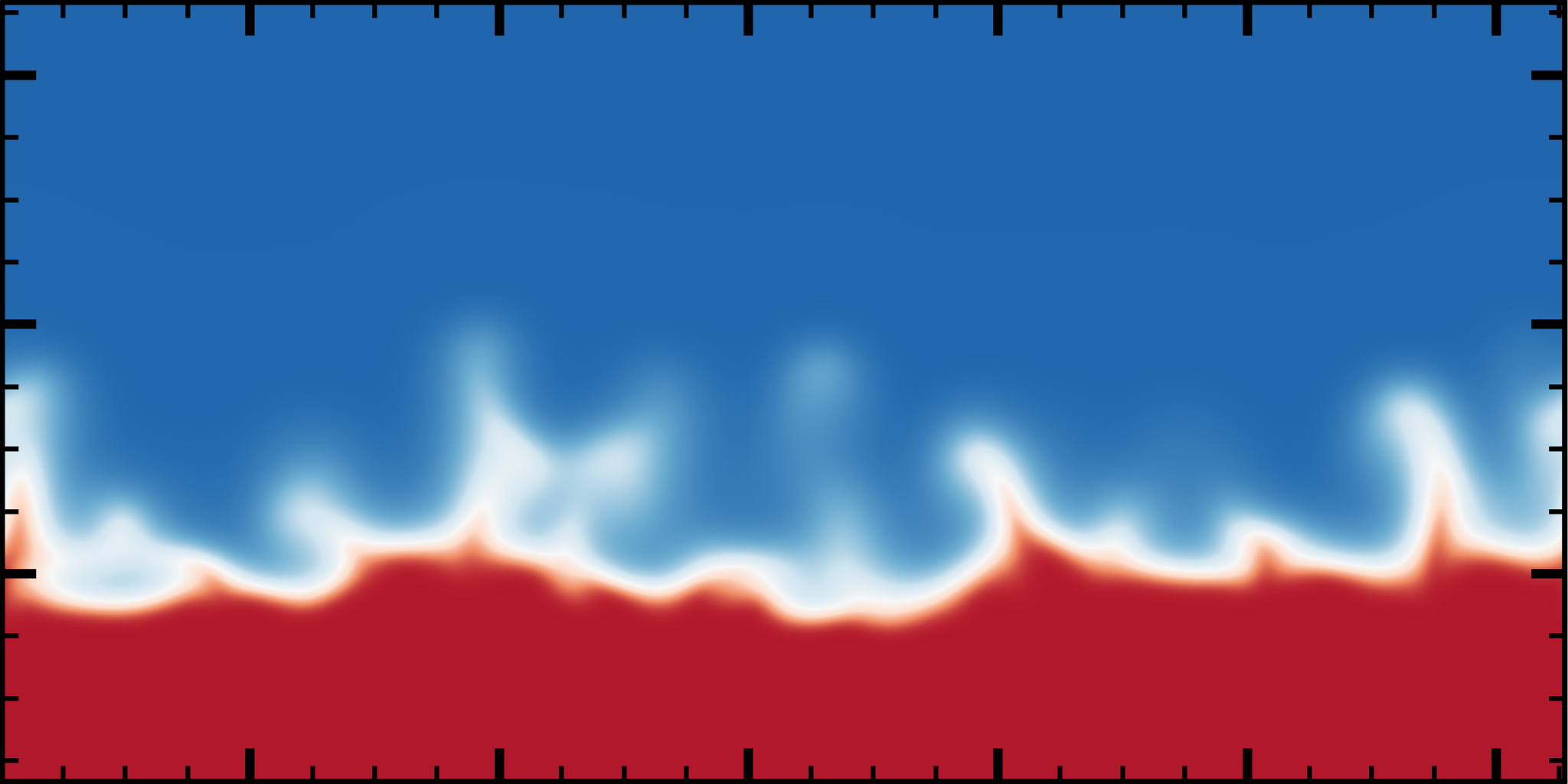}};
			\node[anchor=south west,inner sep=0] at (0.2,2.35) {\includegraphics[width=0.08\textwidth]{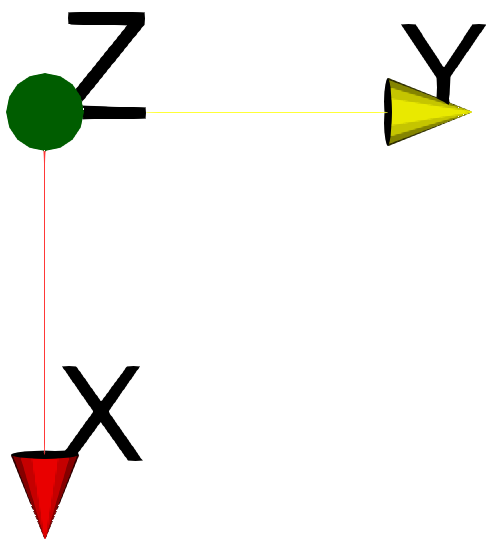}};
		\end{tikzpicture}
		\caption{$m=-1$, $\Rey_0=261$.}
	\end{subfigure}
	\begin{subfigure}{0.45\textwidth}
		\begin{tikzpicture}
			\node[anchor=south west,inner sep=0] at (0,0) {\includegraphics[width=\textwidth]{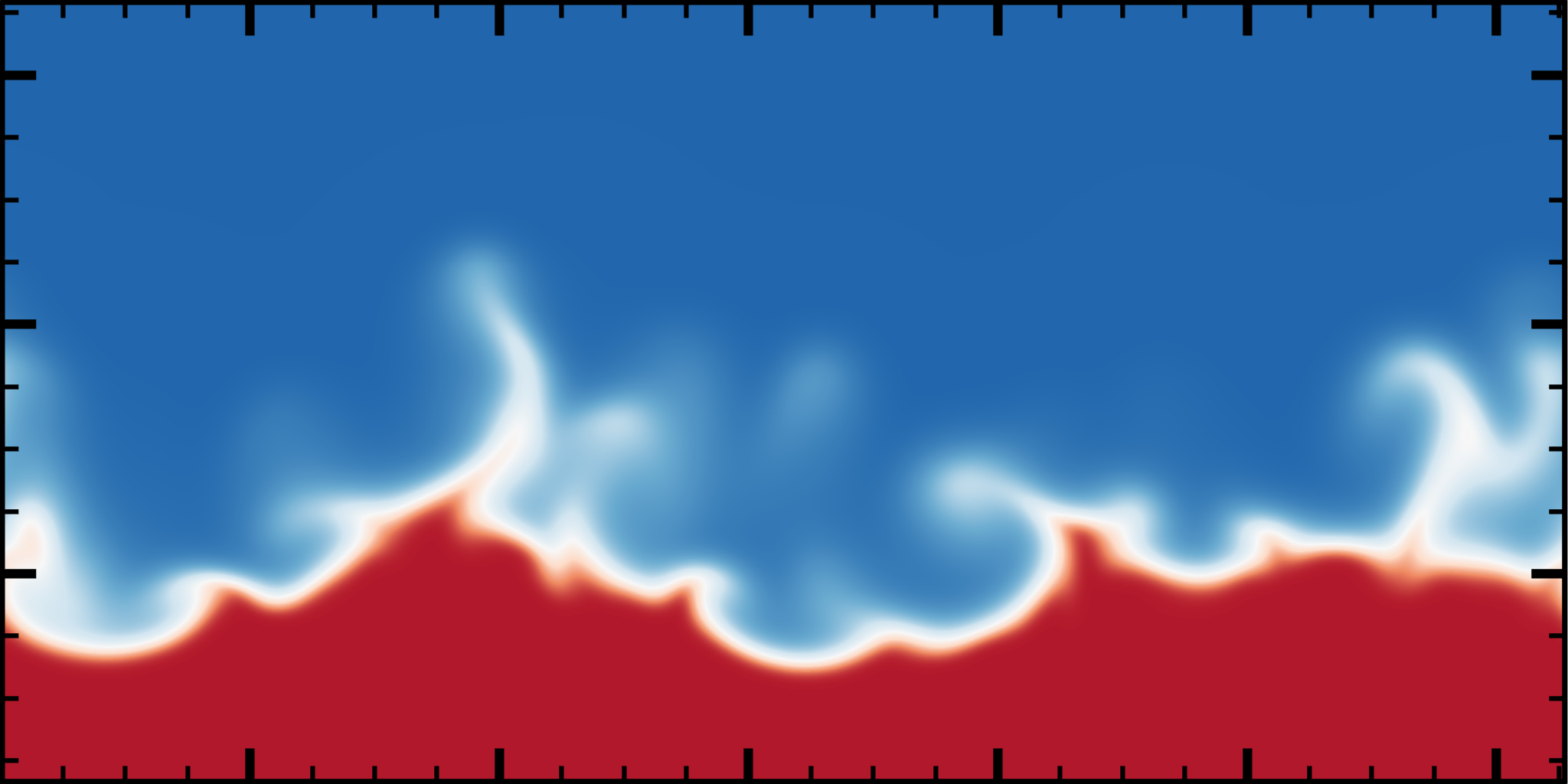}};
			\node[anchor=south west,inner sep=0] at (0.2,2.35) {\includegraphics[width=0.08\textwidth]{Figures/fig5c.png}};
		\end{tikzpicture}
		\caption{$m=-2$, $\Rey_0=526$.}
	\end{subfigure}
	\caption{Contours of volume fraction $f_1$ for the DNS cases at $t=0.1$ \firstrev{s} \firstrev{and $z=0$}. The major ticks on both axes correspond to a grid spacing of $\Delta x=\firstrev{\Delta y=} 1$ m. \label{fig:f1-DNS}}
\end{figure}

\begin{figure} 
	\centering
		\includegraphics[width=0.48\textwidth]{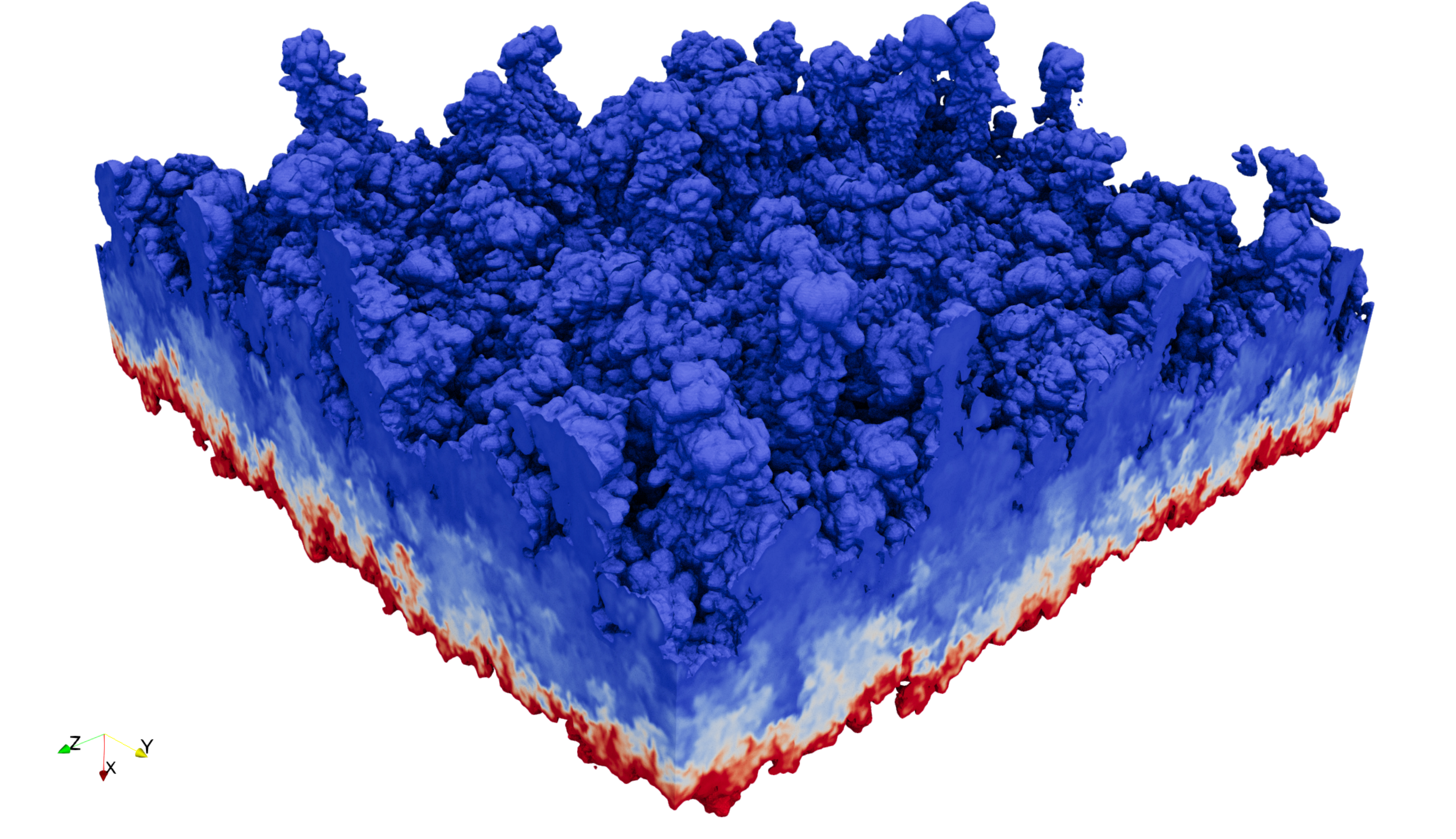}
		\includegraphics[width=0.48\textwidth]{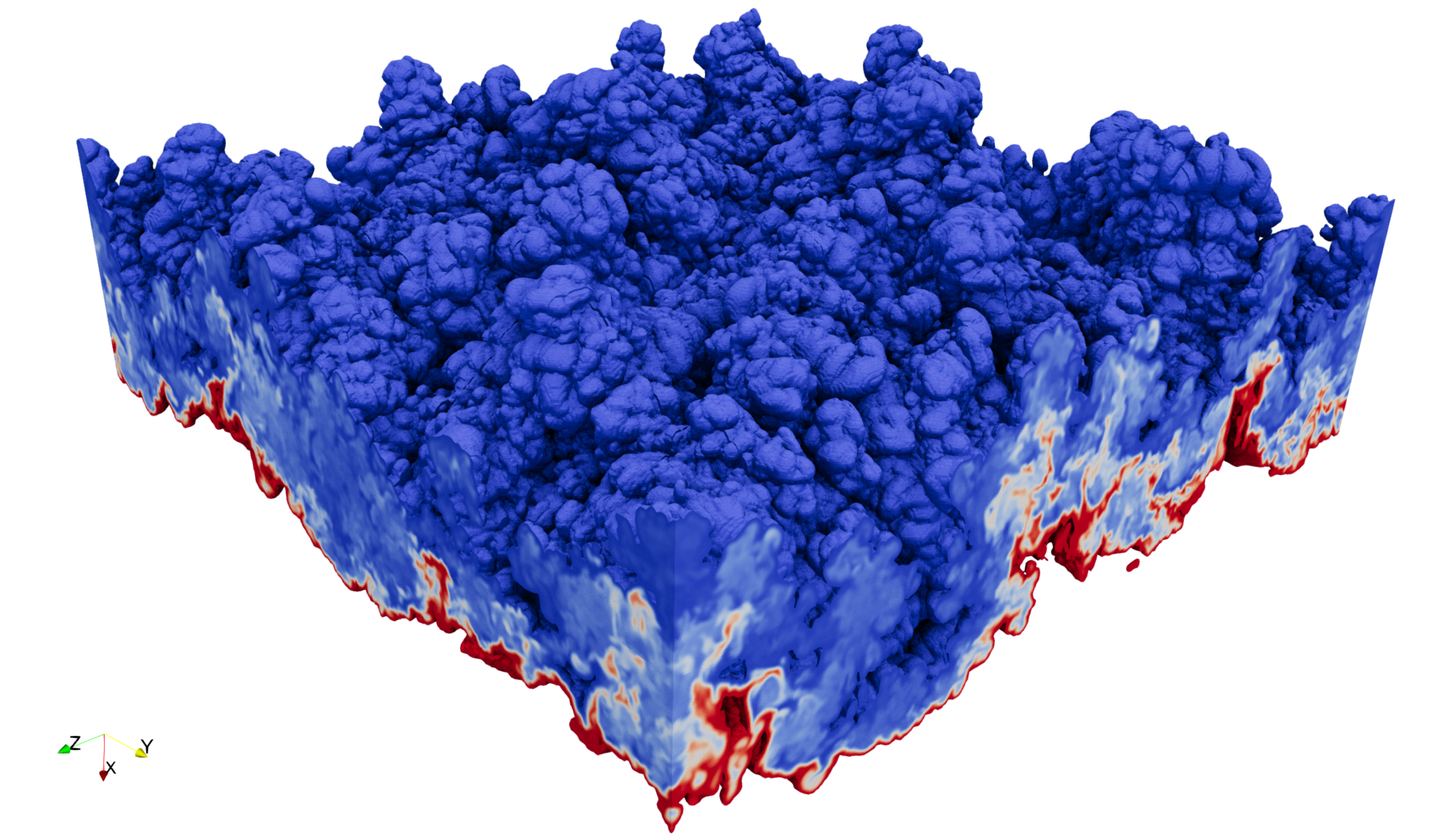}
	\caption{\secondrev{Isosurfaces of volume fraction $f_1$ for the $m=0$ (left) and $m=-2$ (right) ILES cases at $t=0.1$ s.} \label{fig:f1-ILES-m0}}
\end{figure}

Figure \ref{fig:f1-DNS} shows visualisations at the same physical time for the two DNS cases. As discussed in \S\ref{subsec:DNS}, these DNS are at quite low Reynolds number so as to be able to fully resolve the wide range of initial length scales. They are therefore quite diffuse, however good agreement can still be observed in the largest scales of motion with the corresponding ILES cases. The fluctuating kinetic energy spectra presented in \S\ref{subsec:spectra} also corroborate this observation.

\subsection{Non-dimensionalisation}
\label{subsec:nondimensionalisation}
\begin{figure}
	\centering
	\includegraphics[width=0.65\textwidth]{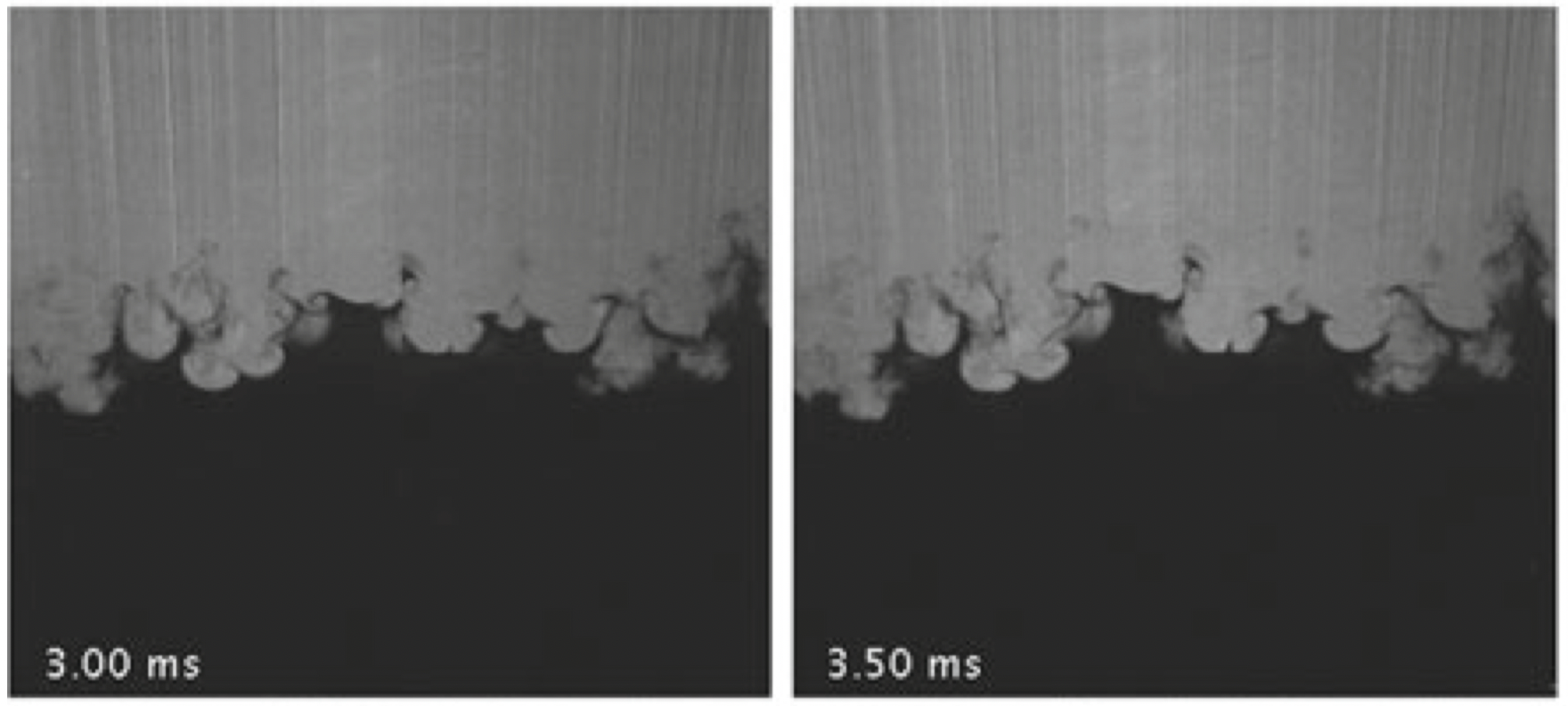}
	\caption{Image sequence taken from a typical vertical shock tube experiment using the Mie diagnostic. Times relative to shock impact are shown in each image. Reshock occurs at $t=6.50$ ms. \textit{Source}: Figure 3 of \citet{Jacobs2013}. \label{fig:Jacobs2013}}
\end{figure}

The results in the following sections are appropriately non-dimensionalised to allow for direct comparisons with the experiments in \citet{Jacobs2013} and \citet{Sewell2021}. All length scales are normalised by $\lambda_{min}$, which is equal to $0.196$ m in the simulations and is estimated to lie between $2.9$ mm and $3.2$ mm in the experiments. As the effects of different initial impulses are of primary interest, it does not makes sense to use $\dot{W_0}$ as the normalising velocity scale, therefore all velocities are normalised by $A^+\Delta u$ instead. In the simulations $A^+=0.72$ and $\Delta u=158.08$ m/s, while in the experiments $A^+=0.71$ and $\Delta u=74$ m/s. Therefore the non-dimensional time is given by
\begin{equation}
\tau=\frac{(t-t_0)A^+\Delta u}{\lambda_{min}}
\end{equation}
where $t_0=0.0011$ s is the shock arrival time. This equates to a dimensionless time of $\tau=57.4$ at the latest time considered in the simulations ($t=0.1$ s), $107\le\tau\le118$ at the latest time prior to reshock in the experiments of \citet{Jacobs2013} ($t-t_0=6.5$ ms) and $73.9\le\tau\le81.5$ at the latest time prior to reshock in the experiments of \citet{Sewell2021} ($t-t_0=4.5$ ms), assuming the same range of values for $\lambda_{min}$ \firstrev{of $2.9$ to $3.2$ mm}. Figure \ref{fig:Jacobs2013} shows a subset of the image sequence taken from a typical vertical shock tube experiment in \citet{Jacobs2013} using the Mie diagnostic . For comparison with the present simulations, a dimensionless time of $\tau=57.4$ corresponds to a physical time in the range of $t=3.17$ ms to $t=3.50$ ms, \firstrev{which may be compared with the images shown for times $t=3.00$ ms and $t=3.50$ ms} in figure \ref{fig:Jacobs2013}.

\subsection{Turbulent Kinetic Energy and Mix Width}
\label{subsec:TKE}
In this section comparisons are made both between the present simulation results and those of the experiments, as well as between the methods for calculating those results in the experiments with methods that have been commonly employed in previous simulation studies of RMI. To measure the mixing layer width, \citet{Jacobs2013} used Mie scattering over a single plane, with each image then row-averaged to obtain the mean smoke concentration in the streamwise direction. For each concentration profile, the mixing layer width is defined as the distance between the 10\% and 90\% threshold locations. This is similar to the definition of visual width used in simulation studies of both RMI and RTI \citep[see][]{Cook2001,Cook2002,Zhou2019a}, where the plane-averaged mole fraction or volume fraction profile is used along with a typical threshold cuttoff of 1\% and 99\%, e.g.
\begin{equation}
h=x\left(\langle f_1\rangle=0.01\right)-x\left(\langle f_1\rangle=0.99\right).
\label{eqn:outer-length}
\end{equation}
This is a useful definition of the outer length scale of the mixing layer, however \thirdrev{the choice of cutoff location is somewhat arbitrary and when used to estimate growth rates the results are influenced by both the the choice of cutoff location as well as statistical fluctuations \citep{Zhou2019a}.} For that purpose, an integral definition is typically used such as the integral width \citep{Andrews1990}
\begin{equation}
W=\int\langle f_1\rangle\langle f_2\rangle\:\mathrm{d} x.
\label{eqn:integral-width}
\end{equation}
If $f_1$ varies linearly with $x$ then $h=6W$ \citep{Youngs1994}. See also the recent paper by \citet{Youngs2020b} where integral definitions of the bubble and spike heights are proposed that are of similar magnitude to the visual width. \thirdrev{These are presented in Appendix \ref{app:integral} and are discussed in \S \ref{subsec:heights} below.}

\begin{figure}
	\centering
	\begin{subfigure}{0.45\textwidth}
		\includegraphics[width=\textwidth]{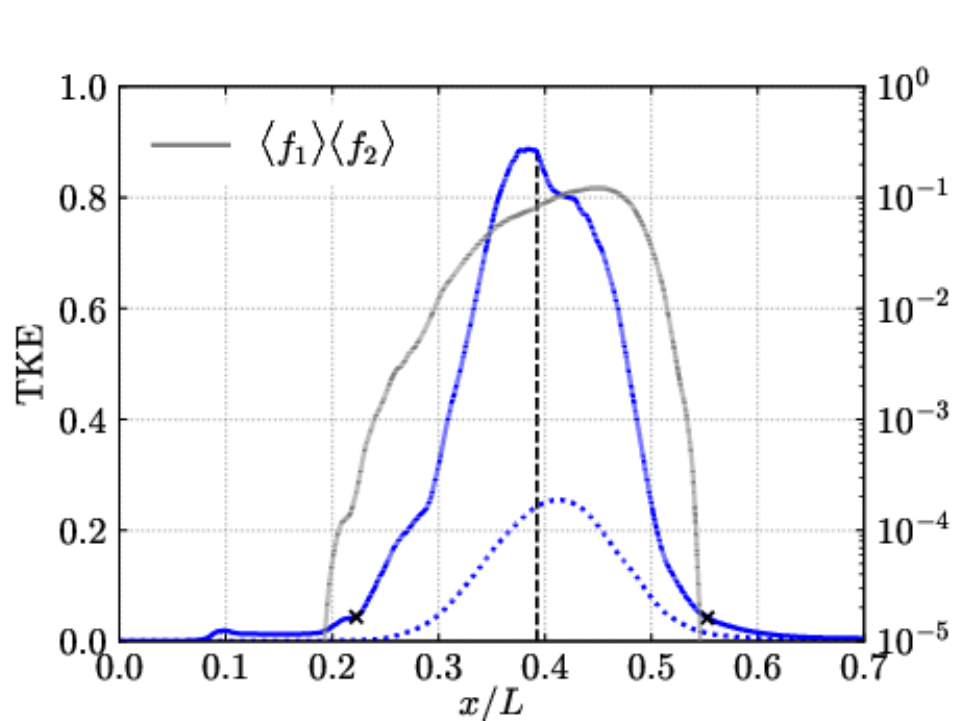}
		\caption{$m=-1$.}
	\end{subfigure}
	\begin{subfigure}{0.45\textwidth}
		\includegraphics[width=\textwidth]{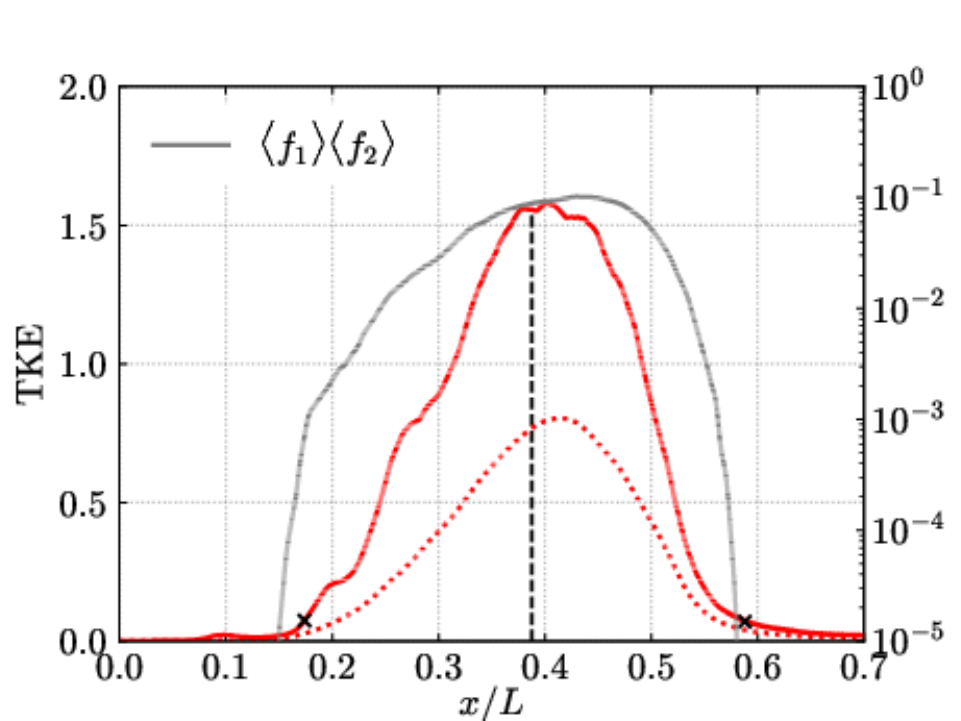}
		\caption{$m=-2$.}
	\end{subfigure}
	\begin{subfigure}{0.45\textwidth}
		\includegraphics[width=\textwidth]{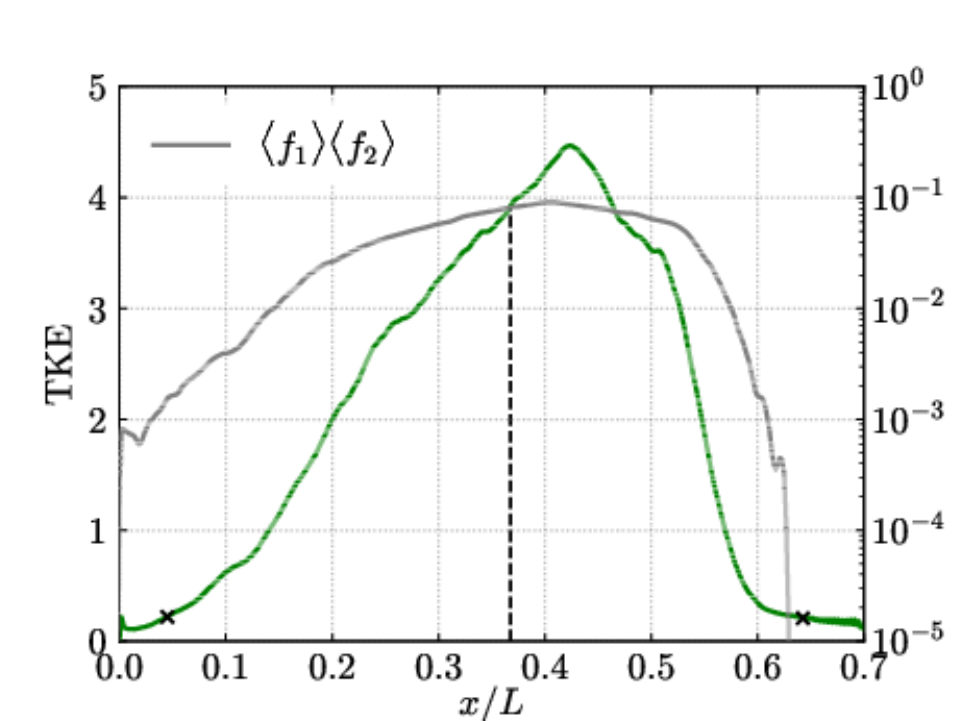}
		\caption{$m=-3$.}
	\end{subfigure}
	\begin{subfigure}{0.45\textwidth}
		\includegraphics[width=\textwidth]{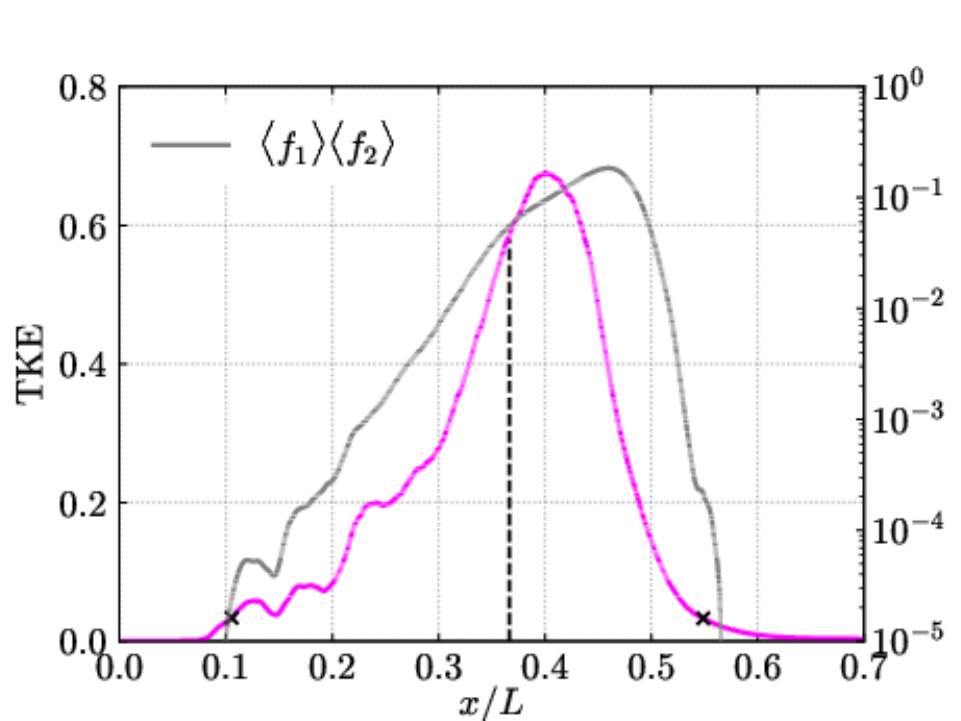}
		\caption{$m=0$ ($R=2$).}
	\end{subfigure}
	\caption{Plane-averaged profiles of $\mathrm{TKE}$ for each initial condition at time $\tau=57.4$. Solid lines indicate ILES results and dotted lines indicate DNS results. Also shown are the volume fraction profiles (grey solid lines), along with the 5\% cutoff locations (crosses) and the TKE centroid (black dashed lines). \label{fig:TKE-x}}
\end{figure}

In the experiments of \citet{Sewell2021}, PIV was used as the main diagnostic and therefore an alternate definition of the mixing layer width was required. In that study, the row-averaged turbulent kinetic energy was used and a mixing layer width defined as the distance between the $x$-locations at which the TKE is 5\% of its peak value. This definition assumes that the turbulent velocity field spreads at the same rate as the mixing layer. Figure \ref{fig:TKE-x} shows streamwise profiles of mean turbulent kinetic energy for each of the four initial conditions, defined as 
\begin{equation}
\mathrm{TKE} = \frac{1}{2}\overline{u_i^{\prime}u_i^{\prime}}
\label{eqn:TKE}
\end{equation} 
where $\psi^{\prime}=\psi-\overline{\psi}$ indicates a fluctuating quantity and the ensemble average $\overline{\psi}=\langle\psi\rangle$ is calculated as a plane average taken over the statistically homogeneous directions (in this case $y$ and $z$). The volume fraction profile $\langle f_1\rangle\langle f_2\rangle$ is also shown on the right axis of each plot, as well as the (outermost) $x$-locations at which the TKE is 5\% of its peak value. An important feature worth noting when comparing the narrowband case with the other broadband cases is that the 5\% cutoff on the spike side \firstrev{($x<x_c$)} is further from the mixing layer centre \firstrev{$x_c$} than in the $m=-1$ and $m=-2$ cases, despite these cases having a greater overall amplitude in the initial perturbation. There is also a greater amount of \firstrev{mixed} material, as measured by the product $\langle f_1\rangle\langle f_2\rangle$, at this location than in those two broadband cases, which is in line with the observations made in figure \ref{fig:f1-ILES} about \secondrev{the greater penetration distances of spikes} from the main layer \secondrev{in the narrowband case}. \thirdrev{In all cases the TKE profile is asymmetric, with the 5\% cutoff on the spike side being located further away from the mixing layer centre than the corresponding 5\% cutoff on the bubble side. This asymmetry, along with the implications it has for the growth rate exponent $\theta$, is discussed in further detail \S \ref{subsec:heights}.}

\begin{figure}
	\centering
	\begin{subfigure}{0.45\textwidth}
		\includegraphics[width=\textwidth]{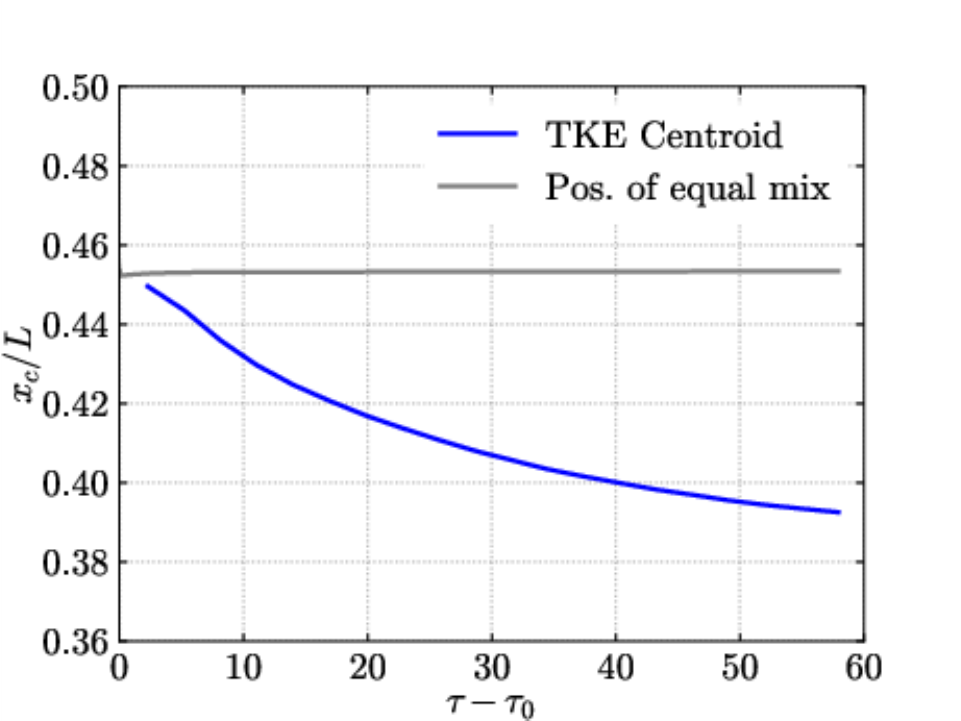}
		\caption{$m=-1$.}
	\end{subfigure}
	\begin{subfigure}{0.45\textwidth}
		\includegraphics[width=\textwidth]{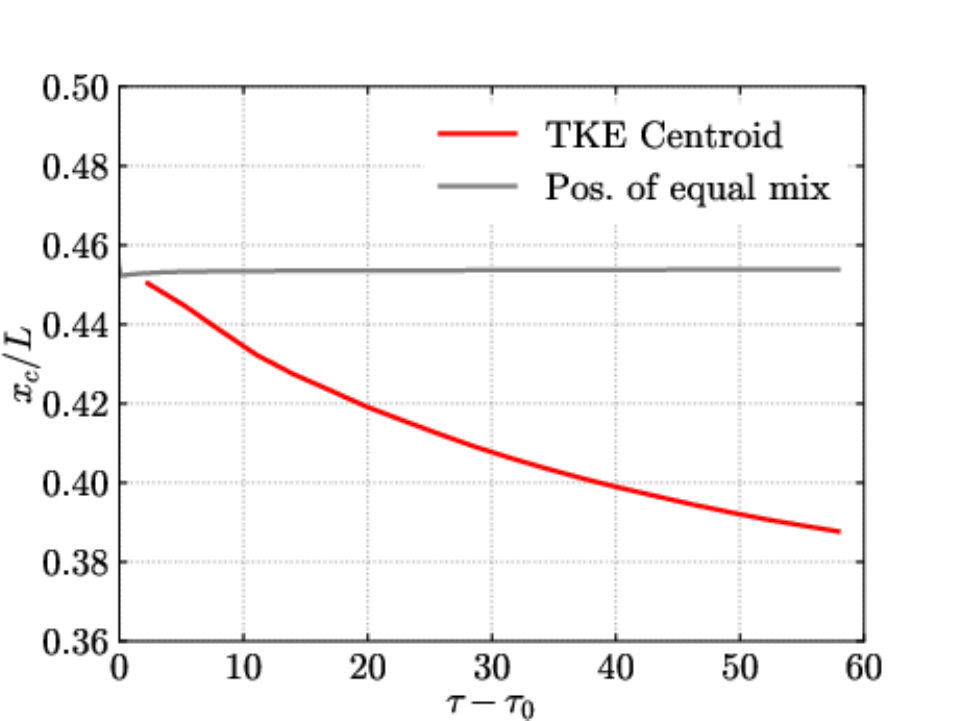}
		\caption{$m=-2$.}
	\end{subfigure}
	\begin{subfigure}{0.45\textwidth}
		\includegraphics[width=\textwidth]{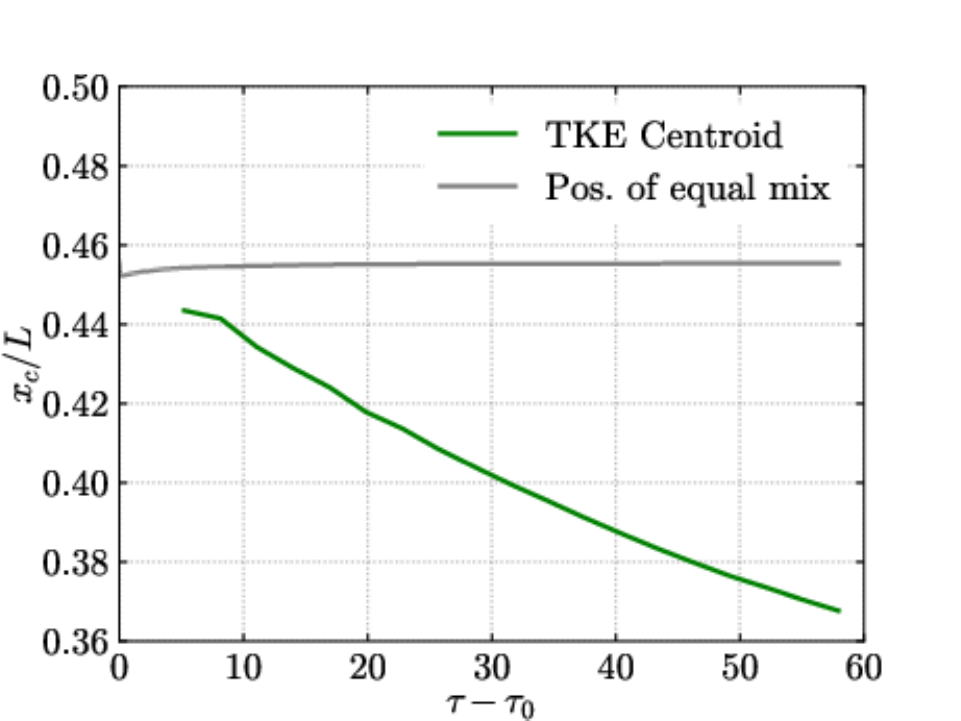}
		\caption{$m=-3$.}
	\end{subfigure}
	\begin{subfigure}{0.45\textwidth}
		\includegraphics[width=\textwidth]{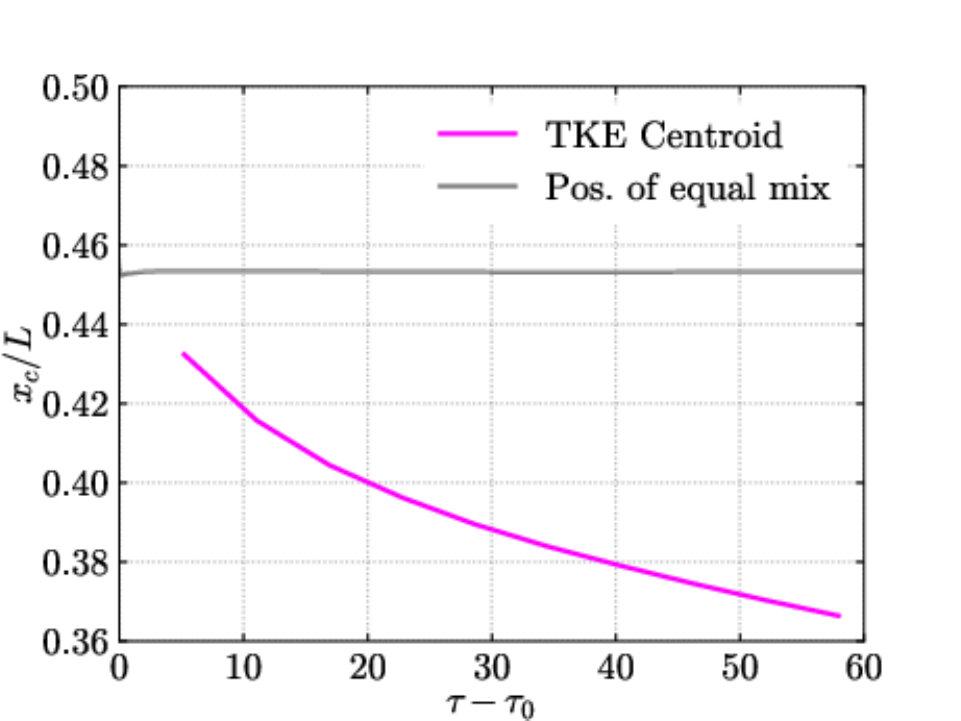}
		\caption{$m=0$ ($R=2$).}
	\end{subfigure}
	\caption{Temporal evolution of the mixing layer centre $x_c$, comparing the definition based on the centroid of the mean turbulent kinetic energy profile with the definition based on the $x$-location of equal mixed volumes. \label{fig:xc-tau}}
\end{figure}

In \citet{Sewell2021} a definition for the mixing layer centre is given as the centroid of the mean turbulent kinetic energy profile, i.e.
\begin{equation}
	x_c = \frac{\displaystyle\int xf(x)\:\mathrm{d} x}{\displaystyle\int f(x)\:\mathrm{d} x}
	\label{eqn:tke-centroid}
\end{equation}
where $f(x)$ is the mean turbulent kinetic energy profile. This centroid is also shown in figure \ref{fig:TKE-x}. This definition is compared with an alternate definition in terms of the $x$-location of equal mixed volumes,
\begin{equation}
\int_{-\infty}^{x_c}\langle f_2\rangle\:\mathrm{d}x=\int_{x_c}^{\infty}\langle f_1\rangle\:\mathrm{d}x
\label{eqn:xc}
\end{equation}
which has been used previously in both computational \citep{Walchli2017,Groom2021} and experimental \citep{Krivets2017} studies of RMI. Figure \ref{fig:xc-tau} plots the temporal evolution of both of these definitions for $x_c$ for each initial condition, showing that the TKE centroid consistently drifts towards the spike side of the layer as time progresses. The definition in terms of position of equal mixed volumes is much more robust and remains virtually constant throughout the simulation. There is also little variation between cases for this definition, unlike the TKE centroid which is more biased towards the spike side in the $m=-3$ and $m=0$ cases. The choice of definition for the mixing layer centre is important as it will influence the bubble and spike heights that are based off it (as well as their ratio), along with any quantities that are plotted at the mixing layer centre over various points in time.

\begin{figure}
	\centering
	\begin{subfigure}{0.45\textwidth}
		\includegraphics[width=\textwidth]{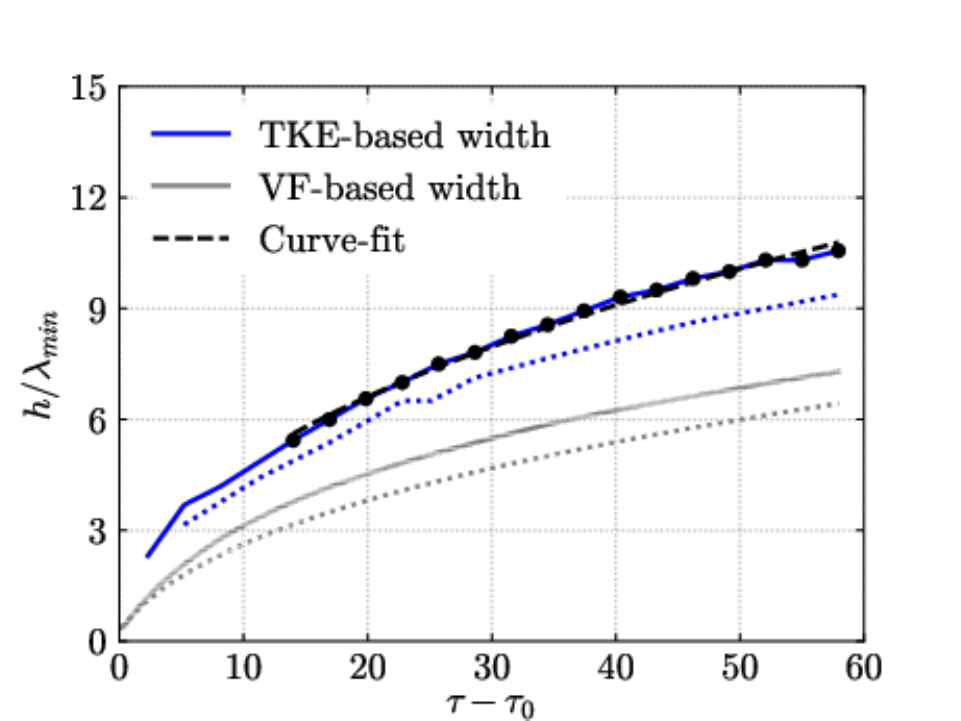}
		\caption{$m=-1$.}
	\end{subfigure}
	\begin{subfigure}{0.45\textwidth}
		\includegraphics[width=\textwidth]{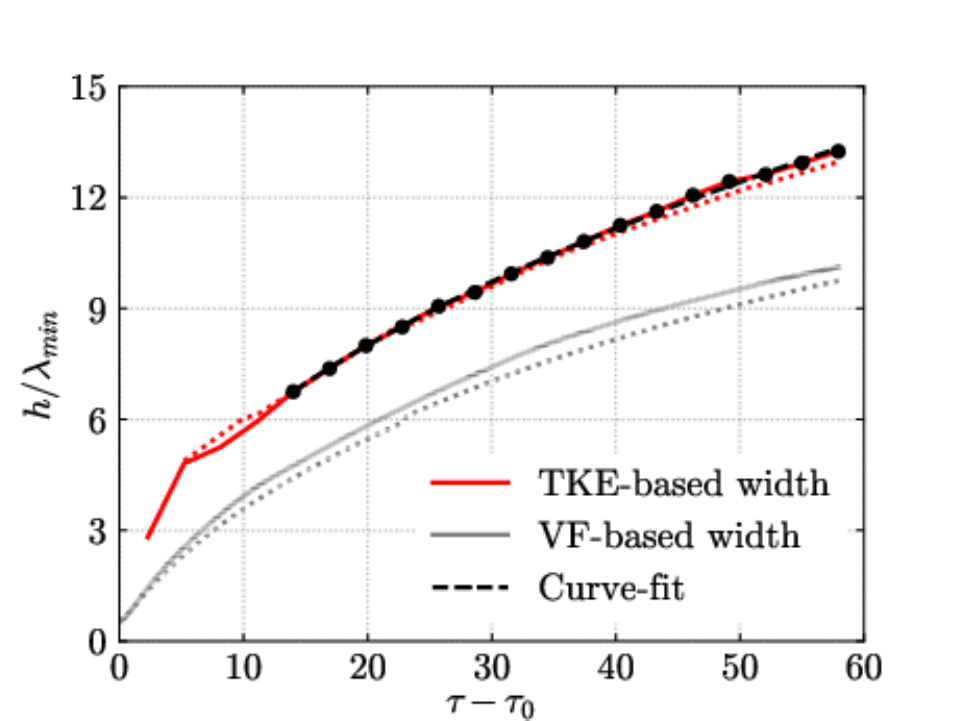}
		\caption{$m=-2$.}
	\end{subfigure}
	\begin{subfigure}{0.45\textwidth}
		\includegraphics[width=\textwidth]{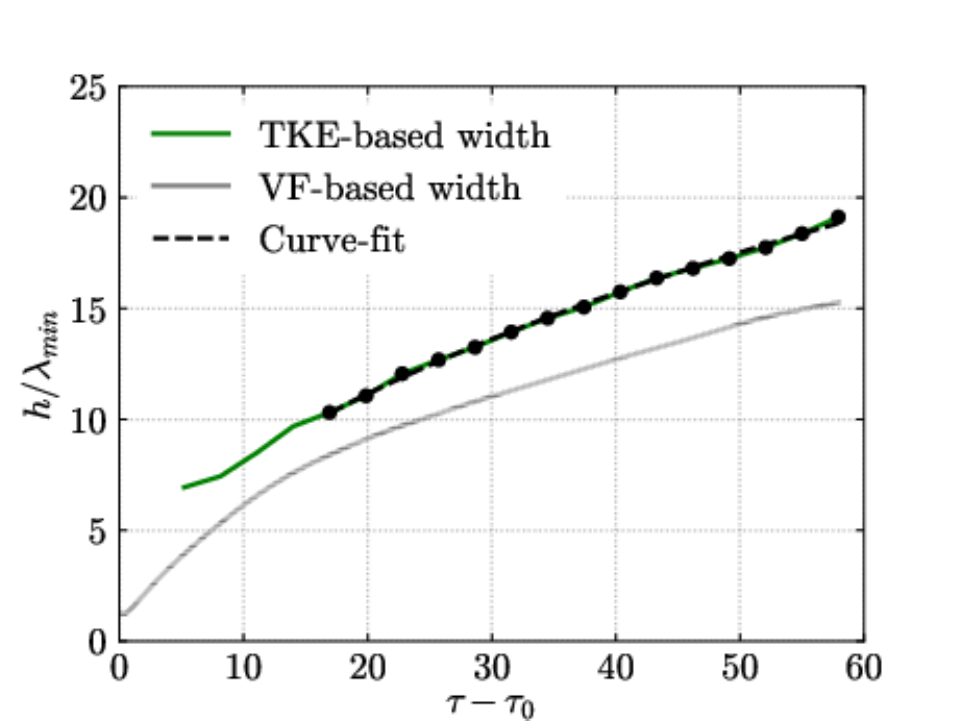}
		\caption{$m=-3$.}
	\end{subfigure}
	\begin{subfigure}{0.45\textwidth}
		\includegraphics[width=\textwidth]{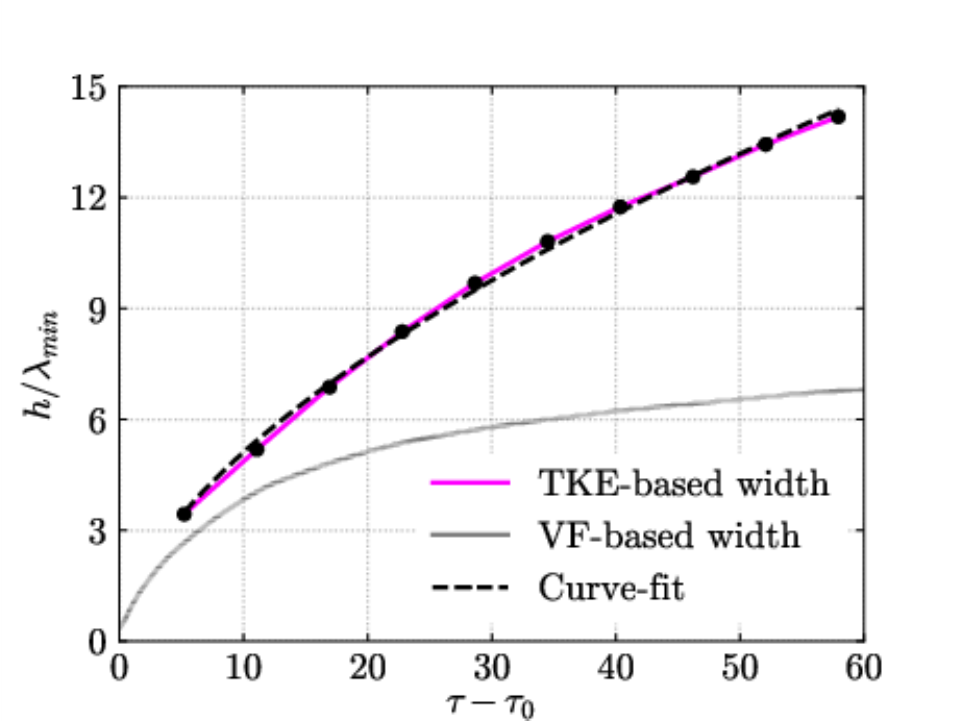}
		\caption{$m=0$ ($R=2$).}
	\end{subfigure}
	\caption{Temporal evolution of mixing layer width $h$ based on the distance between cutoff locations using either the mean turbulent kinetic energy or mean volume fraction profiles. Solid lines indicate ILES results and dotted lines indicate DNS results. Curve-fits to the data are also shown, with the relevant data points used given by the symbols in each plot.  \label{fig:h-tau}}
\end{figure}

Figure \ref{fig:h-tau} shows the temporal evolution of the mixing layer width, using both the visual width definition based on the mean volume fraction profile (referred to as the VF-based width) as well as the definition from \citet{Sewell2021} based on the distance between the 5\% cutoff locations in the mean turbulent kinetic energy profile (referred to as the TKE-based width). The mean volume fraction $f_1$ at these 5\% cutoff locations is $\ge 0.997$ on the spike side \firstrev{($x<x_c$)} and $\le 0.003$ on the bubble side \firstrev{($x>x_c$)} in all cases, hence why the TKE-based width is larger than the VF-based width in each of the plots \firstrev{as the VF-based width is defined using a 1\% and 99\% cutoff in the volume fraction profile}. Using nonlinear regression to fit a function of the form $h=\beta(\tau-\tau_0)^\theta$, the growth rate exponent $\theta$ can be obtained for the TKE-based width, VF-based width and the integral width (not shown in figure \ref{fig:h-tau}) for each case. Following \citet{Sewell2021}, the fit is performed only for times satisfying $\bar{k}\dot{\sigma_0}t>1$ so that the flow is sufficiently developed. The estimated value of $\theta$ for each case is given in table \ref{tab:theta}. \thirdrev{Note that the uncertainties reported are merely taken from the variance of the curve-fit and do not represent uncertainties in the true value of $\theta$.}

Analysing the results in table \ref{tab:theta}, there is good agreement between the values of $\theta$ obtained from the visual and integral widths for all cases. This is mainly a verification that the results are not severely impacted by a lack of statistical resolution at the lowest wavenumbers, which would result in the visual width measurements being dependent on the specific realisation. \thirdrev{The small differences in the values of $\theta$ reported indicate that there is still some influence of statistical fluctuations, therefore the estimates made using the integral width should be regarded as the most accurate.} When comparing the TKE-based and VF-based threshold widths, there is good agreement for the broadband ILES cases and in particular for the $m=-3$ ILES case. For the narrowband ILES case however, the VF-based (and integral) width is growing at close to the theoretical value of $\theta=1/3$ for self-similar decay proposed by \cite{Elbaz2018}, whereas the TKE-based width is growing at a much fast rate of $\theta=0.589$. This is even faster than any of the broadband cases and is due to the sensitivity of the TKE-based width to spikes \secondrev{located far from the mixing layer centre} in the narrowband case, which contain very little material but are quite energetic and which grow at a faster rate than the rest of the mixing layer. For the broadband DNS, the growth rate of the TKE-based width is slightly lower than that of the VF-based width for \thirdrev{both} cases, indicating that turbulent fluctuations are more confined to the \secondrev{core} of the mixing layer. \thirdrev{In the $m=-1$ case, the value of $\theta$ obtained from the integral width is Reynolds number independent, while for $m=-2$ the value of $\theta$ obtained from the integral width in the DNS case is converging towards the high Reynolds number limit given by the ILES case.} Given that the broadband perturbations, specifically the $m=-3$ perturbation, are the most relevant to the experiments in \citet{Jacobs2013} and \citet{Sewell2021}, it is reassuring to note that estimates of $\theta$ made using TKE-based widths measured with PIV correspond well with estimates based off the concentration field.

\begin{table}
	\begin{center}
		\def~{\hphantom{0}}
		\begin{tabular}{lcccccc}
			Case & $m$ & $\Rey_0$ & TKE-based width $\theta$ & VF-based width $\theta$ & Integral width $\theta$ & TKE decay rate $\theta$ \\[3pt]
			1 & 0  & -    & $0.589\pm 1.20\times 10^{-2}$ & $0.323\pm 4.89\times 10^{-3}$ & $0.330\pm 1.27\times 10^{-3}$ & $0.253\pm 7.00\times 10^{-3}$ \\
			2 & -1 & -    & $0.460\pm 1.03\times 10^{-2}$ & $0.450\pm 1.54\times 10^{-3}$ & $0.442\pm 1.10\times 10^{-4}$ & $0.429\pm 5.65\times 10^{-3}$ \\
			3 & -2 & -    & $0.479\pm 3.92\times 10^{-3}$ & $0.522\pm 3.59\times 10^{-3}$ & $0.514\pm 3.60\times 10^{-4}$ & $0.512\pm 3.47\times 10^{-3}$  \\
			4 & -3 & -    & $0.493\pm 6.25\times 10^{-3}$ & $0.492\pm 1.25\times 10^{-3}$ & $0.510\pm 1.91\times 10^{-3}$ & $0.562\pm 2.22\times 10^{-3}$ \\
			5 & -1 & 261  & $0.444\pm 1.41\times 10^{-2}$ & $0.501\pm 8.40\times 10^{-4}$ & $0.441\pm 1.00\times 10^{-4}$ & $0.492\pm 8.08 \times 10^{-3}$\\
			6 & -2 & 526  & $0.456\pm 3.69\times 10^{-3}$ & $0.556\pm 2.27\times 10^{-3}$ & $0.549\pm 1.52\times 10^{-3}$ & $0.576\pm 4.69 \times 10^{-3}$ \\
		\end{tabular}
		\caption{Estimates of the growth rate \thirdrev{exponent} $\theta$ from curve-fits to the TKE-based, VF-based and integral widths, as well as from the decay rate of total turbulent kinetic energy.}
		\label{tab:theta}
	\end{center}
\end{table}

\begin{figure}
	\centering
	\begin{subfigure}{0.45\textwidth}
		\includegraphics[width=\textwidth]{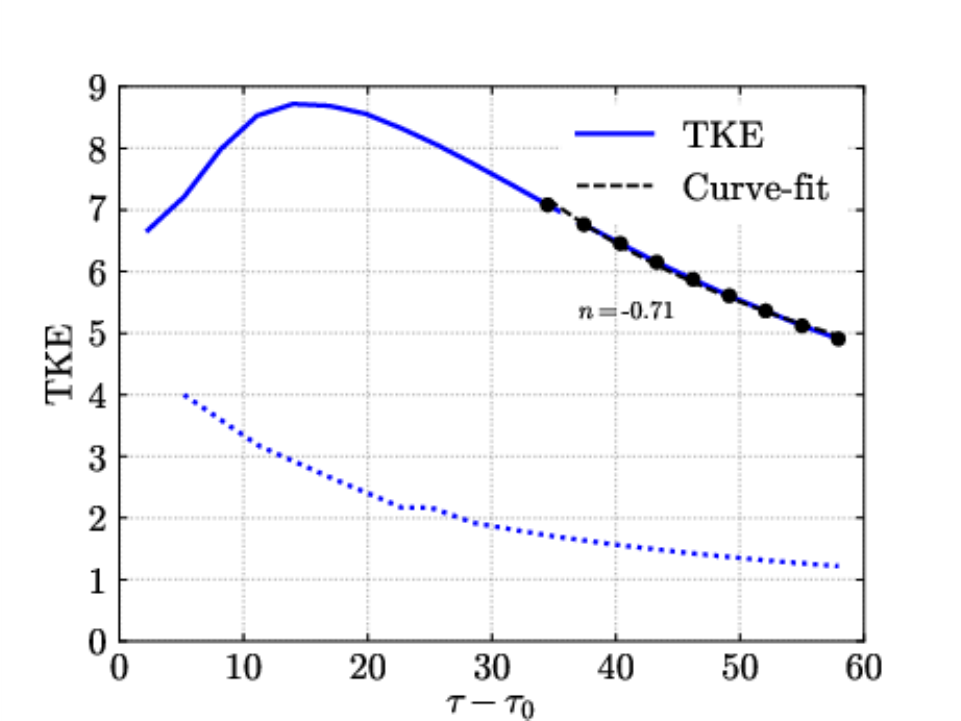}
		\caption{$m=-1$.}
	\end{subfigure}
	\begin{subfigure}{0.45\textwidth}
		\includegraphics[width=\textwidth]{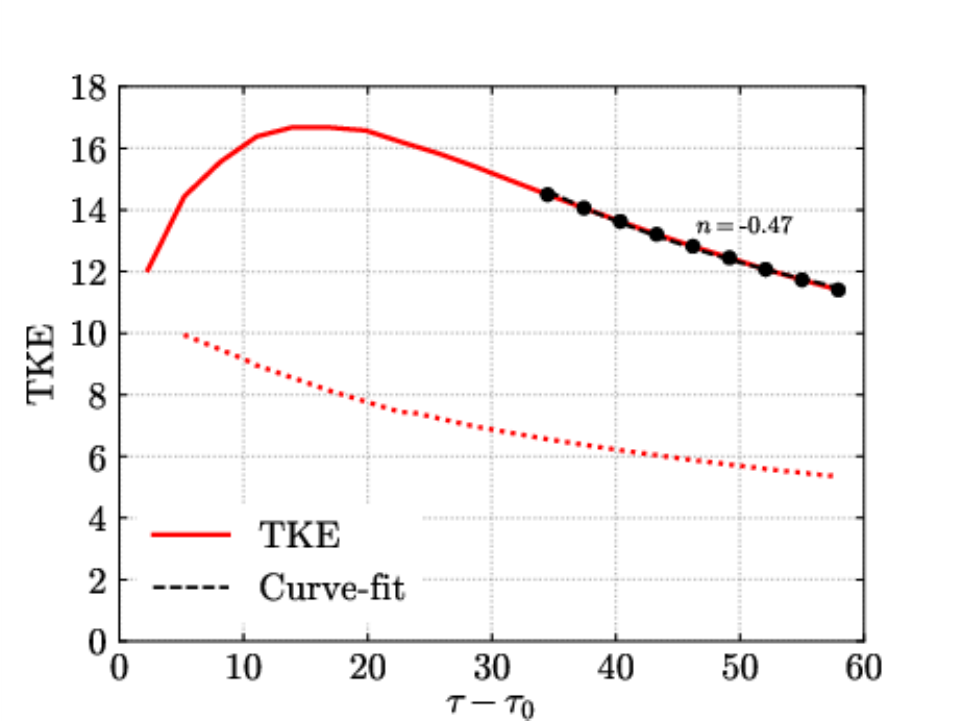}
		\caption{$m=-2$.}
	\end{subfigure}
	\begin{subfigure}{0.45\textwidth}
		\includegraphics[width=\textwidth]{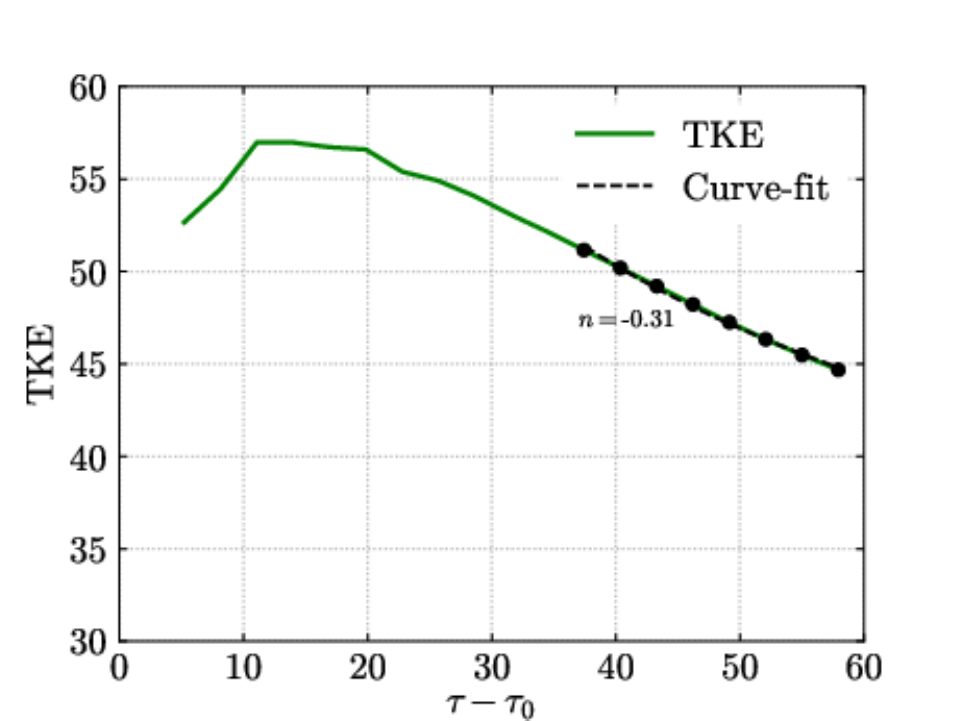}
		\caption{$m=-3$.}
	\end{subfigure}
	\begin{subfigure}{0.45\textwidth}
		\includegraphics[width=\textwidth]{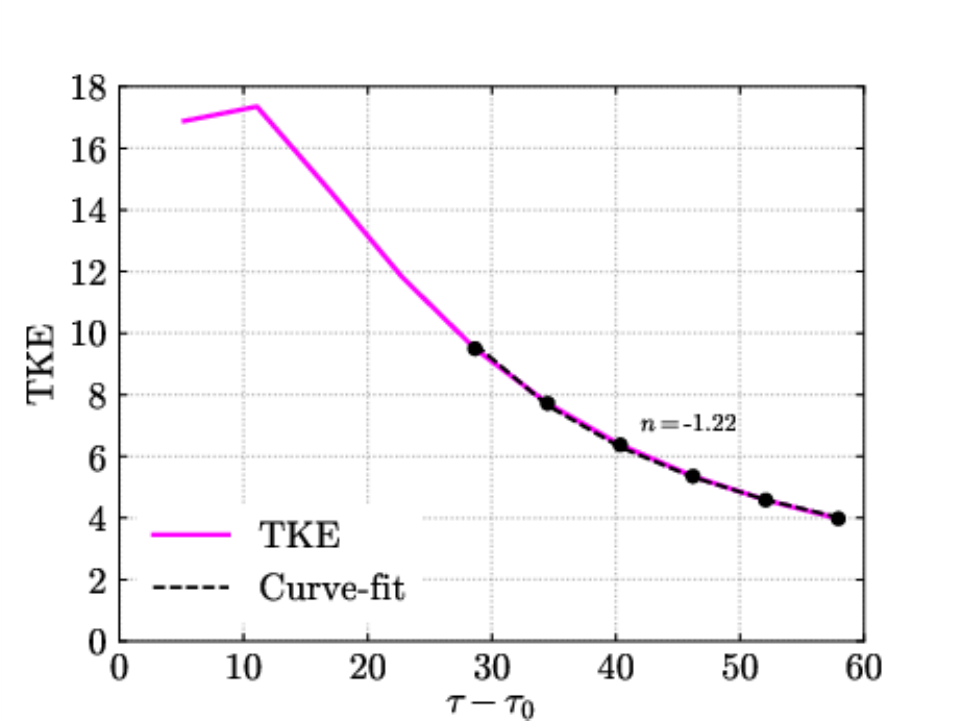}
		\caption{$m=0$ ($R=2$).}
	\end{subfigure}
	\caption{Temporal evolution of total fluctuating kinetic energy, integrated between the 5\% cutoff locations. Solid lines indicate ILES results and dotted lines indicate DNS results. Curve-fits to the data are also shown, with the relevant data points used given by the symbols in each plot. \label{fig:TKE-tau}}
\end{figure}

An alternative method for estimating \firstrev{$\theta$} is also given in \citet{Sewell2021}, which makes use of the decay rate of total fluctuating kinetic energy and a relationship between this decay rate $n$ and the mixing layer growth rate $\theta$ originally derived by \citet{Thornber2010}. Assuming that $h\propto t^\theta$ and the mean fluctuating kinetic energy $q_k \propto \dot{h}^2$ gives the relation $q_k\propto t^{2\theta-2}$. Since the total fluctuating kinetic energy is proportional to the width of the mixing layer multiplied by the mean fluctuating kinetic energy, this gives $\mathrm{TKE} \propto t^{3\theta-2} \propto t^n$. Directly measuring the decay rate $n$ therefore gives an alternative method for estimating $\theta$, which is particularly useful in experimental settings where only velocity field data is available. This predicted value of $\theta = (n+2)/3$ has been found to be in good agreement with the measured growth rate from the integral width in multiple studies of narrowband RMI \cite{Thornber2010,Thornber2017}. However, \citet{Groom2020} showed that for RMI evolving from broadband perturbations with bandwidths as large as $R=128$ the measured values of $\theta$ do not agree with this theoretical prediction, indicating that longer periods \firstrev{of} growth dominated by just-saturating modes are required than can currently be obtained in simulations. Figure \ref{fig:TKE-tau} shows the temporal evolution of TKE, where the integration has been performed between the 5\% cutoff locations used to define the TKE-based width. Nonlinear least squares regression is again used to estimate $n$ for each case, with the fit performed for times greater than the point at which the curvature becomes convex. The corresponding value of $\theta$ for each $n$ using the relation $n=3\theta-2$ is given in table \ref{tab:theta}. 

For the narrowband case the estimate of $\theta$ from the TKE decay rate does not agree with the other estimates, indicating that the mixing layer growth is not sufficiently self-similar (a key assumption in the derivation) and lags the decay in TKE. This is still true even when the range of times used in the curve-fitting procedure is restricted \firstrev{to} be the same as for the curve-fit to the decay rate (not shown). For the broadband cases there is better agreement however, particularly in the $m=-1$ and $m=-2$ ILES cases. In all broadband cases the bandwidth of the initial perturbation is relatively small compared to the perturbations analysed in \citet{Groom2020} and the longest initial wavelength saturates early on in the overall simulation, therefore the conclusions made in that study regarding the $n=3\theta-2$ relation do not necessarily apply here as the current broadband cases are not in the self-similar growth regime. They are also likely not in full self-similar decay however, especially if the narrowband case is not, yet the values of $\theta$ are in better agreement than in the narrowband case. Further work is required to determine why this is indeed the case. 

Comparing the estimates of $\theta$ with those in \citet{Sewell2021} using both the TKE-based width and TKE decay rate, the $m=-3$ simulation results are in between the results of the low-amplitude and high-amplitude experiments. For the low-amplitude experiments (prior to reshock), the TKE-based width measurements gave $\theta=0.45$ and the TKE decay rate measurements gave $\theta=0.68$ (which would correspond to no decay of TKE if the layer was homogeneous \citep{Barenblatt1983}). The equivalent results in the $m=-3$ simulation were $\theta=0.493$ and $\theta=0.562$, i.e. larger and smaller than the respective experimental results but both within the experimental margins of error. Similarly for the high-amplitude experiments, both the TKE-based width measurements and the TKE decay rate measurements gave $\theta=0.51$, indicating that the turbulence in the mixing layer is more developed and closer to self-similar prior to reshock. The $m=-3$ simulation results are also within the experimental margins of error for these results. Overall, the combination of experimental and computational evidence indicates that there are persistent effects of initial conditions when broadband surface perturbations are present for a much greater period of time than just the time to saturation of the longest initial wavelength (as considered in previous simulation studies of broadband RMI) and last for the duration of the first-shock growth in a typical shock tube experiment. Furthermore, a consideration of the impact of finite bandwidth in the initial power spectrum (also referred to as confinement) is required when adapting theoretical results for infinite bandwidth (unconfined, see \citet{Youngs2004,Thornber2010,Soulard2018,Soulard2022}) to a specific application.

\subsection{Bubble and Spike Heights}
\label{subsec:heights}
\begin{figure}
	\centering
	\begin{subfigure}{0.45\textwidth}
		\includegraphics[width=\textwidth]{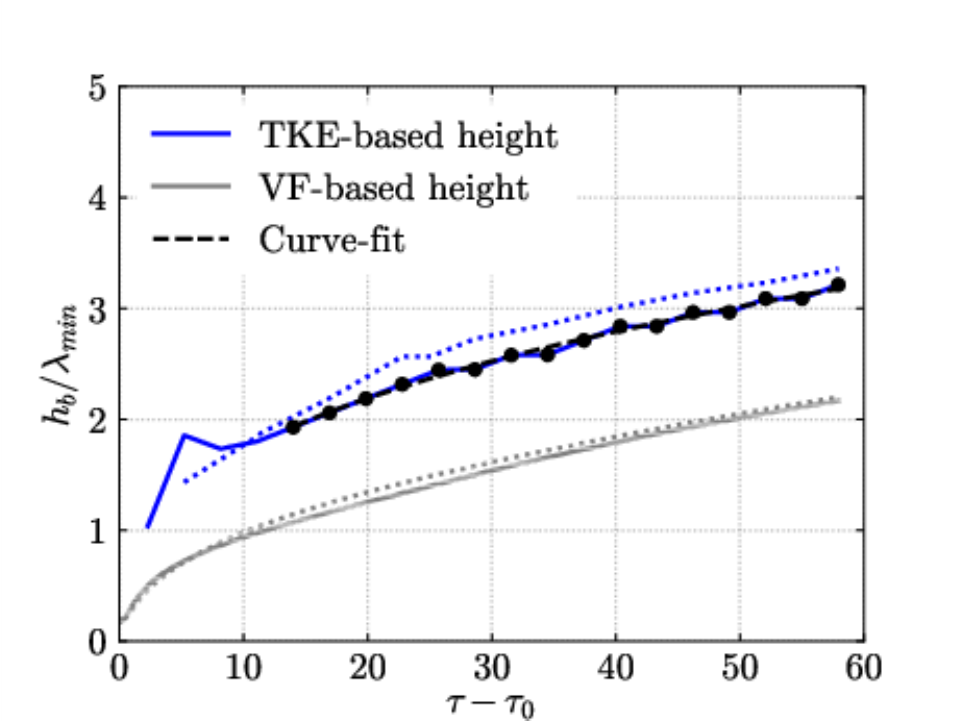}
		\caption{$m=-1$.}
	\end{subfigure}
	\begin{subfigure}{0.45\textwidth}
		\includegraphics[width=\textwidth]{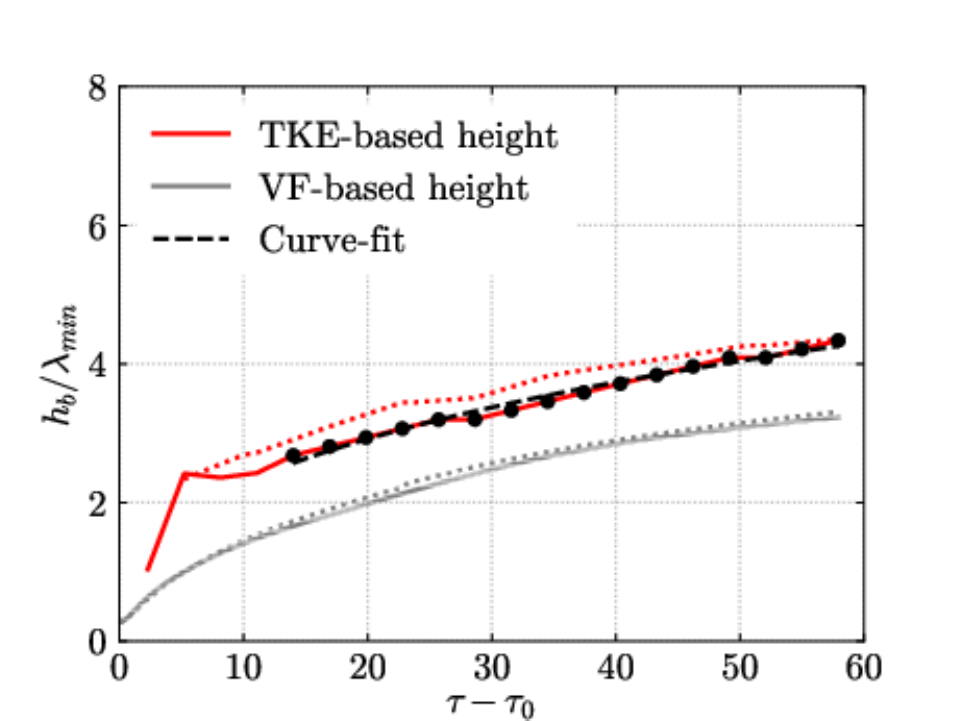}
		\caption{$m=-2$.}
	\end{subfigure}
	\begin{subfigure}{0.45\textwidth}
		\includegraphics[width=\textwidth]{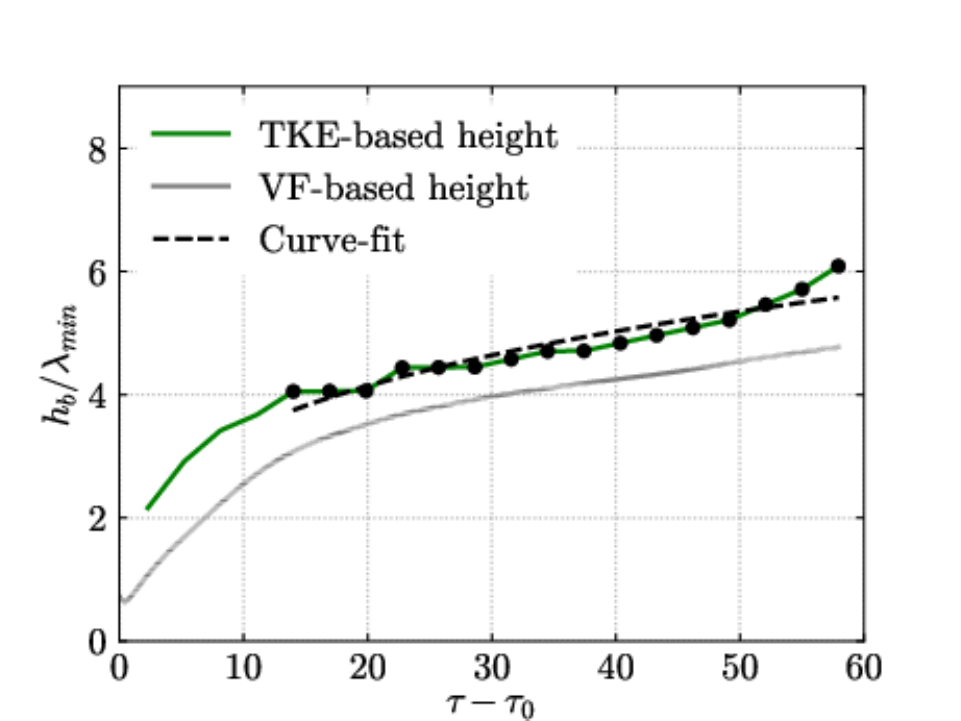}
		\caption{$m=-3$.}
	\end{subfigure}
	\begin{subfigure}{0.45\textwidth}
		\includegraphics[width=\textwidth]{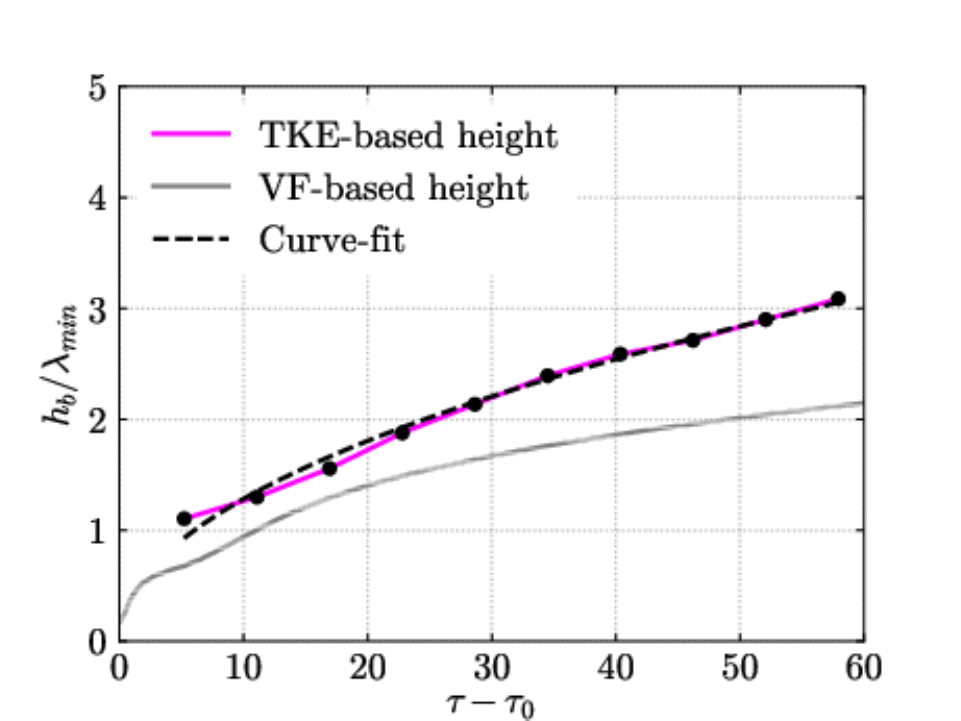}
		\caption{$m=0$ ($R=2$).}
	\end{subfigure}
	\caption{\thirdrev{Temporal evolution of the bubble height $h_b$ based on the distance between cutoff locations using either the mean turbulent kinetic energy or mean volume fraction profiles. Solid lines indicate ILES results and dotted lines indicate DNS results. Curve-fits to the data are also shown, with the relevant data points used given by the symbols in each plot.}  \label{fig:hb-tau}}
\end{figure}

\begin{figure}
	\centering
	\begin{subfigure}{0.45\textwidth}
		\includegraphics[width=\textwidth]{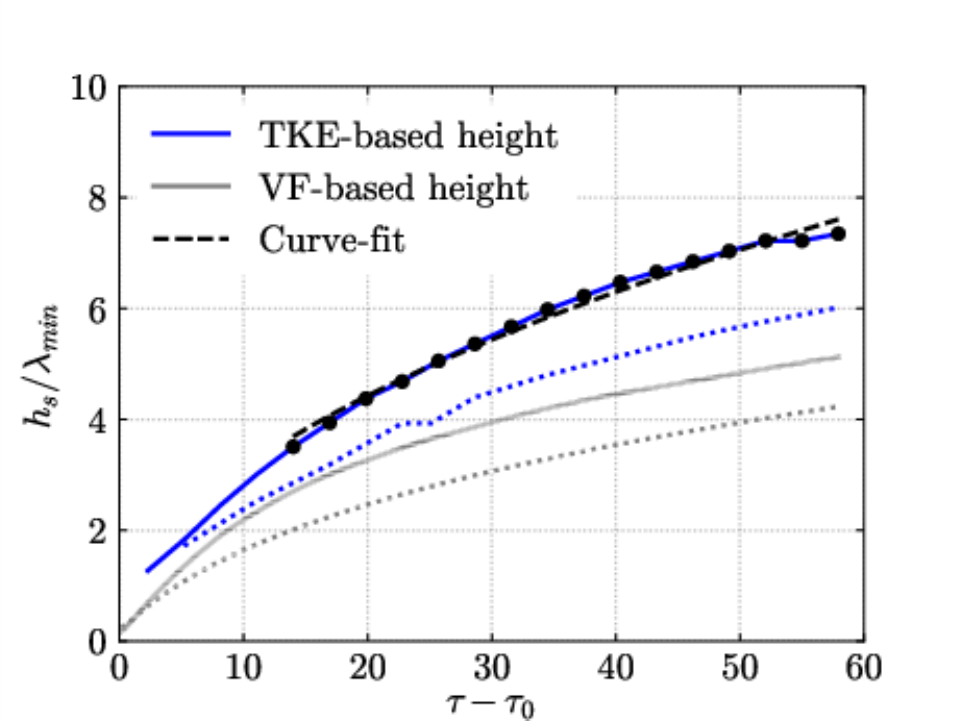}
		\caption{$m=-1$.}
	\end{subfigure}
	\begin{subfigure}{0.45\textwidth}
		\includegraphics[width=\textwidth]{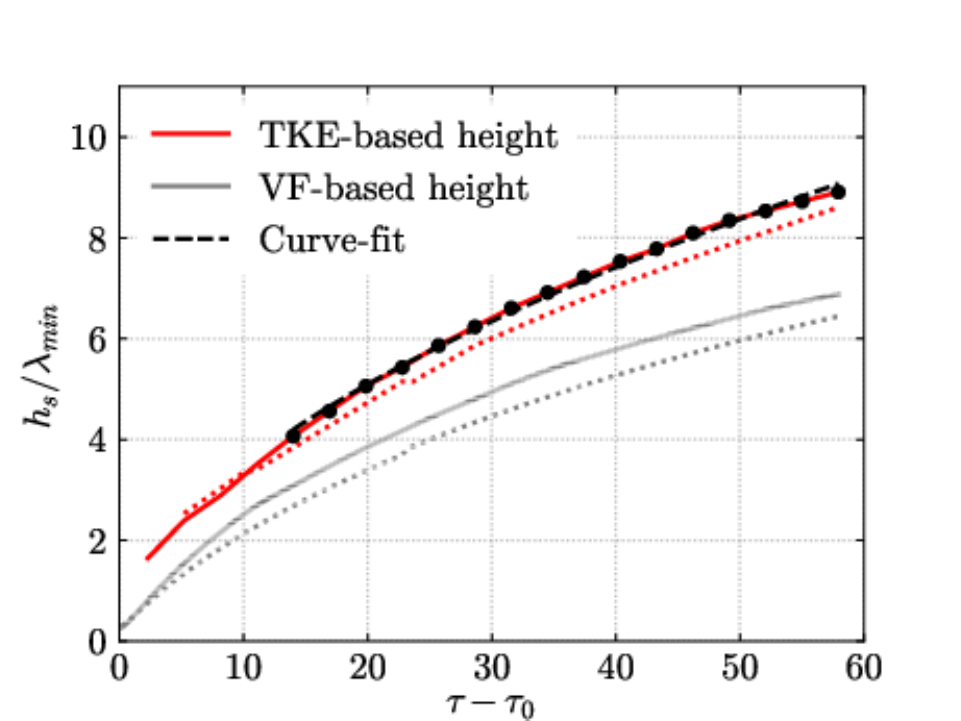}
		\caption{$m=-2$.}
	\end{subfigure}
	\begin{subfigure}{0.45\textwidth}
		\includegraphics[width=\textwidth]{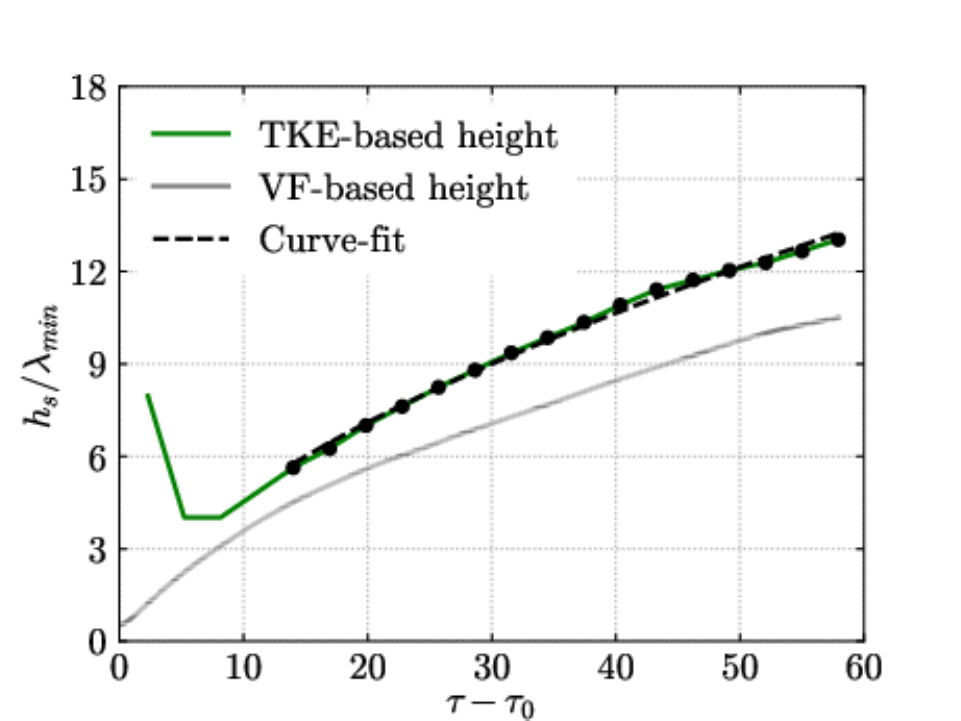}
		\caption{$m=-3$.}
	\end{subfigure}
	\begin{subfigure}{0.45\textwidth}
		\includegraphics[width=\textwidth]{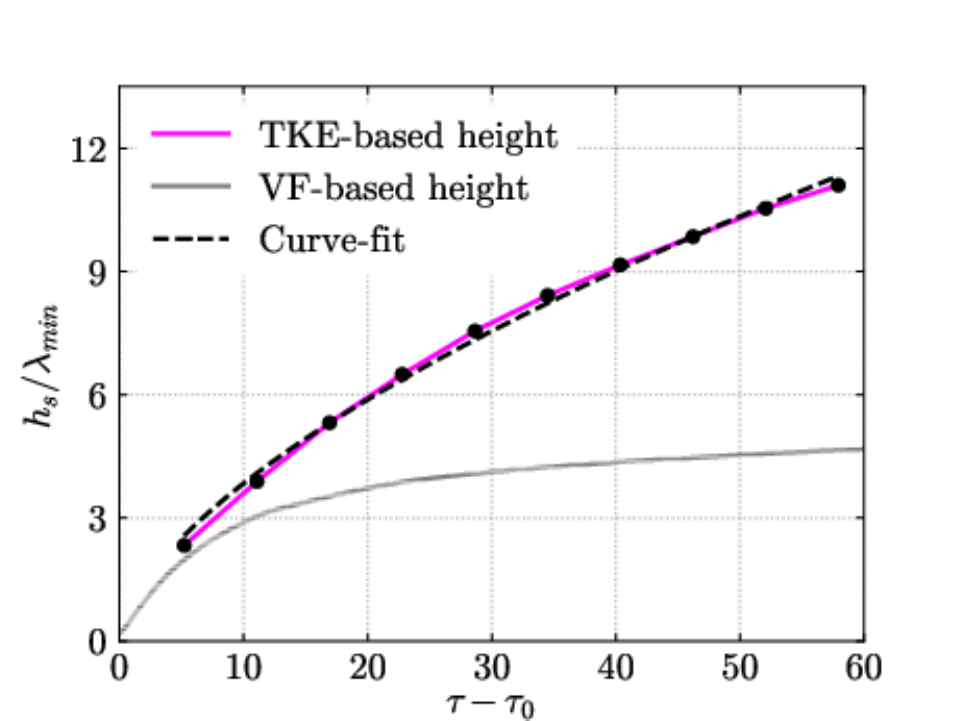}
		\caption{$m=0$ ($R=2$).}
	\end{subfigure}
	\caption{\thirdrev{Temporal evolution of the spike height $h_s$ based on the distance between cutoff locations using either the mean turbulent kinetic energy or mean volume fraction profiles. Solid lines indicate ILES results and dotted lines indicate DNS results. Curve-fits to the data are also shown, with the relevant data points used given by the symbols in each plot.}  \label{fig:hs-tau}}
\end{figure}

\thirdrev{In order to help better explain the estimates for $\theta$ given in table \ref{tab:theta}, it is useful to decompose the TKE-based and VF-based widths into separate bubble and spike heights, $h_b$ and $h_s$, defined as the distance from the mixing layer centre $x_c$ to the relevant cutoff location on the bubble and spike side of the layer respectively. Given the drift in time for the centroid of the TKE profile shown in figure \ref{fig:xc-tau}, the $x$-location of equal mixed volumes is used as the definition of the mixing layer centre for both the VF-based and TKE-based bubble and spike heights. Figures \ref{fig:hb-tau} and \ref{fig:hs-tau} show the evolution in time of $h_b$ and $h_s$ respectively for heights based off both the 5\% TKE cutoff (referred to as TKE-based heights) and the 1\% and 99\% volume fraction cutoff (referred to as VF-based heights).}

\begin{figure}
	\centering
	\begin{subfigure}{0.45\textwidth}
		\includegraphics[width=\textwidth]{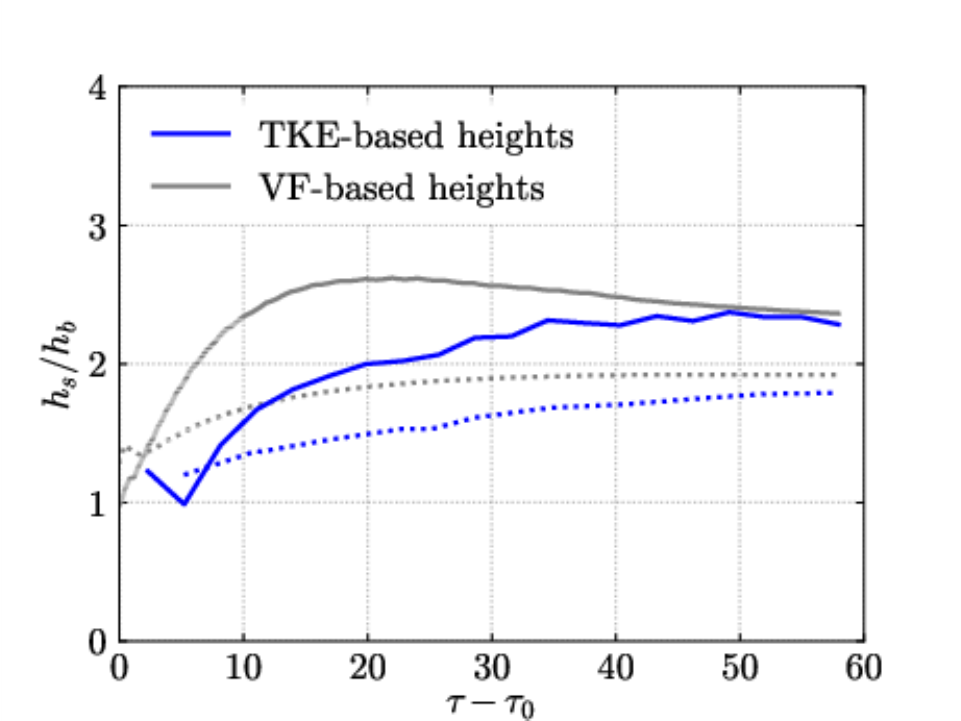}
		\caption{$m=-1$.}
	\end{subfigure}
	\begin{subfigure}{0.45\textwidth}
		\includegraphics[width=\textwidth]{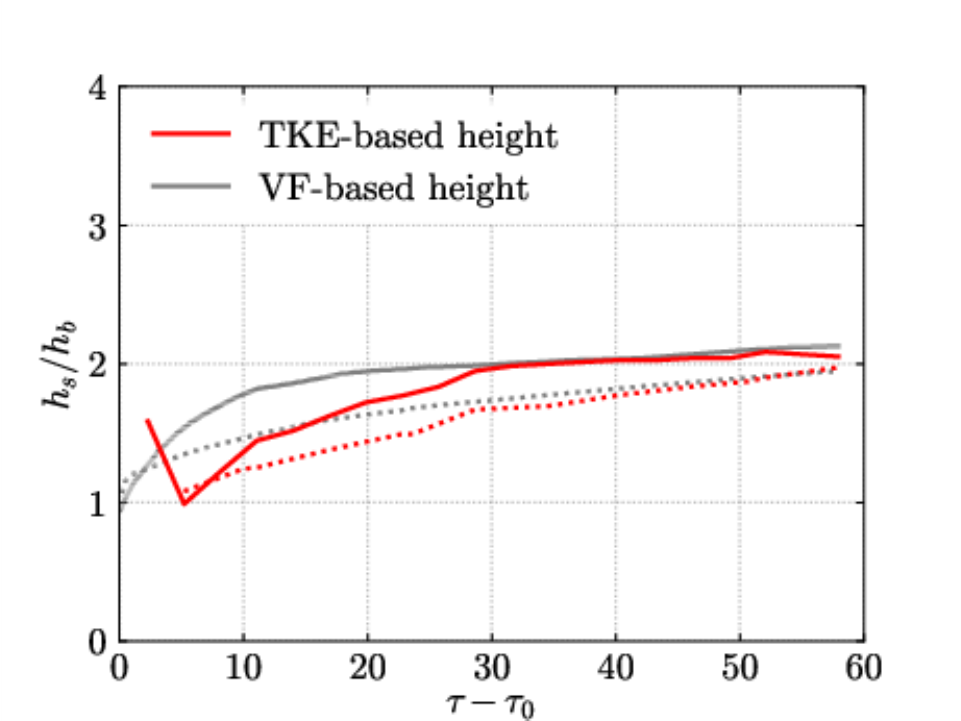}
		\caption{$m=-2$.}
	\end{subfigure}
	\begin{subfigure}{0.45\textwidth}
		\includegraphics[width=\textwidth]{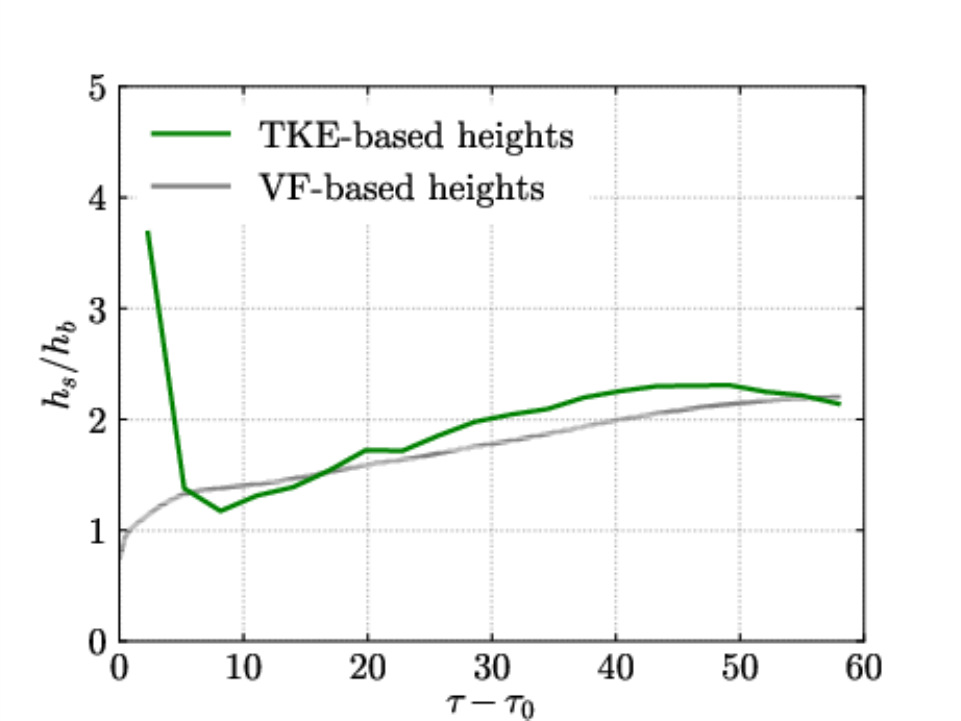}
		\caption{$m=-3$.}
	\end{subfigure}
	\begin{subfigure}{0.45\textwidth}
		\includegraphics[width=\textwidth]{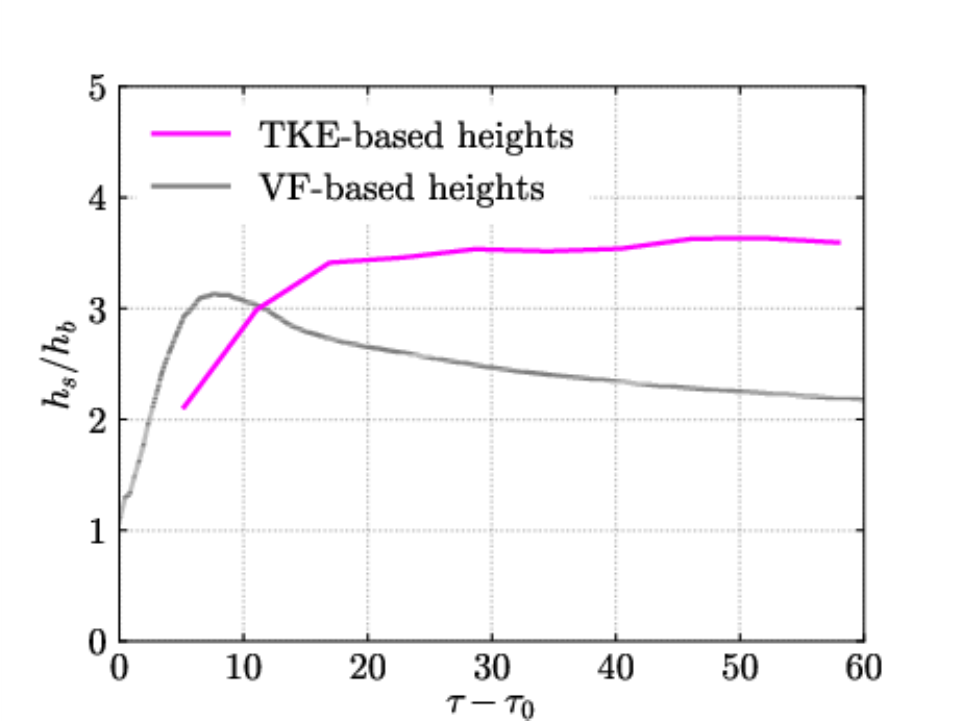}
		\caption{$m=0$ ($R=2$).}
	\end{subfigure}
	\caption{\thirdrev{Temporal evolution of the ratio of spike to bubble height. Solid lines indicate ILES results and dotted lines indicate DNS results. Curve-fits to the data are also shown, with the relevant data points used given by the symbols in each plot.}  \label{fig:ratio-tau}}
\end{figure}

\thirdrev{Some important trends can be observed. Firstly, the VF-based heights are smoother than the corresponding TKE-based heights indicating that they are less sensitive to statistical fluctuations. Secondly, the TKE-based $h_b$ and $h_s$ are greater than the corresponding VF-based heights in all cases and for both measures the spike height is greater than the bubble height. This can also be seen in figure \ref{fig:ratio-tau}, which plots the ratio $h_s/h_b$ vs. time and shows that $h_s/h_b>1$ for all cases. The same trend was observed in \citet{Youngs2020b} for both $At=0.5$ and $At=0.9$ but in a heavy-light configuration where the heavy spikes are being driven into the lighter fluid in the same direction as the shock wave. Appendix \ref{app:integral} plots the same integral definitions of the bubble and spike heights used in \citet{Youngs2020b}, verifying that the behaviour is very similar to the VF-based heights presented here. The ratio of spike to bubble heights using both threshold measures is also very similar at late time in all cases with the exception of the narrowband case. The ratio $h_s/h_b$ also appears to be converging to the same value at late-time in all cases except for the TKE-based heights in the narrowband case, suggesting it is only dependent on the Atwood number.}

\begin{table}
	\begin{center}
		\def~{\hphantom{0}}
		\begin{tabular}{lcccccc}
			Case & $m$ & $\Rey_0$ & TKE-based $\theta_b$ & VF-based $\theta_b$ & TKE-based $\theta_s$ & VF-based $\theta_s$ \\[3pt]
			1 & 0  & -    & $0.493\pm 2.43\times 10^{-2}$ & $0.441\pm 2.43\times 10^{-2}$  & $0.615\pm 1.72\times 10^{-2}$ & $0.277\pm 5.38\times 10^{-3}$\\
			2 & -1 & -    & $0.350\pm 8.94\times 10^{-3}$ & $0.514\pm 1.04\times 10^{-3}$  & $0.509\pm 1.57\times 10^{-2}$ & $0.425\pm 2.46\times 10^{-3}$\\
			3 & -2 & -    & $0.355\pm 1.31\times 10^{-2}$ & $0.466\pm 4.29\times 10^{-3}$  & $0.543\pm 8.58\times 10^{-3}$ & $0.550\pm 3.31\times 10^{-3}$\\
			4 & -3 & -    & $0.280\pm 2.95\times 10^{-2}$ & $0.282\pm 1.79\times 10^{-3}$  & $0.586\pm 1.07\times 10^{-2}$ & $0.606\pm 1.49\times 10^{-3}$\\
			5 & -1 & 261  & $0.338\pm 1.36\times 10^{-2}$ & $0.461\pm 3.20 \times 10^{-4}$ & $0.509\pm 1.61\times 10^{-2}$ & $0.523\pm 1.46\times 10^{-3}$\\
			6 & -2 & 526  & $0.284\pm 8.39\times 10^{-3}$ & $0.458\pm 2.57 \times 10^{-3}$ & $0.561\pm 4.89\times 10^{-3}$ & $0.613\pm 2.23\times 10^{-3}$\\
		\end{tabular}
		\caption{\thirdrev{Estimates of the growth rate exponents $\theta_b$ and $\theta_s$ from curve-fits to the TKE-based and VF-based bubble and spike heights.}}
		\label{tab:thetabs}
	\end{center}
\end{table}

\thirdrev{Figure \ref{fig:ratio-tau} shows that the ratio of $h_s/h_b$ is approximately constant by the end of the simulations. This indicates that a single $\theta$ is appropriate for describing the growth of the mixing layer beyond this point. However, prior to that $h_b$ and $h_s$ do grow at different rates as shown in table \ref{tab:thetabs}, where the bubble growth rate exponent is denoted by $\theta_b$ and the spike growth rate exponent is denoted by $\theta_s$. Two key trends can be observed; the VF-based $\theta_b$ is greater than the TKE-based $\theta_b$ in all cases other than the narrowband ($m=0$) case, while the VF-based $\theta_s$ is greater than the TKE-based $\theta_s$ in all cases other than the $m=-1$ ILES case and the narrowband case. The $m=-3$ case also has the smallest difference in $\theta_b$ and $\theta_s$ for both threshold measures. Comparing the DNS cases with their respective ILES cases, the VF-based $h_b$ is almost independent of the Reynolds number in both the $m=-1$ and $m=-2$ cases. This is also true for the TKE-based $h_s$ in the $m=-2$ cases. A higher degree of Reynolds number dependence is observed for both definitions of $h_s$, which is consistent with previous observations made about turbulence developing preferentially on the spike side of the mixing layer \cite{Groom2021}. This can also be observed for the integral definitions of $h_b$ and $h_s$ given in Appendix \ref{app:integral}.}

\thirdrev{This analysis provides evidence that, prior to reshock, $h_b$ and $h_s$ do grow at different rates in a typical shock tube experiment. However, their growth rate exponents have equalised by the time reshock arrives. This is a complicating factor when estimating a single value for $\theta$ at early times and points to the difficulties in obtaining self-similar growth for RMI in both experiments and simulations. This also suggests that the ratio of spike to bubble heights could be used to determine when it is appropriate to start curve-fitting for estimating a single value of $\theta$, and that measurements based on the concentration field are likely more accurate in this regard than those made using the velocity field.}

\subsection{Anisotropy}
\label{subsec:anisotropy}

The anisotropy of the fluctuating velocity field is explored using the same two measures presented in \citet{Sewell2021}. The first is a global measure of anisotropy, defined as
\begin{equation}
	\mathrm{TKR}=\frac{2\times\mathrm{TKX}}{\mathrm{TKY}+\mathrm{TKZ}}
\end{equation}
where $\mathrm{TKX}=\frac{1}{2}\overline{u^{\prime}u^{\prime}}$, $\mathrm{TKY}=\frac{1}{2}\overline{v^{\prime}v^{\prime}}$ and $\mathrm{TKZ}=\frac{1}{2}\overline{w^{\prime}w^{\prime}}$, with each quantity integrated between the cutoff locations based on 5\% of the maximum TKE. The second measure is the Reynolds stress anisotropy tensor, whose components are defined by
\begin{equation}
	b_{ij}=\frac{\overline{u_i^{\prime}u_j^{\prime}}}{\frac{1}{2}\overline{u_i^{\prime}u_i^{\prime}}}-\frac{1}{3}\delta_{ij}.
\end{equation}
This tensor, specifically the $x$-direction principal component $\boldsymbol{b}_{11}$ for this particular flow, is a measure of anisotropy in the energy-containing scales of the fluctuating velocity field with a value of 0 indicating isotropy in the direction of that component. The local version of TKR (i.e. with TKX, TKY and TKZ not integrated in the $x$-direction) can be written in terms of $\boldsymbol{b}_{11}$ as
\begin{equation}
	\frac{2\overline{u^{\prime}u^{\prime}}}{\overline{v^{\prime}v^{\prime}}+\overline{w^{\prime}w^{\prime}}}=\frac{2\boldsymbol{b}_{11}+2/3}{2/3-\boldsymbol{b}_{11}}
\end{equation}
allowing the two measures to be related to one another. 

\begin{figure}
	\centering
	\begin{subfigure}{0.45\textwidth}
		\includegraphics[width=\textwidth]{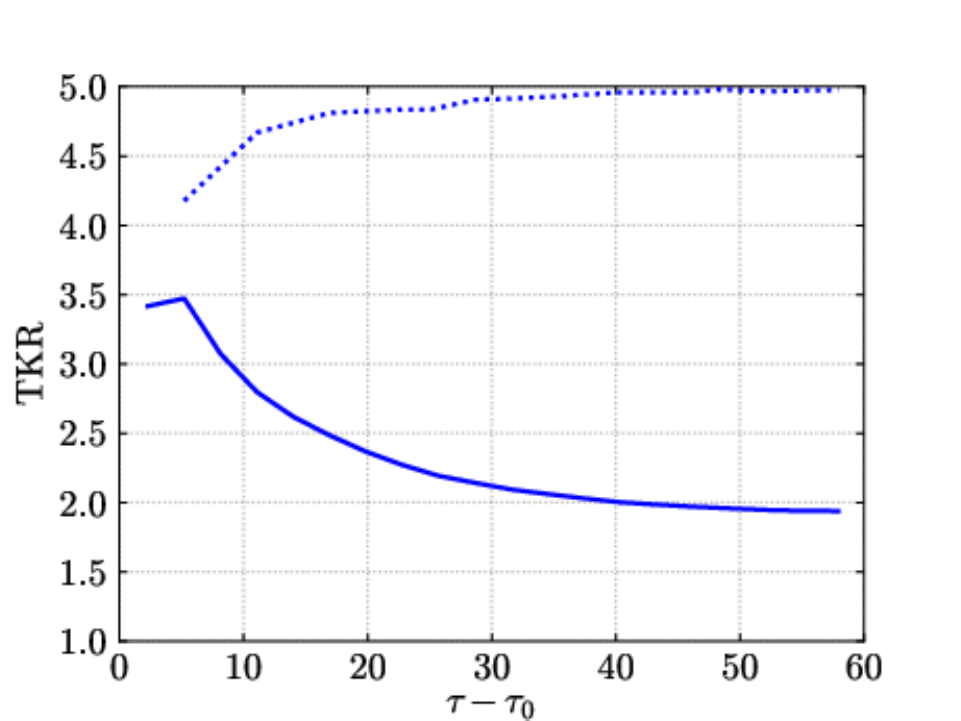}
		\caption{$m=-1$.}
	\end{subfigure}
	\begin{subfigure}{0.45\textwidth}
		\includegraphics[width=\textwidth]{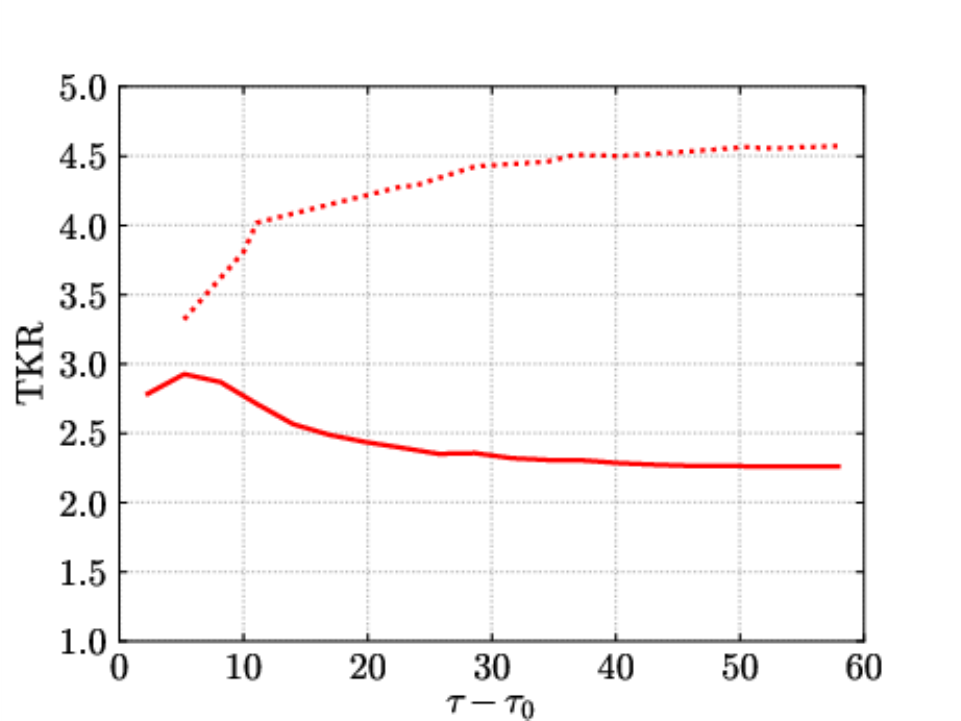}
		\caption{$m=-2$.}
	\end{subfigure}
	\begin{subfigure}{0.45\textwidth}
		\includegraphics[width=\textwidth]{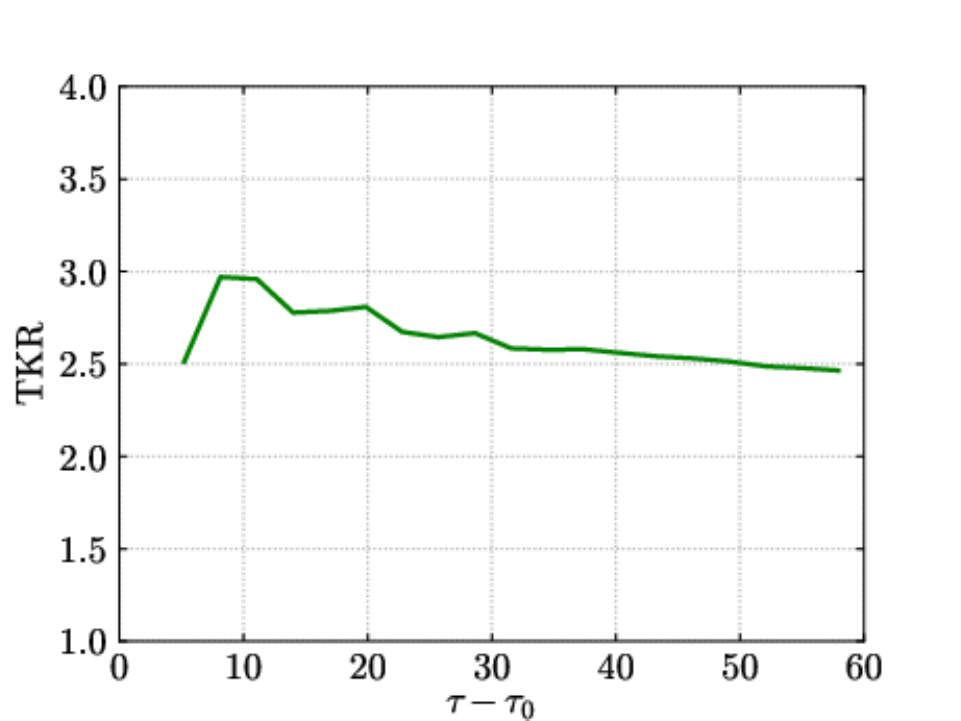}
		\caption{$m=-3$.}
	\end{subfigure}
	\begin{subfigure}{0.45\textwidth}
		\includegraphics[width=\textwidth]{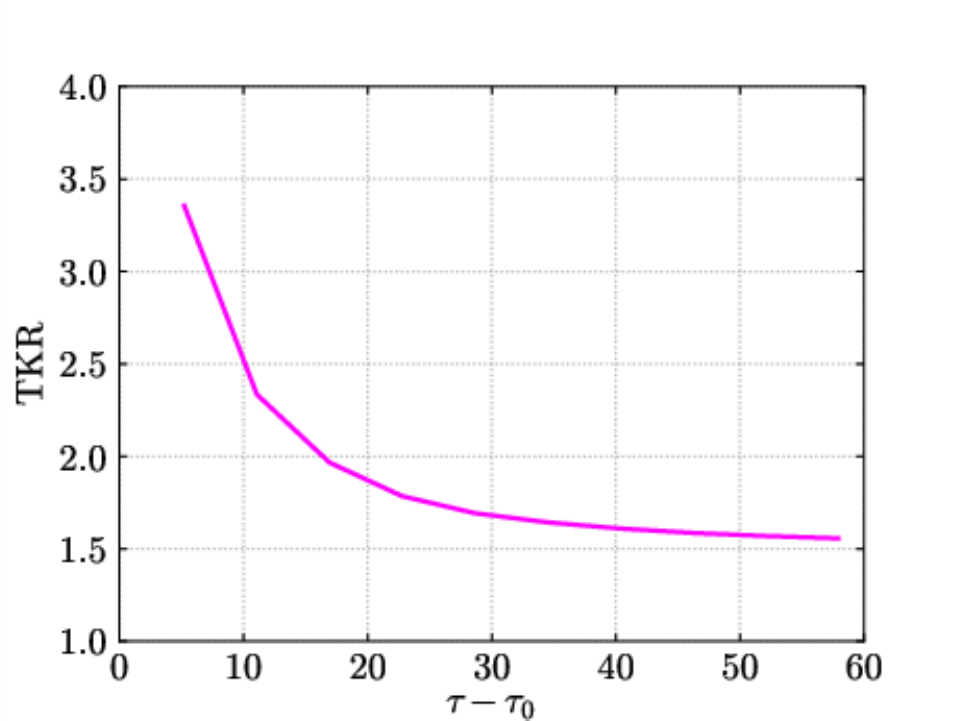}
		\caption{$m=0$ ($R=2$).}
	\end{subfigure}
	\caption{Temporal evolution of the global anisotropy measure, with each component integrated between the 5\% cutoff locations. Solid lines indicate ILES results and dotted lines indicate DNS results. \label{fig:TKR-tau}}
\end{figure}

Figure \ref{fig:TKR-tau} shows the temporal evolution of the global anisotropy measure TKR for each case. Compared to the equivalent figure 13 in \citet{Sewell2021} the peak in anisotropy at early time is less pronounced, however this is due to only integrating TKX, TKY and TKZ between the 5\% cutoff locations. Figure 10 in \citet{Groom2019} shows the same measure without this limit on the integration for a similar case, with the peak in anisotropy much closer to that observed in \citet{Sewell2021}. This indicates that much of the anisotropy observed at very early times is due to the shock wave. At an equivalent dimensionless time to the latest time simulated here, the anisotropy ratio presented in \citet{Sewell2021} is approximately 2 for the high-amplitude experiments and 3 for the low-amplitude experiments. For the $m=-3$ perturbation that most closely matches those experiments the TKR at the latest time is 2.46, while for the other ILES cases the late-time TKR decreases as $m$ increases. For the $m=0$ narrowband case the late-time value is 1.55, which is within the range of 1.49-1.66 observed across codes on the $\theta$-group quarter-scale case \citep{Thornber2017}; a case which is essentially the same perturbation but at a lower Atwood number. For the DNS cases a very different trend is observed where the anisotropy continually grows as time progresses. This is due to the very low Reynolds numbers of these simulations, with the lack of turbulence preventing energy from being transferred to the transverse directions. 

\begin{figure}
	\centering
	\begin{subfigure}{0.45\textwidth}
		\includegraphics[width=\textwidth]{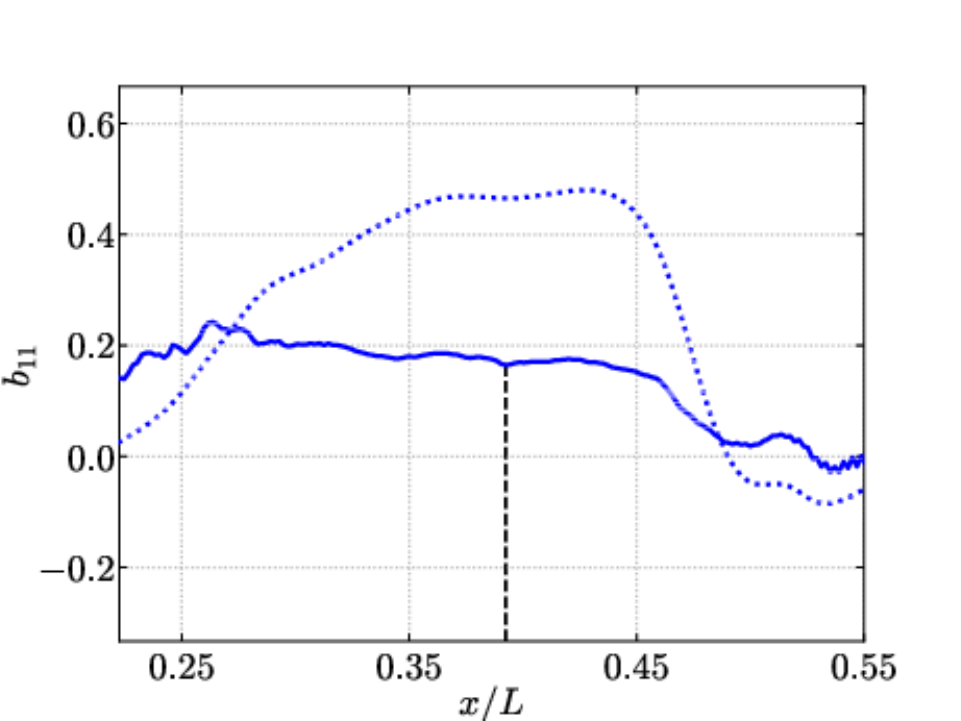}
		\caption{$m=-1$.}
	\end{subfigure}
	\begin{subfigure}{0.45\textwidth}
		\includegraphics[width=\textwidth]{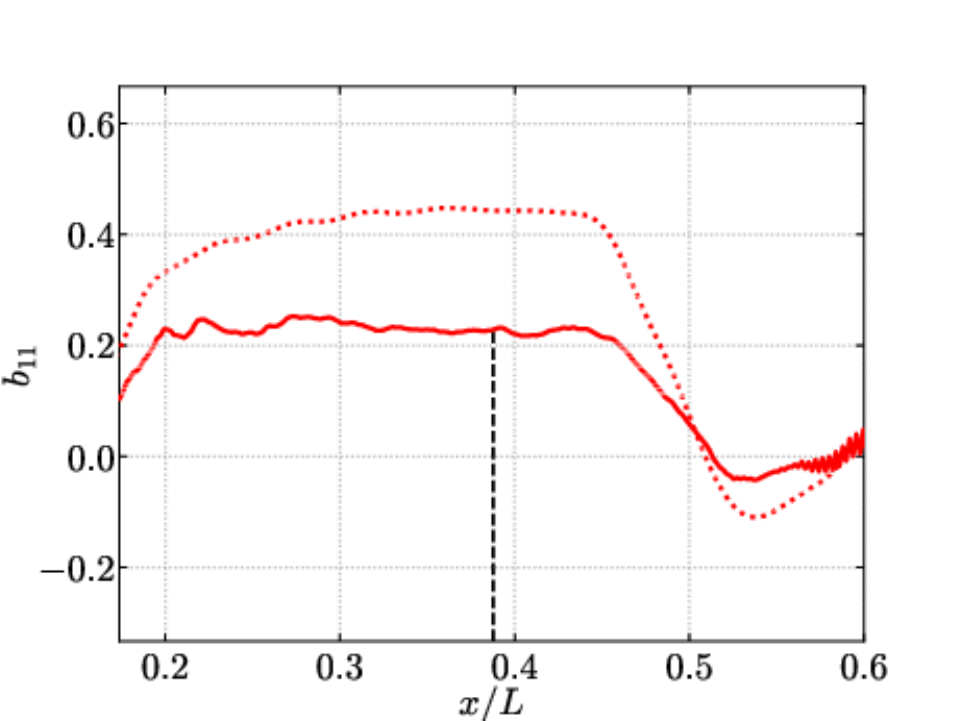}
		\caption{$m=-2$.}
	\end{subfigure}
	\begin{subfigure}{0.45\textwidth}
		\includegraphics[width=\textwidth]{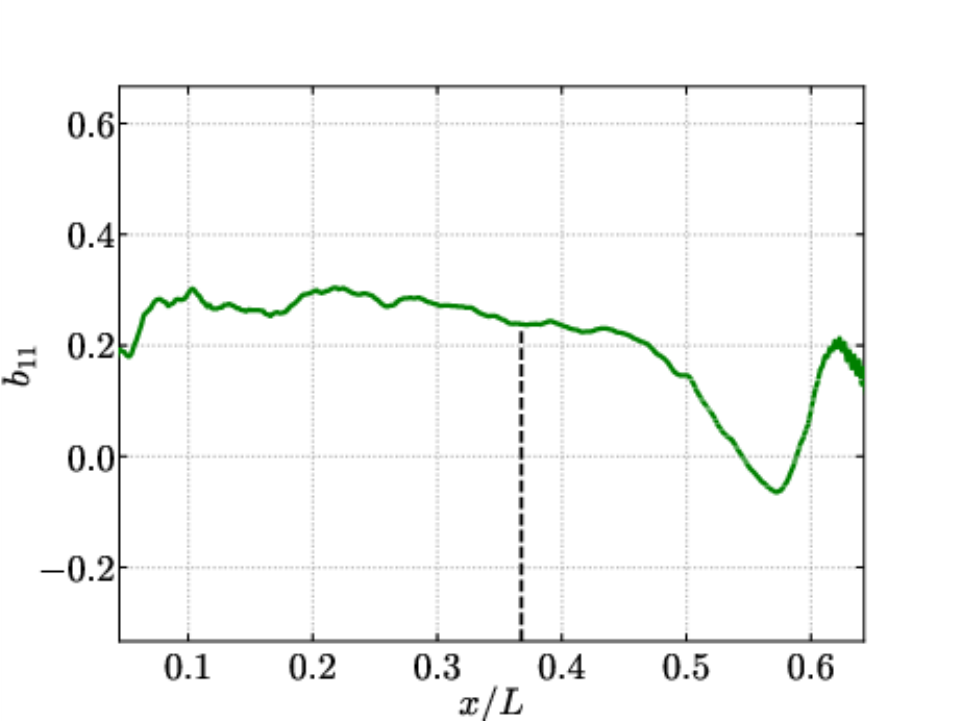}
		\caption{$m=-3$.}
	\end{subfigure}
	\begin{subfigure}{0.45\textwidth}
		\includegraphics[width=\textwidth]{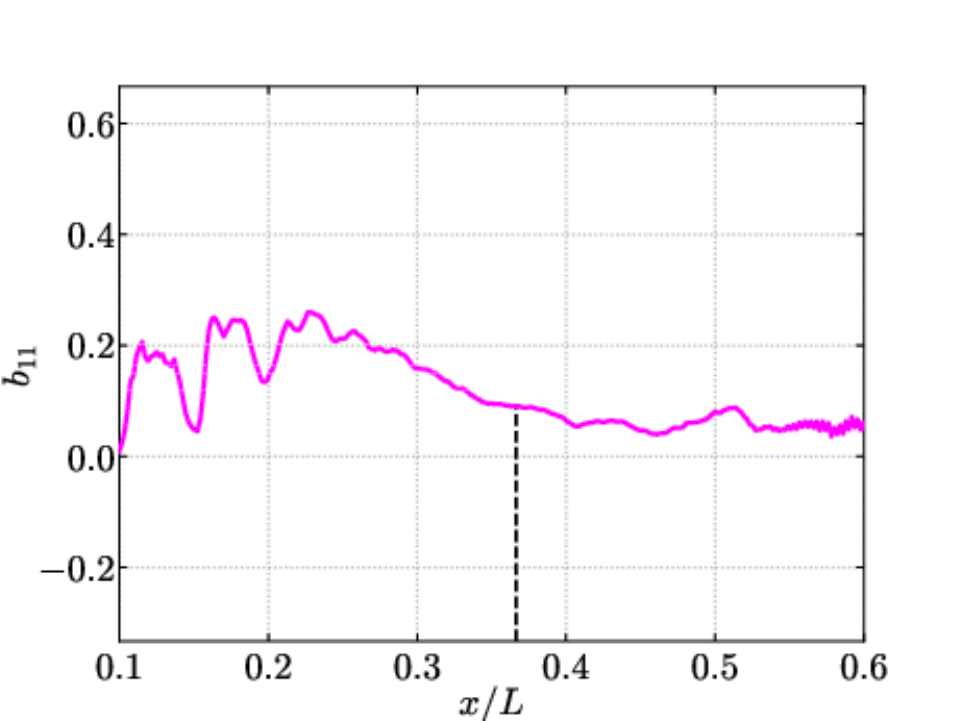}
		\caption{$m=0$ ($R=2$).}
	\end{subfigure}
	\caption{Spatial distribution of the $x$-direction principal component of the Reynolds stress anisotropy tensor at time $\tau=57.4$. Solid lines indicate ILES results and dotted lines indicate DNS results. Also shown is the mixing layer centre defined by the TKE centroid (black dashed lines). \label{fig:Bii-x}}
\end{figure}

The spatial variation in anisotropy is shown in figure \ref{fig:Bii-x}, plotted between the 5\% cutoff locations for each case. For the broadband cases the anisotropy is slightly higher on the spike side of the layer, with the greatest increase in the $m=-3$ case. This mirrors the results shown in \citet{Sewell2021} for $\boldsymbol{b}_{11}$, with quite good agreement observed between the $m=-3$ case at the latest time and the low-amplitude experiments just prior to reshock. In the narrowband case the increase in anisotropy from the mixing layer centre to the spike side is greater but the overall magnitude of $\boldsymbol{b}_{11}$ is lower, consistent with what was observed for TKR. The DNS results show that the biggest increase in anisotropy at low Reynolds numbers is in the centre of the mixing layer; there is a smaller difference in anisotropy between the DNS and ILES cases at either edge.

\begin{figure}
	\centering
	\begin{subfigure}{0.45\textwidth}
		\includegraphics[width=\textwidth]{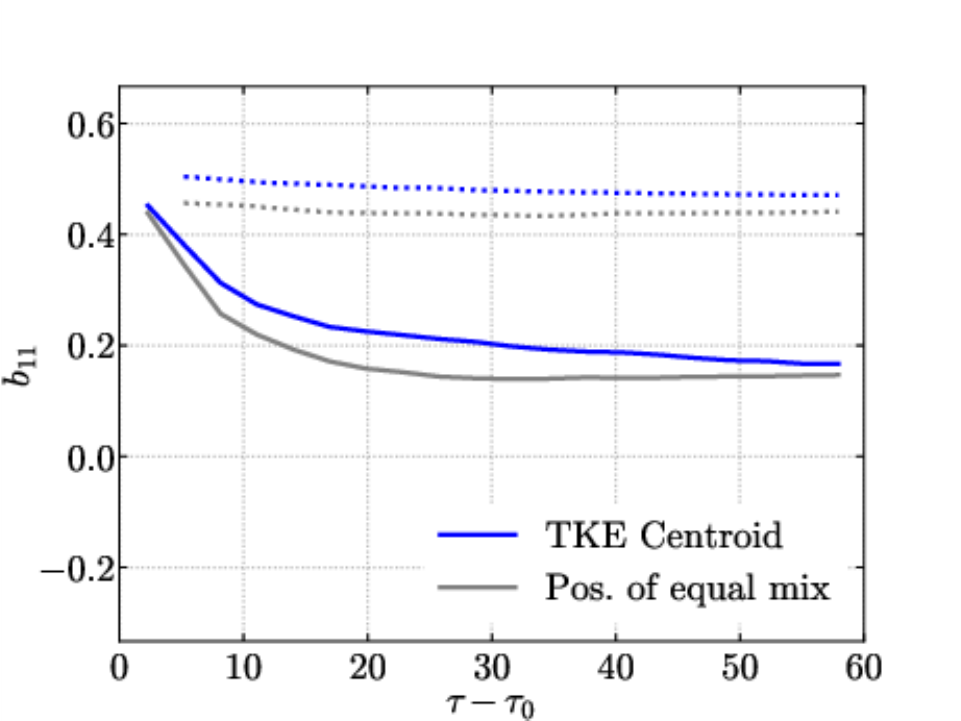}
		\caption{$m=-1$.}
	\end{subfigure}
	\begin{subfigure}{0.45\textwidth}
		\includegraphics[width=\textwidth]{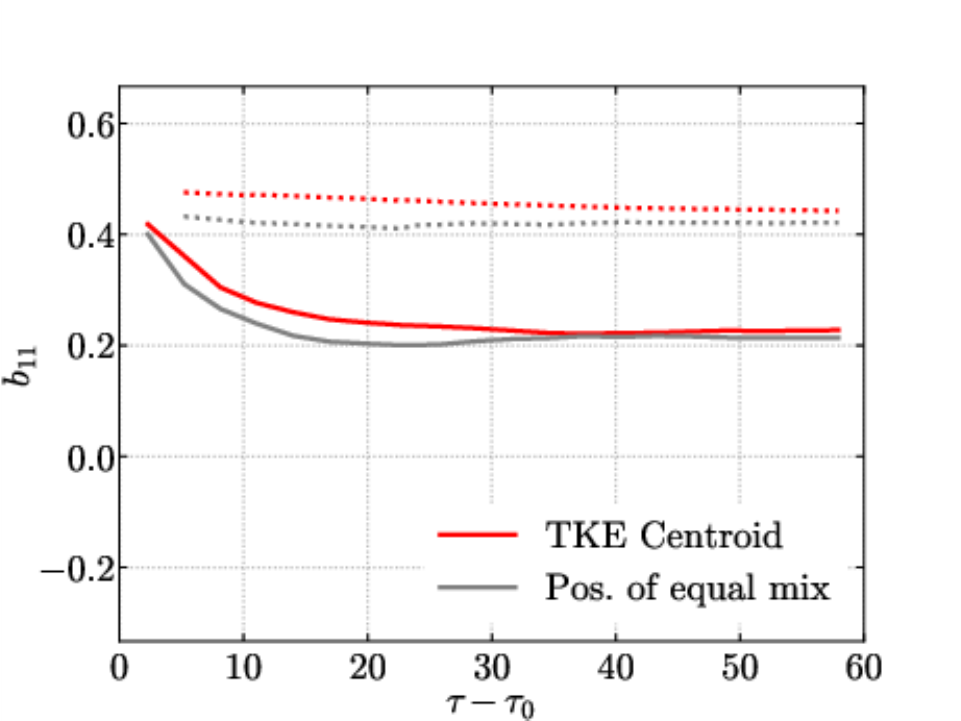}
		\caption{$m=-2$.}
	\end{subfigure}
	\begin{subfigure}{0.45\textwidth}
		\includegraphics[width=\textwidth]{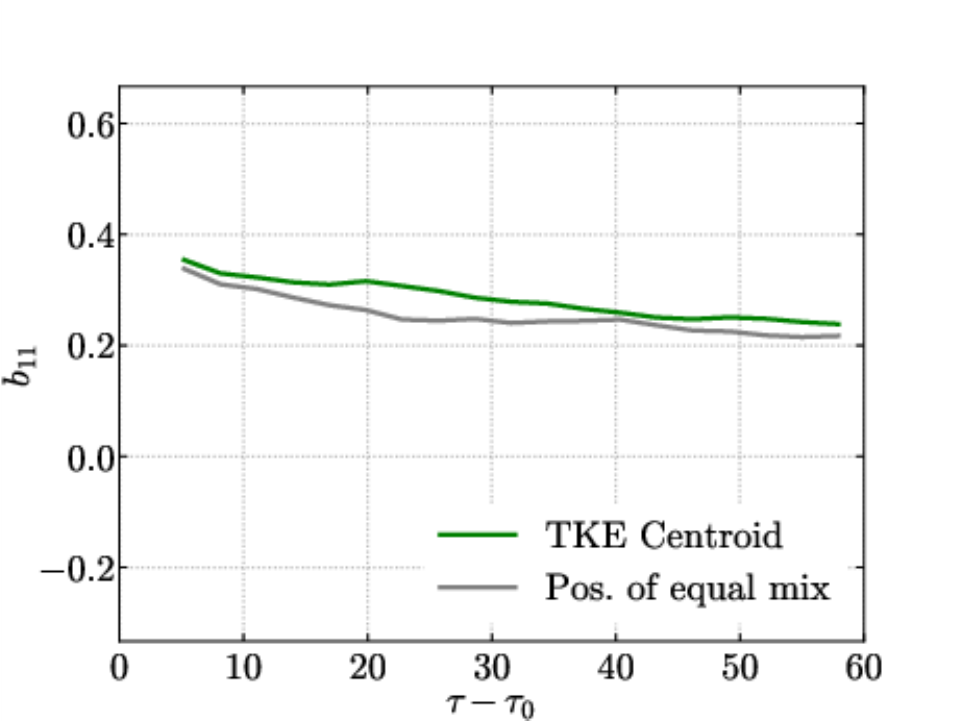}
		\caption{$m=-3$.}
	\end{subfigure}
	\begin{subfigure}{0.45\textwidth}
		\includegraphics[width=\textwidth]{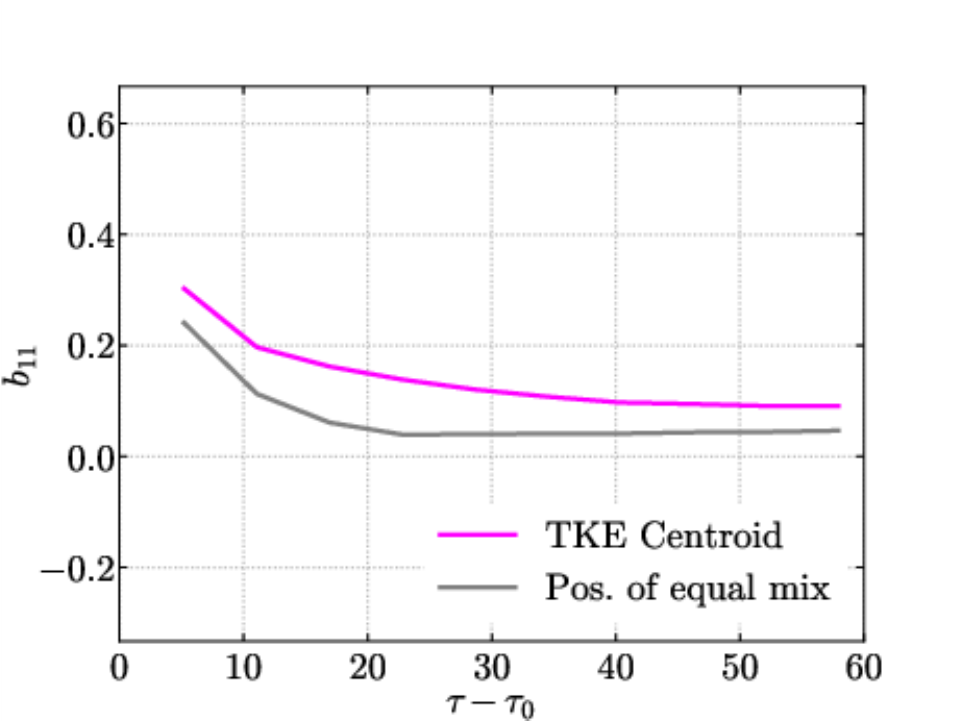}
		\caption{$m=0$ ($R=2$).}
	\end{subfigure}
	\caption{Temporal evolution of $x$-direction principal component of the Reynolds stress anisotropy tensor at the mixing layer centre plane. Solid lines indicate ILES results and dotted lines indicate DNS results. \label{fig:B11-tau}}
\end{figure}

Figure \ref{fig:B11-tau} shows the temporal evolution of $\boldsymbol{b}_{11}$ at the mixing layer centre, both for the definition of $x_c$ in terms of the TKE centroid (shown in figure \ref{fig:Bii-x}) as well as the alternate definition in terms of the position of equal mixed volumes. The results for both definitions are similar across all cases, with the anisotropy at the position of equal mix being slightly lower in all cases. In the DNS cases $\boldsymbol{b}_{11}$ is approximately constant in time, indicating that the growth in anisotropy that was observed for TKR in figure \ref{fig:TKR-tau} is occurring \firstrev{on} either side of the mixing layer centre. \thirdrev{The range of values are also comparable to those given in \citet{Wong2019} prior to reshock.}

\subsection{Spectra}
\label{subsec:spectra}

The distribution of fluctuating kinetic energy per unit mass across the different scales of motion is examined using radial power spectra of the transverse and normal components, calculated as
\begin{equation}
	E_i(\kappa)=\widehat{u^\prime_i}^\dagger\widehat{u^\prime_i}
\end{equation}
where $\kappa=\sqrt{\kappa_y+\kappa_z}$ is the (dimensionless) radial wavenumber in the $y$-$z$ plane at $x=x_c$ (given by the $x$-location of equal mixed volumes), $\widehat{(\ldots)}$ denotes the 2D Fourier transform taken over this plane and $\widehat{(\ldots)}^\dagger$ is the complex conjugate of this transform. As isotropy is expected in the transverse directions, the $E_y(\kappa)$ and $E_z(\kappa)$ spectra are averaged to give a single transverse spectrum $E_{yz}(\kappa)$. 

\begin{figure}
	\centering
	\begin{subfigure}{0.45\textwidth}
		\includegraphics[width=\textwidth]{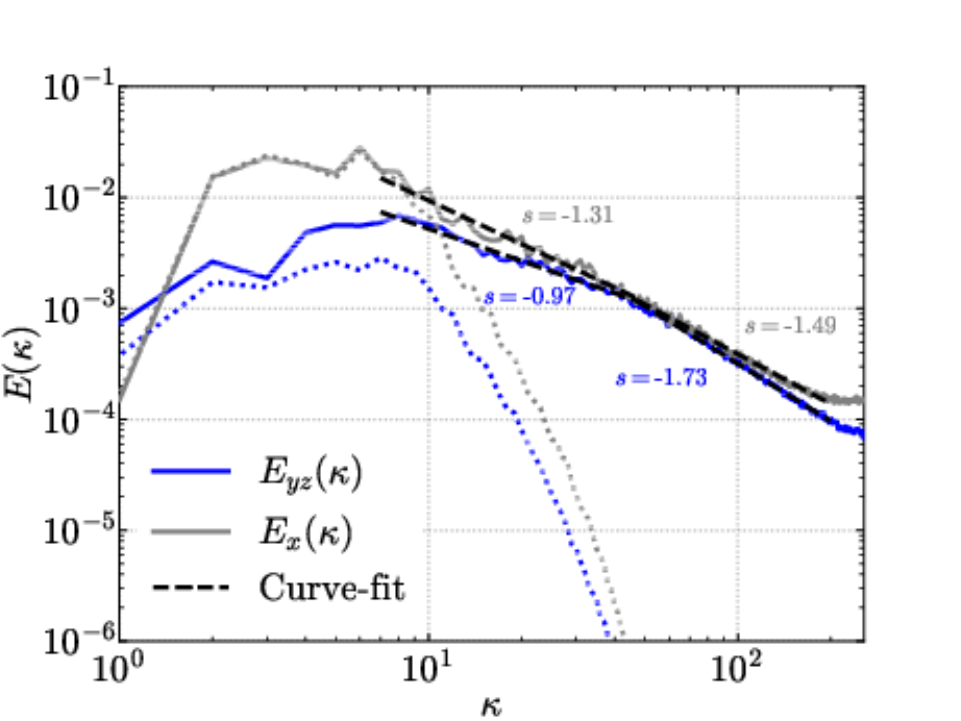}
		\caption{$m=-1$.}
	\end{subfigure}
	\begin{subfigure}{0.45\textwidth}
		\includegraphics[width=\textwidth]{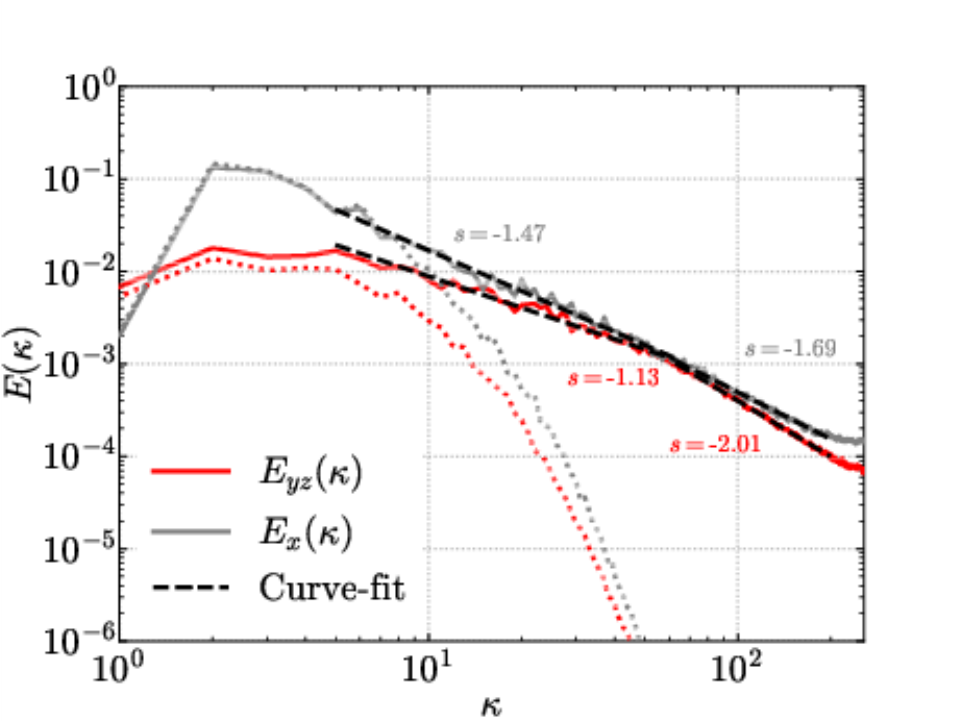}
		\caption{$m=-2$.}
	\end{subfigure}
	\begin{subfigure}{0.45\textwidth}
		\includegraphics[width=\textwidth]{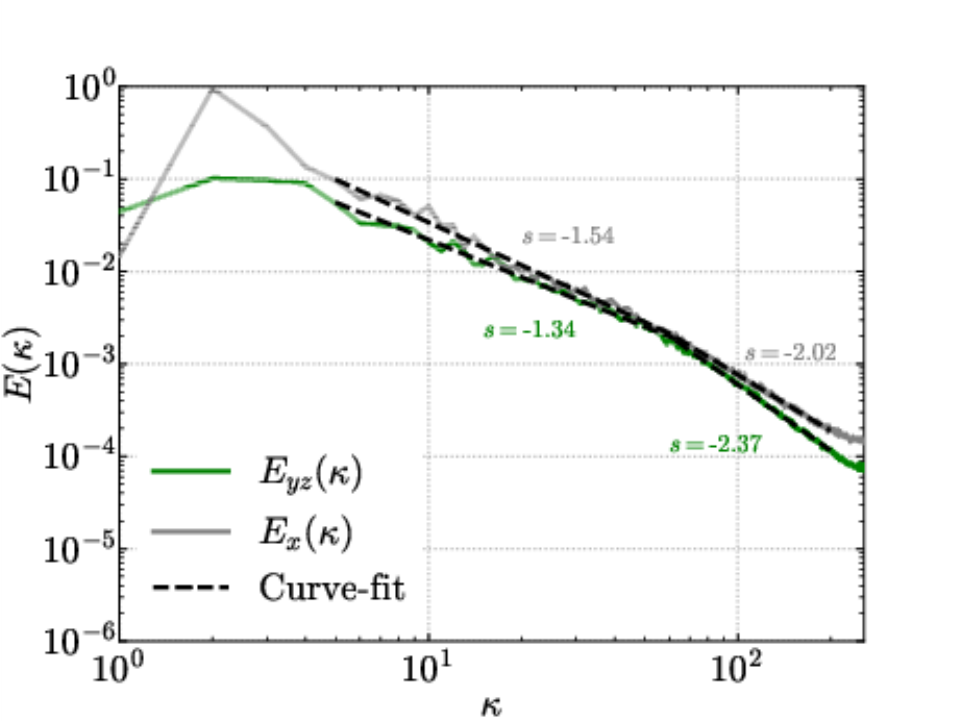}
		\caption{$m=-3$.}
	\end{subfigure}
	\begin{subfigure}{0.45\textwidth}
		\includegraphics[width=\textwidth]{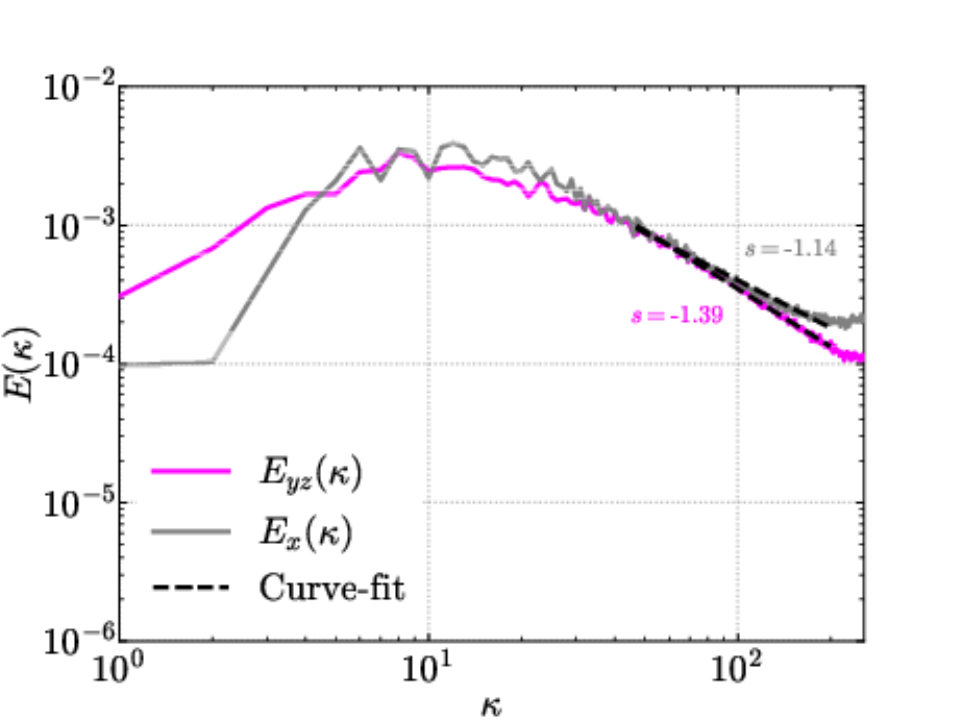}
		\caption{$m=0$ ($R=2$).}
	\end{subfigure}
	\caption{Transverse and normal components of fluctuating kinetic energy per unit mass at the mixing layer centre plane at time $\tau=57.4$. Solid lines indicate ILES results and dotted lines indicate DNS results. \label{fig:KEM-k}}
\end{figure}

The normal and transverse spectra are shown in figure \ref{fig:KEM-k} for each of the ILES and DNS cases at the latest simulated time. Curve-fits are made to the data to determine the scaling of each spectrum, with some interesting trends observed. For broadband cases evolving from perturbations of the form given in (\ref{eqn:P-k}), a scaling of $E(\kappa)\sim\kappa^{(m+2)/2}$ is expected for the low wavenumbers at early time while the growth of the mixing layer is being dominated by the just-saturating mode \citep{Groom2020}. This is not observed in figure \ref{fig:KEM-k} since saturation of the longest wavelength occurs quite early relative to the end time of the simulations, however some lingering effects can still be seen at the lowest wavenumbers. For all three broadband ILES cases there are two distinct ranges in both the normal and transverse spectra, which approximately correspond to wavenumbers lower and higher than $\kappa_{max}=k_{max}(L/2\upi)=32$. \citet{Thornber2010} modified the analysis of \citet{Zhou2001} to take into account the effects of the initial perturbation spectrum, resulting in an expected scaling for broadband perturbations of the form $E(\kappa)\sim\kappa^{(m-6)/4}$. This scaling is observed for the transverse spectra at wavenumbers greater than $\kappa_{max}$, while for the normal spectra a scaling of $E(\kappa)\sim\kappa^{(m-5)/4}$ is observed, the reason for which is currently unclear. 

For wavenumbers less than $\kappa_{max}$ the normal spectra scale as $\kappa^{-3/2}$ in the $m=-2$ and $m=-3$ cases, which is in good agreement with previous calculations for narrowband perturbations \citep{Thornber2016,Groom2019}. The narrowband case presented here has a slightly less steep scaling for both the normal and transverse spectra, although it has not been run to as late of a dimensionless time as in previous studies such as \citet{Thornber2017}. The normal spectrum in the $m=-1$ case also has a scaling that is less steep than $\kappa^{-3/2}$. A possible explanation for this is that saturation occurs a lot later in this case than the other broadband cases and therefore it may still be transitioning between \firstrev{an} $E(\kappa)\sim\kappa^{(m+2)/2}$ and a $\kappa^{-3/2}$ scaling. For the transverse spectra in each of the broadband cases at wavenumbers less than $\kappa_{max}$ a similar trend is observed, with each spectrum having a scaling that is shallower than $\kappa^{-3/2}$. The same argument of transition between \firstrev{an} $E(\kappa)\sim\kappa^{(m+2)/2}$ and a $\kappa^{-3/2}$ scaling may also be applied here, however simulations to later time would be required to confirm this. 

Finally, for the DNS cases no inertial range is observed due to the low Reynolds numbers that are simulated. For the normal spectra there is quite good agreement between the DNS and ILES data in the energy-containing scales at low wavenumbers. The transverse spectra contain less energy at these wavenumbers in the DNS cases due to suppression of secondary instabilities that transfer energy from the normal to transverse directions. \citet{Sewell2021} did not observe an inertial range in their TKE spectra prior to reshock, however they noted that there is likely some attenuation of the spectra at scales smaller than the effective window size of their PIV method, which is equivalent to a dimensionless wavenumber of $\kappa=47$. This makes it difficult to compare and verify the current findings with their existing experimental setup.

\subsection{Turbulent Length Scales and Reynolds Numbers}
\label{subsec:length-scales}
\begin{figure}
	\centering
	\begin{subfigure}{0.45\textwidth}
		\includegraphics[width=\textwidth]{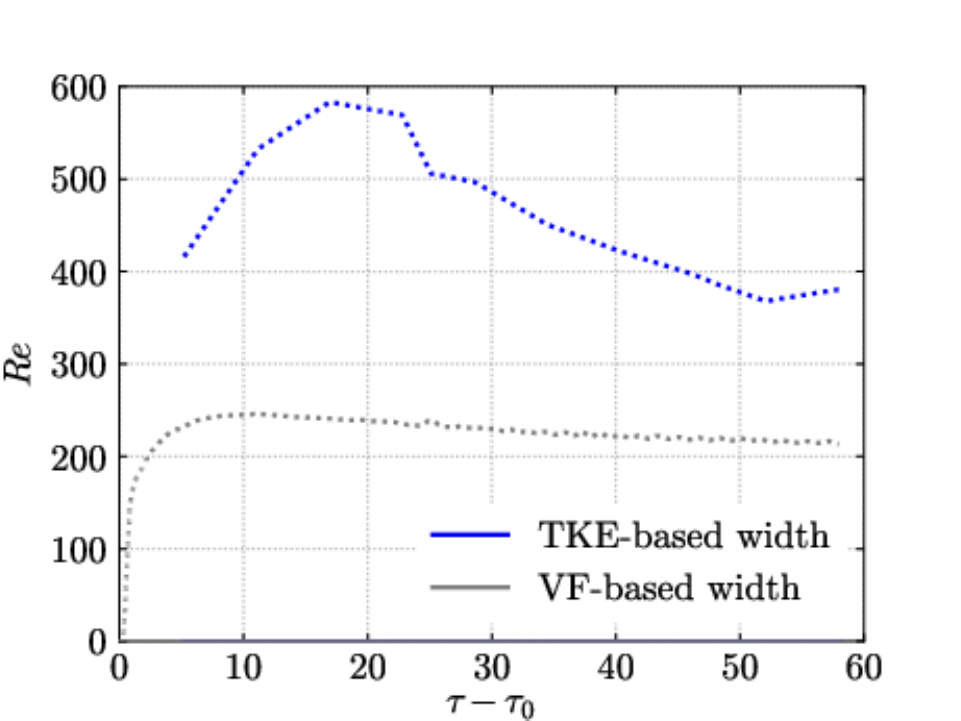}
		\caption{$m=-1$.}
	\end{subfigure}
	\begin{subfigure}{0.45\textwidth}
		\includegraphics[width=\textwidth]{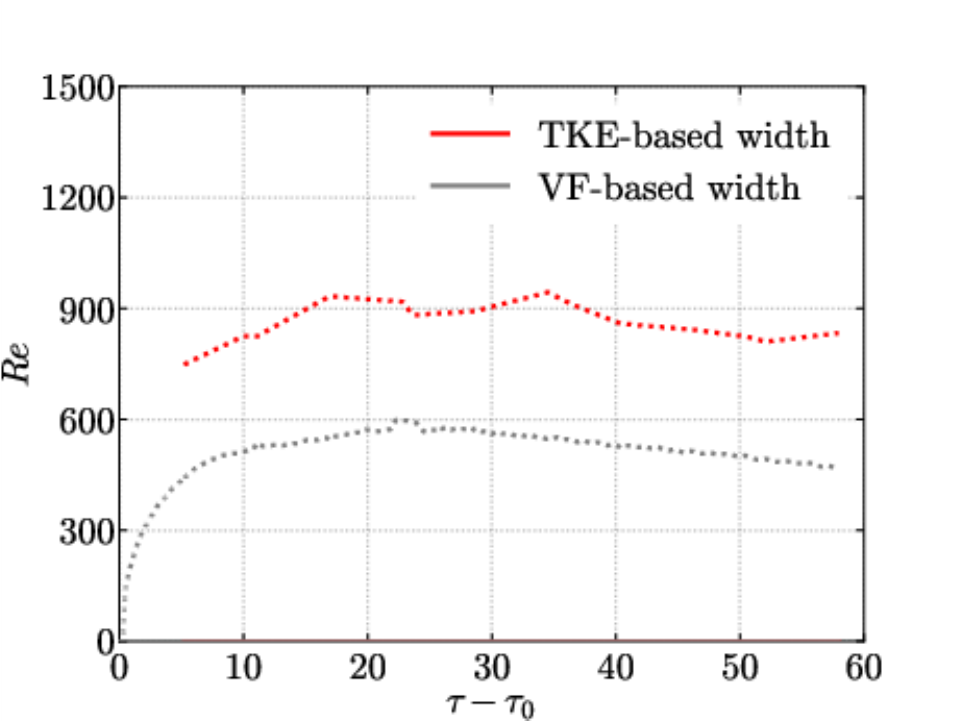}
		\caption{$m=-2$.}
	\end{subfigure}
	\caption{Outer-scale Reynolds numbers vs. time. \label{fig:Re-tau}}
\end{figure}

In order to give a better indication of how the present set of results compare with the experiments of \citet{Jacobs2013} and \citet{Sewell2021}, the outer-scale Reynolds numbers and key turbulent length scales used to evaluate whether a flow has transitioned to turbulence are computed using the DNS data. For the purposes of comparison, both the TKE-based and VF-based threshold widths are used as the outer length scale $h$ from which to compute the outer-scale Reynolds number as
\begin{equation}
\Rey_h=\frac{\overline{\rho^+} h \dot{h}}{\overline{\mu}}.
\label{eqn:outer-reynolds}
\end{equation}
Figure \ref{fig:Re-tau} shows the temporal variation for both definitions of the outer-scale Reynolds number. The outer-scale Reynolds numbers using the TKE-based definition for $h$ are roughly a factor of 2 larger, mostly due to the TKE-based width being a lot larger than the VF-based width in all cases, with neither definition close to reaching the critical value of $\Rey_h\gtrsim1$-$2\times10^4$ for fully developed turbulence \citep{Dimotakis2000}. For both the $m=-1$ and $m=-2$ perturbations the VF-based Reynolds number is approximately constant in time, consistent with the measured values of $\theta$ given in table \ref{tab:theta}.

\begin{figure}
	\centering
	\begin{subfigure}{0.45\textwidth}
		\includegraphics[width=\textwidth]{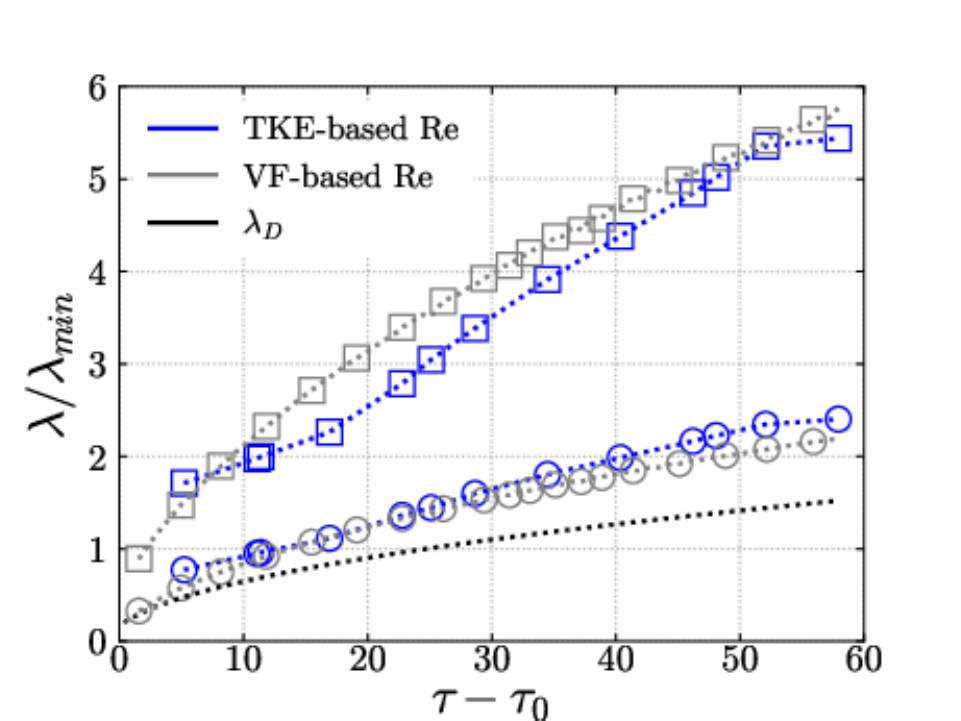}
		\caption{$m=-1$.}
	\end{subfigure}
	\begin{subfigure}{0.45\textwidth}
		\includegraphics[width=\textwidth]{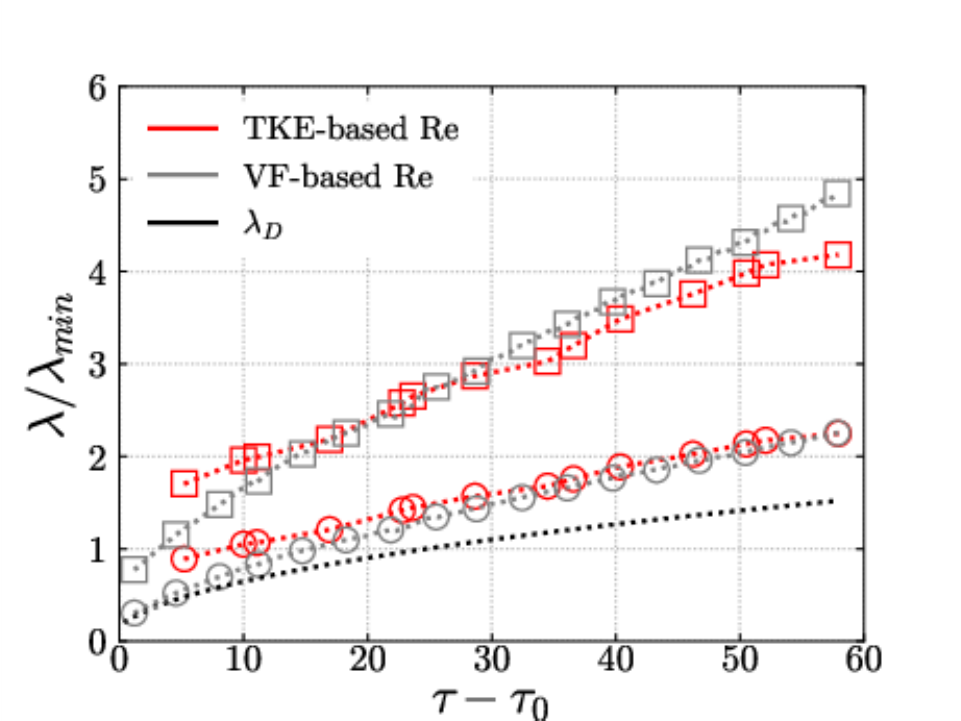}
		\caption{$m=-2$.}
	\end{subfigure}
	\caption{The Liepmann--Taylor (circles), inner-viscous (squares) and diffusion length scales vs. time for both definitions of the outer-scale Reynolds number. \label{fig:lambda-tau}}
\end{figure}

\citet{Dimotakis2000} showed that for stationary flows, fully developed turbulence is obtained when $\lambda_L/\lambda_V\ge1$ where $\lambda_L=5\lambda_T$ is the Liepmann--Taylor length scale and $\lambda_V=50\lambda_K$ is the inner-viscous length scale, with $\lambda_T$ and $\lambda_K$ the Taylor and Kolmogorov length scales respectively. These length scales may be related to the outer-scale Reynolds number by
\begin{eqnarray}
	\lambda_L & = 5 \Rey_h^{-1/2}h \\
	\lambda_V & = 50\Rey_h^{-3/4}h
	\label{eqn:turbulent-length-scales}
\end{eqnarray}
from which it can be shown that $\Rey_h\geq10^4$ for fully developed turbulence. For a time-dependent flow, \citet{Zhou2003b} showed that an additional length scale $\lambda_D = 5(\nu t)^{1/2}$ that characterises the growth rate of shear-generated vorticity must be considered, referred to as the diffusion layer scale. The condition for fully developed turbulence then becomes
\begin{equation}
	\textrm{min}(\lambda_L,\lambda_D)>\lambda_V.
\label{eqn:unsteady-mixing-transition}
\end{equation}
Figure \ref{fig:lambda-tau} shows the temporal variation of each length scale in (\ref{eqn:unsteady-mixing-transition}), with $\lambda_L$ and $\lambda_V$ calculated \firstrev{from} the outer-scale Reynolds number \firstrev{using both definitions for $h$}. In \thirdrev{both} cases there is good agreement between the length scales calculated from either definition of $\Rey_h$. The inner-viscous length scale is greater than the Liepmann--Taylor scale at all times in \thirdrev{both} cases, consistent with other observations in this paper on the lack of fully developed turbulence in the DNS cases at the Reynolds numbers capable of being simulated currently. 

\citet{Sewell2021} also observed $\lambda_L<\lambda_V$ at all times prior to reshock in their low-amplitude experiments. The authors note that, because of the different dependence of each length scale on $\Rey_h$, for $\theta\le0.5$ the flow can never transition to turbulence as $\lambda_V$ will grow faster than $\lambda_D$. Furthermore, the definition for $\lambda_D$ implies that it will be 0 at time $t=0$, which would seem to imply that an RMI-induced flow with $\theta\le0.5$ can never become turbulent. However, the virtual time origin is neglected in the original definition for $\lambda_D$; if it is included then this allows for the possibility that $\lambda_V < \lambda_D$ at early time. In that situation, transition to turbulence will occur provided the initial \firstrev{velocity jump} is strong enough to produce $\lambda_L>\lambda_V$ for some period of time. The turbulence will still be decaying over time if $\theta\le0.5$ though and will eventually no longer be fully developed, reflecting a fundamental difficulty to obtaining universal behaviour in experiments or numerical simulations of RMI.

\section{Conclusions}
\label{sec:conclusions}
This paper has presented simulations of an idealised shock tube experiment between air and sulphur hexafluoride that builds upon the previous results and analysis presented in \citet{Groom2020, Groom2021}. In particular, the effects of additional long wavelength modes in the initial perturbation were explored by comparing the results obtained using a narrowband surface perturbation (similar to the one presented in \citet{Groom2021}) and three broadband perturbations (similar to those presented in \citet{Groom2020}). Both implicit large-eddy simulations (ILES) of the high-Reynolds number limit as well as direct numerical simulations (DNS) at Reynolds numbers lower than those observed in the experiments were performed with the \texttt{Flamenco} finite-volume code. 

Various measures of the mixing layer width, based on both the plane-averaged turbulent kinetic energy and volume fraction profiles, were compared in order to explore the effects of initial conditions as well as the validity of using measurements based on the velocity field to draw conclusions about the concentration field (and vice versa) as is commonly done in experiments due to the difficulties of using diagnostics for both fields simultaneously. The effects of initial conditions on the growth rate exponent $\theta$ were analysed by curve-fitting the expected power law behaviour for the mixing layer width $h$ to two different definitions of $h$; one based on a threshold of 5\% of the peak turbulent kinetic energy (TKE) and the other based on 1\% and 99\% of the mean volume fraction (VF). A third method for estimating $\theta$ was also considered, based on the relationship between the total fluctuating kinetic energy decay rate $n$ and $\theta$ that is derived under the assumption that the mixing layer growth is self-similar. 

In general, estimates of $\theta$ using either definition for $h$ were found to be in good agreement with one another, particularly for the $m=-3$ broadband perturbation that is the most representative of the initial conditions used in the experiments of \citet{Sewell2021}. The estimates of $\theta$ based on $h$ for all three broadband cases were between 0.44 and 0.52, which is in very good agreement with the experimental estimates in \citet{Sewell2021}, who found $\theta=0.45\pm0.08$ for their low-amplitude cases and $\theta=0.51\pm0.04$ for their high-amplitude cases prior to reshock. When the TKE decay rate was used to estimate $\theta$ the results were generally close to the estimates based on $h$, indicating \firstrev{that} the mixing layer growth is close to self-similar \thirdrev{by the end of the simulation}. Comparing the ILES and DNS results also shows that there is only a small Reynolds number dependence, which is consistent with previous observations in \citet{Groom2019} that the integral quantities are mostly determined by the largest scales of motion. \thirdrev{When the mixing widths were decomposed into individual bubble and spike heights $h_b$ and $h_s$, it was found that $h_b\sim t^{\theta_b}$ and $h_s\sim t^{\theta_s}$ with $\theta_b\ne\theta_s$ at early time. However, it was shown that $\theta_b\approx\theta_s$ by the end of each simulation by examining the ratio of $h_s/h_b$ and showing this to be tending towards a constant at late time.}

The particular regime being analysed here is different to the self-similar growth regime analysed in \citet{Groom2020} as the current set of broadband perturbations have a much smaller bandwidth and therefore saturate quite early relative to the total simulation time. The present findings, which are supported by the experiments, are that while the growth rate  in the saturated regime is less sensitive to the specific power spectrum of the initial conditions, the effects of additional long wavelength modes are quite persistent over the duration of a typical shock tube experiment and give rise to growth rates much higher than for narrowband perturbations. 

Comparing $\theta$ for the two definitions of $h$ in the narrowband case also leads to some interesting observations. For the TKE-based mixing layer width the value of $\theta$ that is measured is almost a factor of two higher than the value that is measured for the VF-based width. This \secondrev{is} due to spikes that \secondrev{penetrate} further into the lighter fluid and in some cases are ejected from the main layer. These \secondrev{spikes} have been observed in previous studies of similar cases, such as \citet{Thornber2012,Youngs2020b}, \secondrev{and} are quite energetic but contain very little \secondrev{heavy} material. Therefore they affect the TKE-based \firstrev{width} much more than the VF-based width, \thirdrev{which can be seen in the greater relative difference between the two measures for the spike height $h_s$ than the bubble height $h_b$.} Presumably if such spikes are ejected at early time in the broadband cases then they get overtaken by the linear growth of the low wavelength modes; future work \secondrev{will} investigate this in further detail as it is potentially quite an important phenomenon for applications where multiple interfaces are located in close proximity to one another. Future work \secondrev{will} also aim to further quantify the effects of finite bandwidth on $\theta$ and other important integral quantities, see \citet{Soulard2022} for an initial discussion in this direction.

\thirdrev{Analysing the anisotropy of the fluctuating velocity field showed that the mixing layer is persistently anisotropic in the direction of the shock wave in all cases, in good agreement with previous experiments (prior to reshock) as well as numerical studies. For the broadband ILES cases, the energy spectra in both the normal and transverse directions showed two distinct scalings either side of the highest wavenumber $k_{max}$ in the initial perturbation and which were dependent on the specific initial condition. These scalings were also different for the normal vs. transverse energy spectrum in each case. This was also observed in the narrowband case but only for wavenumbers higher than $k_{max}$. Finally, calculations of outer-scale Reynolds numbers and turbulent length scales in the DNS cases showed that the outer-scale Reynolds numbers are approximately constant throughout the simulations, as expected from the estimates of $\theta\approx0.5$, and that good agreement was obtained between the turbulent length scales calculated using either the TKE-based or VF-based width as the outer length scale.}

\thirdrev{Overall the results of this study show that, in general, care needs to be taken when using measurements based on the velocity field to infer properties of the concentration field such as the growth rate $\theta$. This is particularly true when using thresholds rather than integral quantities to represent the mixing layer width. At early times (i.e. prior to reshock in a typical shock tube experiment) the mixing layer is not growing self-similarly, which makes it difficult to determine the value for the growth rate exponent $\theta$ as a single value may not even be appropriate. However, at the latest time simulated here (just prior to reshock in the experiments of \citet{Jacobs2013,Sewell2021}) the mixing layer is tending toward self-similarity and good agreement was able to be obtained with the experimental results across a wide range of quantities, providing additional insight on how to correctly interpret such results and when it is valid to use a single growth rate to describe the mixing layer.}

\backsection[Acknowledgements]{The authors would like to acknowledge David Youngs for providing useful advice and insight on the formulation of the initial conditions, as well as Jeffrey Jacobs for helpful discussions on the computational setup and interpreting experimental results. 

The authors would also like to acknowledge the computational resources at the National Computational Infrastructure provided through the National Computational Merit Allocation Scheme, as well as the Sydney Informatics Hub and the University of Sydney’s high performance computing cluster Artemis, which were employed for all cases presented
here.}

\backsection[Declaration of interests]{The authors report no conflict of interest.}

\backsection[Author ORCID]{M. Groom, https://orcid.org/0000-0003-2473-7229; B. Thornber, https://orcid.org/0000-0002-7665-089X}

\appendix

\section{Grid Convergence of Direct Numerical Simulations}\label{app:convergence}
\begin{figure}
	\centering
	\begin{subfigure}{0.45\textwidth}
		\includegraphics[width=\textwidth]{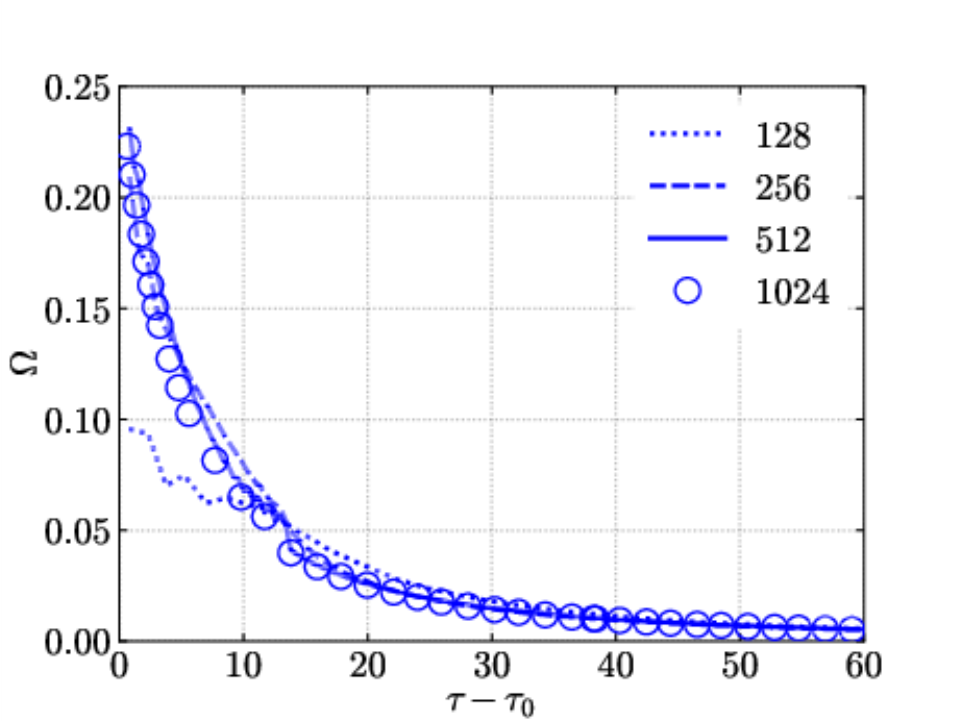}
		\caption{$Re_0=261$.}
	\end{subfigure}
	\begin{subfigure}{0.45\textwidth}
		\includegraphics[width=\textwidth]{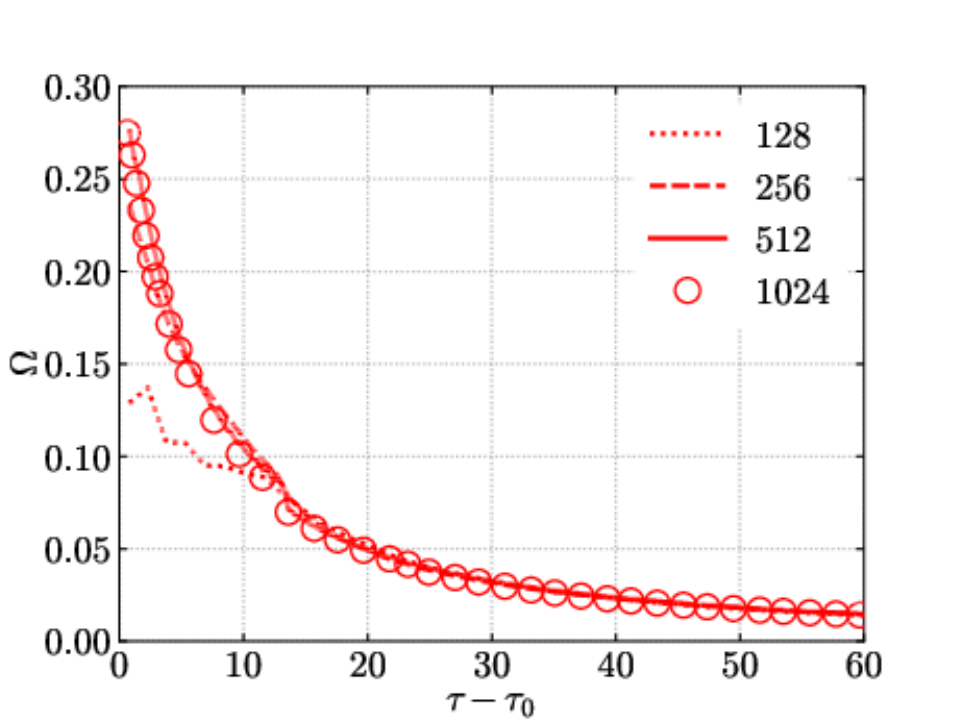}
		\caption{$Re_0=526$.}
	\end{subfigure}
	\caption{\thirdrev{Temporal evolution of domain integrated enstrophy for each grid resolution employed in the DNS cases.} \label{fig:omega-tau}}
\end{figure}

\begin{figure}
	\centering
	\begin{subfigure}{0.45\textwidth}
		\includegraphics[width=\textwidth]{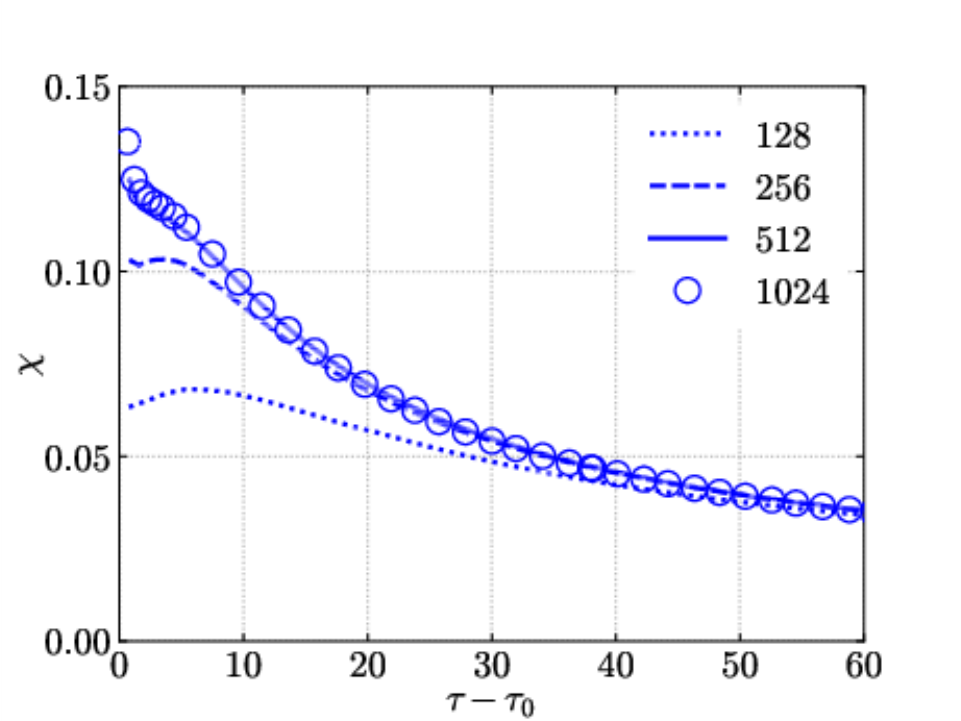}
		\caption{$Re_0=261$.}
	\end{subfigure}
	\begin{subfigure}{0.45\textwidth}
		\includegraphics[width=\textwidth]{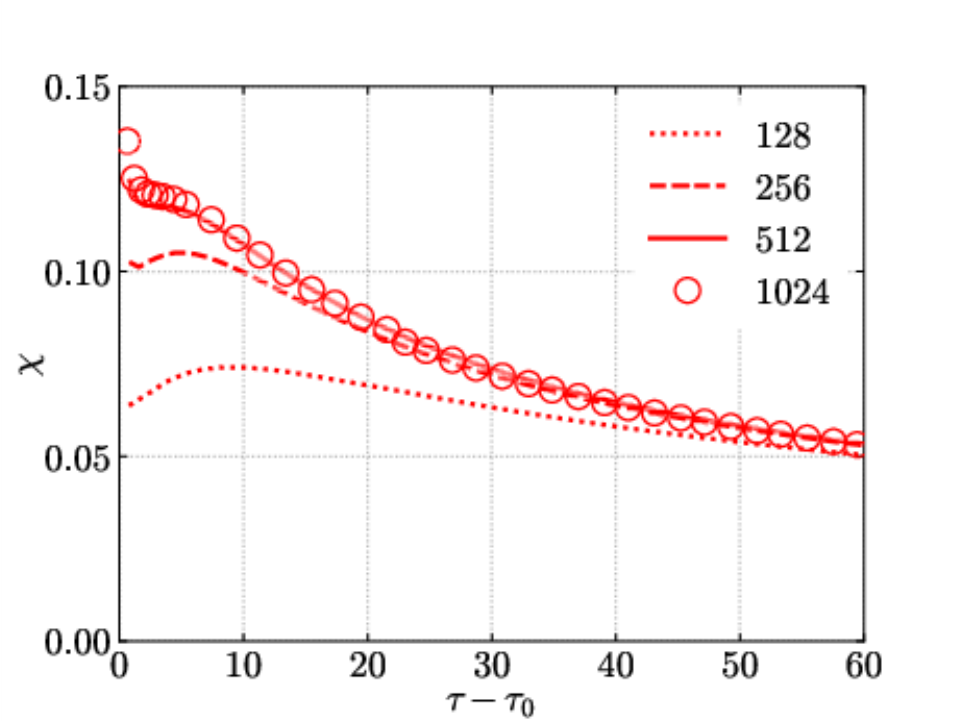}
		\caption{$Re_0=526$.}
	\end{subfigure}
	\caption{\thirdrev{Temporal evolution of domain integrated scalar dissipation rate for each grid resolution employed in the DNS cases.} \label{fig:chi-tau}}
\end{figure}

\thirdrev{Following the methodology presented in \citet{Olson2014} for demonstrating grid convergence, two quantities are used that depend on gradients of the velocity and concentration fields and which are therefore sensitive to the smallest scales in the flow. These are the domain-integrated enstrophy $\Omega$ and scalar dissipation rate $\chi$, given by}
\begin{equation}
\Omega(t)=\iiint \rho \omega_i\omega_i\:\mathrm{d} x\:\mathrm{d} y\:\mathrm{d} z
\end{equation}
\thirdrev{and}
\begin{equation}
	\chi(t)=\iiint D_{12} \frac{\partial Y_1}{\partial x_i}\frac{\partial Y_1}{\partial x_i}\:\mathrm{d} x\:\mathrm{d} y\:\mathrm{d} z
\end{equation}
\thirdrev{where $\omega_i$ is the vorticity in direction $i$ (summation over $i$ is implied). Figures \ref{fig:omega-tau} and \ref{fig:chi-tau} demonstrate grid convergence in the domain-integrated enstrophy and scalar dissipation rate for both DNS cases. Each case is shown to be suitably converged for both of these integral quantities at the finest grid resolution considered, even during the early-time period prior to the shock exiting the domain.} 

\newpage

\section{Integral Definitions of Bubble and Spike Heights}\label{app:integral}
\begin{figure}
	\centering
	\begin{subfigure}{0.32\textwidth}
		\includegraphics[width=\textwidth]{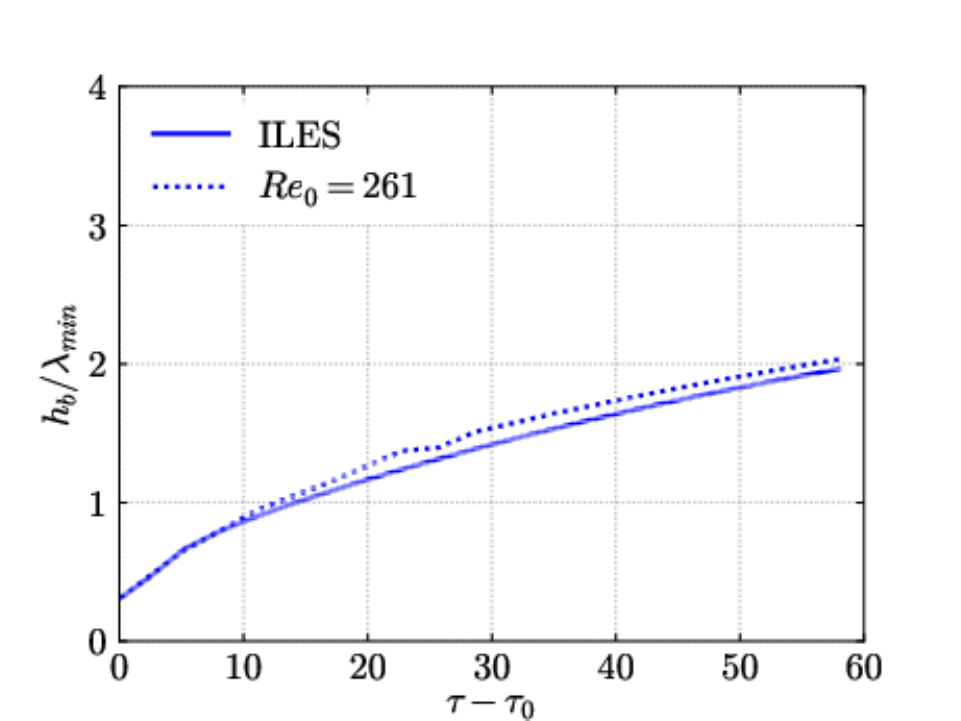}
		\caption{$m=-1$.}
	\end{subfigure}
	\begin{subfigure}{0.32\textwidth}
		\includegraphics[width=\textwidth]{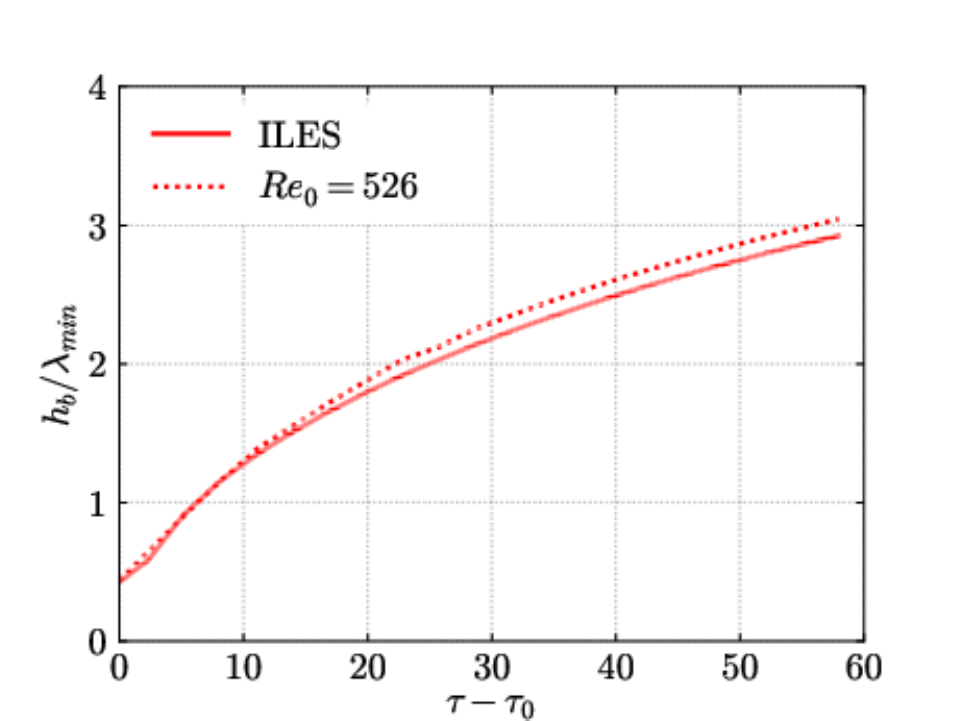}
		\caption{$m=-2$.}
	\end{subfigure}
		\begin{subfigure}{0.32\textwidth}
		\includegraphics[width=\textwidth]{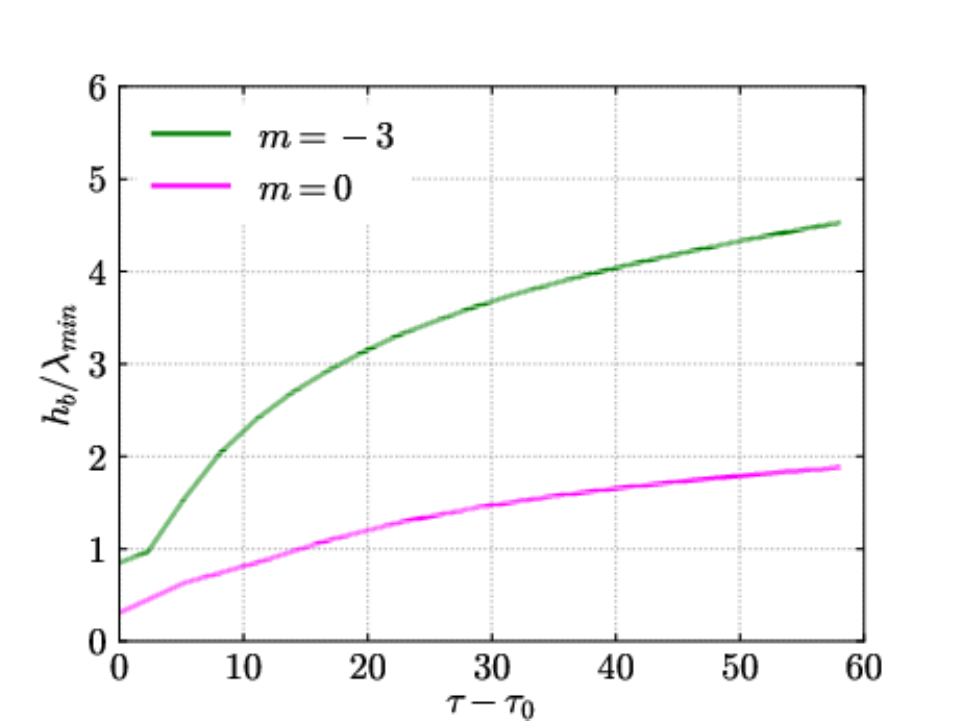}
		\caption{$m=-3$ and $m=0$.}
	\end{subfigure}
	\caption{\thirdrev{Temporal evolution of the bubble height $h_b$ based on the integral definitions of \cite{Youngs2020b}.} \label{fig:hb_int-tau}}
\end{figure}

\begin{figure}
	\centering
	\begin{subfigure}{0.32\textwidth}
		\includegraphics[width=\textwidth]{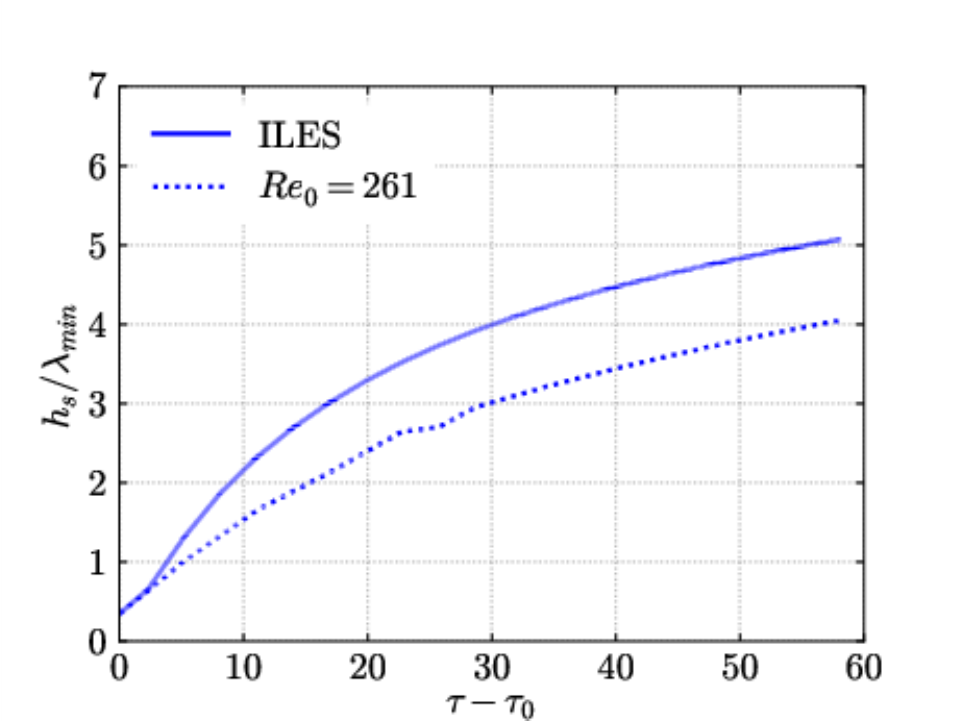}
		\caption{$m=-1$.}
	\end{subfigure}
	\begin{subfigure}{0.32\textwidth}
		\includegraphics[width=\textwidth]{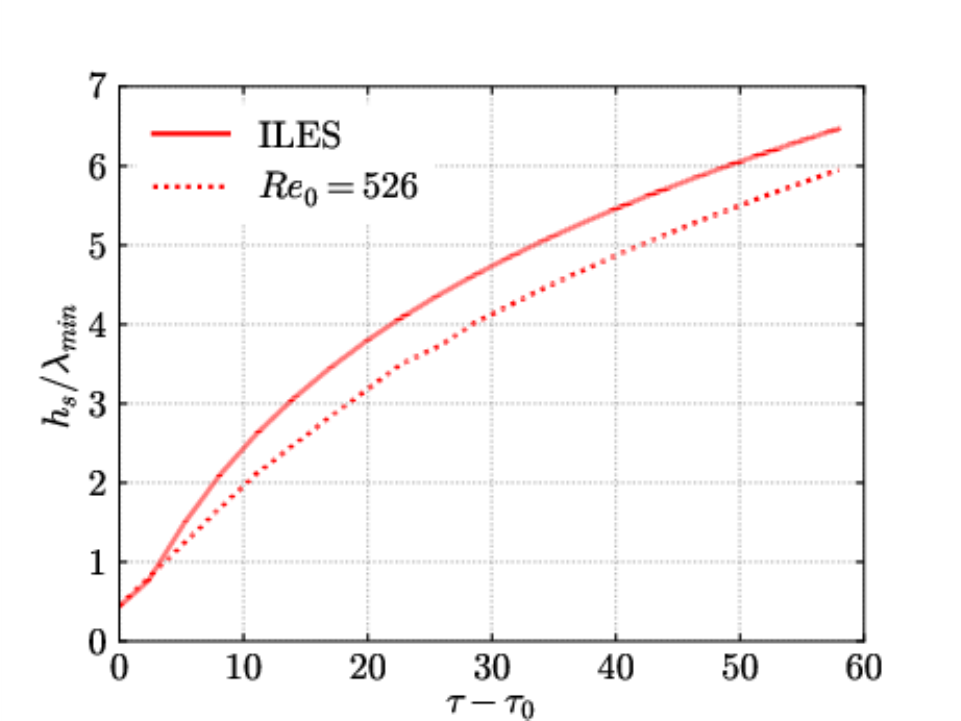}
		\caption{$m=-2$.}
	\end{subfigure}
		\begin{subfigure}{0.32\textwidth}
		\includegraphics[width=\textwidth]{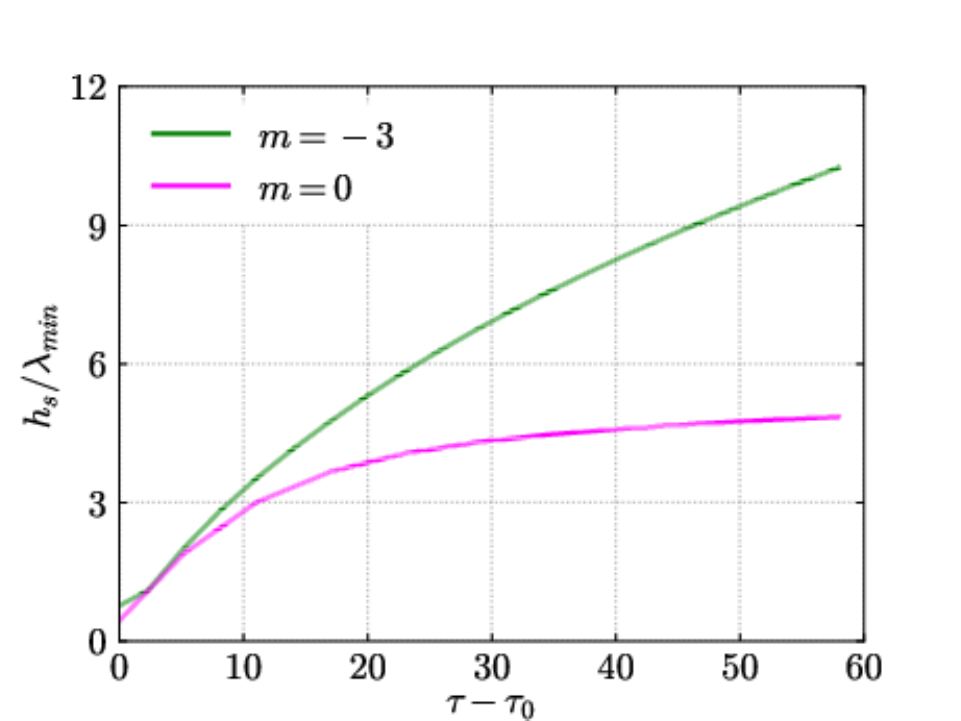}
		\caption{$m=-3$ and $m=0$.}
	\end{subfigure}
	\caption{\thirdrev{Temporal evolution of the spike height $h_s$ based on the integral definitions of \cite{Youngs2020b}.} \label{fig:hs_int-tau}}
\end{figure}

\begin{figure}
	\centering
	\begin{subfigure}{0.32\textwidth}
		\includegraphics[width=\textwidth]{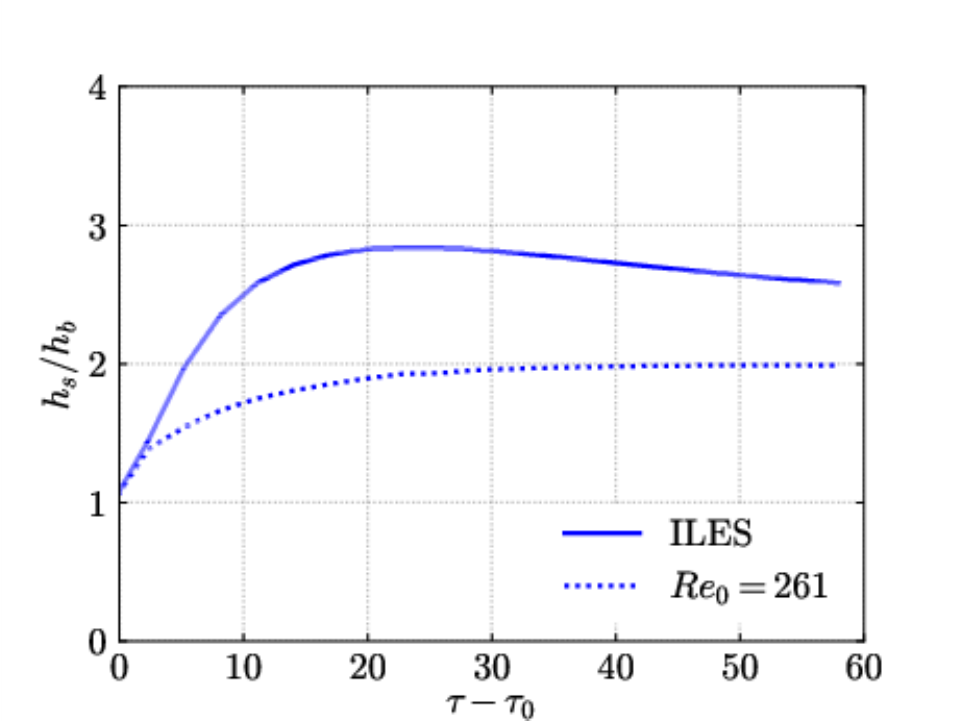}
		\caption{$m=-1$.}
	\end{subfigure}
	\begin{subfigure}{0.32\textwidth}
		\includegraphics[width=\textwidth]{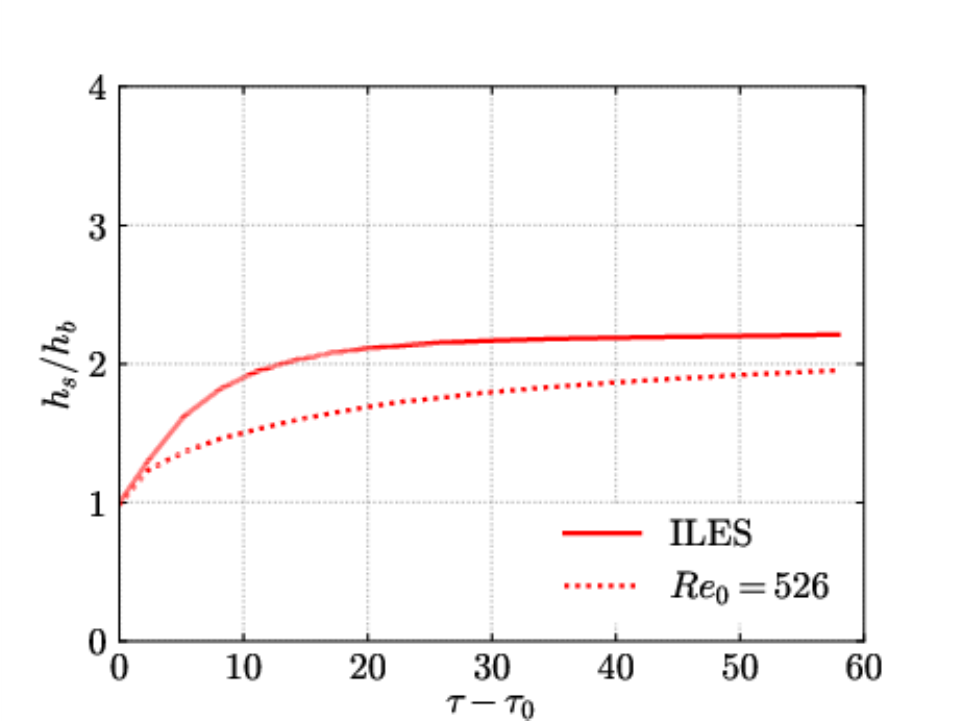}
		\caption{$m=-2$.}
	\end{subfigure}
		\begin{subfigure}{0.32\textwidth}
		\includegraphics[width=\textwidth]{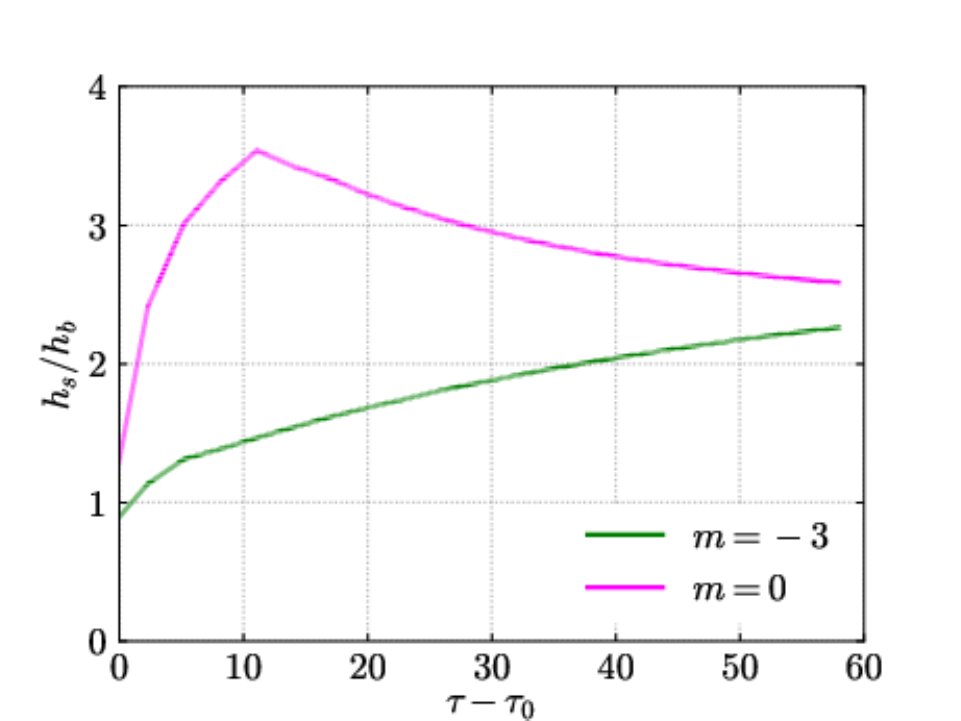}
		\caption{$m=-3$ and $m=0$.}
	\end{subfigure}
	\caption{\thirdrev{Temporal evolution of the ratio of spike to bubble heights based on the integral definitions of \cite{Youngs2020b}.} \label{fig:ratio_int-tau}}
\end{figure}

\thirdrev{In \citet{Youngs2020b} novel definitions were given for the bubble and spike heights $h_b$ and $h_s$ as weighted average distances from the mixing layer centre,}
\begin{subeqnarray}
	\centering
	h_s^{(p)} & = & \left[\frac{(p+1)(p+2)}{2}\frac{\int_{-\infty}^{x_c}|x|^p(1-\langle f_1\rangle)\:\mathrm dx}{\int_{-\infty}^{x_c}(1-\langle f_1\rangle)\:\mathrm dx} \right]^{1/p} \label{subeqn:hbhs1} \\
	h_b^{(p)} & = & \left[\frac{(p+1)(p+2)}{2}\frac{\int_{x_c}^\infty|x|^p\langle f_1\rangle\:\mathrm dx}{\int_{x_c}^\infty\langle f_1\rangle\:\mathrm dx} \right]^{1/p}. \label{subeqn:hbhs2}
    \label{eqn:hbhs}
\end{subeqnarray}
\thirdrev{Figures \ref{fig:hb_int-tau} and \ref{fig:hs_int-tau} plot the bubble and spike heights (with $p=3$), while figure \ref{fig:ratio_int-tau} plots their ratio $h_s/h_b$. The results are quite similar to the VF-based bubble and spike heights shown in figures \ref{fig:hb-tau} to \ref{fig:ratio-tau}, albeit smoother and therefore more suitable for estimating $\theta_b$ and $\theta_s$. While the main purpose of this paper is to compare the quantities typically measured in experiments based on thresholds of the TKE or VF profiles, it is recommended that future studies focus on using integral definitions such as the ones given here.}

\bibliographystyle{jfm}
\bibliography{bibliography}

\end{document}